\documentclass[11pt]{amsart}
\usepackage{amsmath}
\usepackage{amssymb}
\usepackage{palatino}

\def\Mgn{\mathfrak{M}_{g}^{(N)}}

\def\Mgnp{\mathfrak{M}_{g}^{(N),0}}
\def\Ecal{{\mathcal E}}
\def\Eb{\overline{\Ecal}}
\def\Zcal{{\mathcal Z}}
\def\RS{\CC}
\def\Hgk{{{\mathcal H}_g({\bf k}_{n+m})}}

\def\qd{Q}
\def\hs{\hskip0.7cm}
\def\cal{\mathcal}

\def\a{\alpha}
\def\g{\gamma}

\def\b{\beta}

\def\d{\delta}
\def\e{\varepsilon}
\def\ka{\kappa}
\def\l{\lambda}

\def\t{\tau}
\def\th{\vartheta}
\def\Th{\Theta}

\def\O{\Omega}

\def\ah{\hat{a}}
\def\bh{\hat{b}}
\def\wh{\hat{w}}
\def\Oh{\widehat{\Omega}}

\def\qh{\hat{q}}
\def\CP1{{\mathbb C}P^1}
\def\Pcal{{\cal P}}
\def\Sh{\widehat{S}}
\def\Thpq{\Th\left[^\pb_\qb\right]}

\def\C{{\mathbb C}}
\def\Z{{\mathbb Z}}

\def\la{\label}

\def\c{\cite}
\def\f{\frac}
\def\L{{\cal L}}
\def\p{\partial}
\def\pb{{\bf p}}
\def\qb{{\bf q}}

\def\rb{{\bf r}}
\def\sb{{\bf s}}
\def\zb{{\bf z}}
\def\bk{{\bf k}}
\def\tr{{\rm tr}}
\def\lh{\hat{\lambda}}
\def\det{{\rm det}}
\def\gh{\hat{g}}
\def\0{S}
\def\log{\ln}
\def\ol{\overline}
\def\M{{\cal M}}
\def\Lh{\hat{\L}}

\def\la{\label}

\def\c{\cite}
\def\f{\frac}
\def\L{{\mathcal L}}
\def\p{\partial}
\def\res{{\rm res}}

\def\tr{{\rm tr}}
\def\0{S}
\def\1{T}

\def\log{\ln}
\def\ol{\overline}

\def\sl{{\mathfrak{sl}}}

\def\Lh{{\widehat{{\cal L}}}}

\def\bar{\overline}
\def\det{{\rm det}}

\def\dif{Q}
\def\xh{\hat{x}}
\def\Mcal {{\mathcal M}}
\def\Mcomb{\overline \Mcal _{g,n}[{\bf p}]}
\def \eqref #1{ (\ref{#1})}

\newlength{\dinwidth}
\newlength{\dinmargin}
\def\Wfive{W_{5} }
\def\Wbdr{W_{1,1}}

\def \&{\hspace{-18pt}&}
\def\arg{{\rm arg}}

\newtheorem{theorem}{Theorem}[section]

\newtheorem{proposition}[theorem]{Proposition}
\newtheorem{corollary}[theorem]{Corollary}
\newtheorem{remark}[theorem]{Remark}

\def\nzh{{\widehat{m}}}

\def\be{\begin{equation}}
\def\ee{\end{equation}}
\def\ben{\begin{displaymath}}
\def\een{\end{displaymath}}
\def\baa{\begin{eqnarray}}
\def\eaa{\end{eqnarray}}

\def\ba{\begin{array}}
\def\ea{\end{array}}
\def\la{\label}
\def\p{\partial}
\def\nz{{\bf M}}

\def\Pcal{{\mathcal P}}
\def\Hcal{{\mathcal H}}
\def\Ccalh{\widehat{{\mathcal C}}}
\def\zetah{\widehat{\zeta}}
\def\rb{{\bf r}}
\def\sb{{\bf s}}

\def\ah{\hat{a}}
\def\bh{\hat{b}}

\def\Bh{\widehat{B}}
\def\Ch{\hat{C}}

\def\Eh{{\widehat{E}}}
\def\Kh{{\widehat{K}}}

\def\res{{\rm res}}
\def\Acalh{{\widehat{{\mathcal A}}}}
\makeatletter
\@addtoreset{equation}{section}
\makeatother

\def\gm{g_-}
\def\mo{{m_{odd}}}
\def\me{{m_{even}}}

\def\pb{{\bf p}}
\def\zh{\hat{z}}

\def\dh{\hat{d}}

\def\C{{\mathbb C}}
\def\R{{\mathbb R}}
\def\Z{{\mathbb Z}}
\def\Q{{\mathbb Q}}

\def\O{\Omega}

\def\Eb{\overline{{\mathbb E}}}

\def\Qcal{{\mathcal Q}}
\def\Mcal{{\mathcal M}}

\def\Mcal{{\mathcal M}}
\def\Mb{\overline{{\mathcal M}}}
\def\f{\frac}
\def\e{\epsilon} 
\def\d{\delta}
\def\a{\alpha}
\def\b{\beta}
\def\g{\gamma}
\def\s{\sigma}

\def\deg{{\rm deg}}
\def\qd{Q}
\def\CC{{\mathcal R}}

\def\Ch{\widehat{C}}
\def\Ccal{{\mathcal C}}

\def\gh{\hat{g}}

\def\ka{\kappa}
\def\wh{\hat{w}}

\def\Acal{{\mathcal A}}
\def\bk{{\bf k}}
\def\bl{{\bf l}}
\def\l{\lambda}

\def\Bh{\widehat{B}}

\def\tauh{\widehat{\tau}}
\def\Mgn{M^{(N)}_{g}}

\def\dim{{\rm dim}}

\def\be{\begin{equation}}
\def\ee{\end{equation}}
\def\ben{\begin{displaymath}}
\def\een{\end{displaymath}}
\def\baa{\begin{eqnarray}}
\def\eaa{\end{eqnarray}}

\def\ba{\begin{array}}
\def\ea{\end{array}}
\def\la{\label}
\def\p{\partial}
\def\zb{\bar{z}}

\def\tbar{{\bar{t}}}
\def\zbar{{\bar{z}}}

\makeatletter
\@addtoreset{equation}{section}
\makeatother

\def\C{{\mathbb C}}
\def\R{{\mathbb R}}
\def\Z{{\mathbb Z}}
\def\Q{{\mathbb Q}}

\def\Eb{\overline{{\mathbb E}}}

\def\Hb{\overline{{\mathcal H}}}
\def\M{{\mathcal M}}
\def\Mb{\overline{{\mathcal M}}}

\def\2x2{{\left(\!\!\begin{array}{cc}a&b\\c&d\\\end{array}\!\!\right)}}

\def\f{\frac}
\def\e{\epsilon}

\def\d{\delta}
\def\a{\alpha}

\def\deg{{\rm deg}}
\def\L{{\mathcal L}}

\def\p{\partial}

\def\lmj{z_m^{(j)}}
\def\lki{z_k^{(i)}}
\def\rmj{r_m^{(j)}}

\def\rki{r_k^{(i)}}

\def\Ccal{{\mathcal C}}
\def\olMcal\overline{{\mathcal M}}

\def\f{\frac}
\def\l{\lambda}
\def\pb{{\bf p}}
\def\qb{{\bf q}}
\def\Acal{{\mathcal A}}
\def\a{\alpha}
\def\b{\beta}
\def\p{\partial}
\def\e{\epsilon}
\def\Ch{\widehat{\CC}}
\def\Scal{{\mathcal S}}
\def\la{\label}

\def\g{\gamma}
\def\xb{{\bar{\xi}}}

\def\Eb{\bar{{\cal E}}}
\def\3{\ss}
\def\a{\alpha}

\def\b{\beta}

\def\d{\delta}
\def\l{\lambda}

\def\s{\sigma}
\def\t{\tau}
\def\th{\vartheta}
\def\Th{\Theta}

\def\O{\Omega}

\def\phi{\varphi}

\def\Lh{\hat{\L}}

\def\C{\mathbb{C}}
\def\Z{\mathbb{Z}}
\def\R{\mathbb{R}}

\def\t0{\Theta_0}

\def\la{\label}

\def\c{\cite}
\def\f{\frac}
\def\L{{\cal L}}
\def\p{\partial}
\def\pb{{\bf p}}
\def\qb{{\bf q}}

\def\rb{{\bf r}}
\def\sb{{\bf s}}

\def\tr{{\rm tr}}
\def\0{S}
\def\1{T}

\def\log{\ln}
\def\ol{\overline}

\def\f{\frac}

\def\8{\infty}
\def\p{\partial}

\def\tr{{\rm tr}}

\def\sl2{\mathfrak{sl}(2,\R)}
\def\SL2{SL(2,\R)}

\def\Ot{W}
\def\Om{\Omega}
\def\si{\sigma}
\usepackage{amssymb}

\def\vh{\hat{v}}

\title{Bergman tau-function: from Einstein equations and Dubrovin-Frobenius  manifolds to geometry of moduli spaces}
\author{Dmitry Korotkin}

\begin{document}
\hskip8.0cm  Dedicated to 65th birthday of Emma Previato
\vskip0.5cm
\maketitle

\bigskip
\begin{center}
\begin{small}
 {\it   Department of Mathematics and
Statistics, Concordia University\\ 1455 de Maisonneuve W., Montr\'eal, Qu\'ebec,
Canada H3G 1M8} \\
\end{small}
\vspace{0.5cm}
\end{center}
{\bf Abstract.} We review the role played by  tau-functions of special type - called {\it Bergman} tau-functions
in various areas: theory of isomonodromic deformations, solutions of Einstein's equations, theory of Dubrovin-Frobenius manifolds, geometry of moduli spaces and
spectral theory of Riemann surfaces.  These  tau-functions  are   natural generalizations of Dedekind's 
eta-function to higher genus. Study of their properties  allows  to get an explicit form of Einstein's metrics, obtain new relations 
in Picard groups of various moduli spaces and derive holomorphic factorization formulas of determinants of Laplacians in flat singular metrics on Riemann surfaces, among other things.

\tableofcontents

\section{Introduction}

There exist several alternative definitions of tau-function of integrable systems. One of them
 is based on Hirota's equation \cite{Hirota}; the analytical theory of these tau-functions
(which in particular include the tau-functions of the KP hierarchy) was essentially developed in the paper by Segal and Wilson \cite{SigalWilson}. Alternatively the tau-function can be defined   as the partition function of some integrable
quantum model. It was this definition which led Jimbo, Miwa and their collaborators to the notion of the isomonodromic tau-function of Schlesinger system \cite{JimboMiwa} based on the theory of holonomic quantum fields. The divisor of zeros of the Jimbo-Miwa tau-function was then shown by Malgrange \cite{Malg83} to play the main role in the
solvability of matrix Riemann-Hilbert problems.

To describe the class of tau-functions discussed in this paper we consider a Riemann surface $\CC$ of genus $g$ with some choice of
canonical basis of cycles (the Torelli marking) and introduce the canonical bimeromorphic differential  on $\CC$ by the formula
$B(x,y)=d_xd_y\log E(x,y)$ where $E(x,y)$ is the prime-form. This bidifferential has pole of second order on the diagonal with 
biresidue 1. Let $v$ be some (holomorphic or meromorphic) abelian differential on $\CC$. Using  $v$ one can regularize $B(x,y)$ 
near diagonal as follows:
$$
B^v_{reg}(x,x)=\left(B(x,y)-\f{v(x)v(y)}{(\int_x^y v)^2}\right)\Big|_{y=x}\;.
$$

Although  Schlesinger systems are in general  not explicitly solvable, 
they admit explicit solutions corresponding to special monodromy groups. 
The first class of examples of such explicit solutions arises in the theory of Dubrovin-Frobenius manifolds; corresponding monodromy groups don't have any continuous parameters. Most ingredients of  Dubrovin-Frobenius manifold structures on Hurwitz spaces allow an explicit description \cite{Dubr1,Dubr2}. In particular, their  isomonodromic tau-function (which coincides with genus one free energy up to an auxiliary factor)
satisfies the following system of equations with respect to critical values $z_j=f(x_j)$ ($x_j$ is a critical point of $f$ which we assume to be simple) of the meromorphic function $f$ \cite{KokKor03}:
\be
\f{\p\log \tau_B(\CC,df)}{\p z_j}=-{\rm res}\Big|_{x_k}\frac{B^{df}_{reg}(x,x)}{df(x)}\;.
\la{tauint}\ee
We call $\tau_B$ the {\it Bergman tau-function} on the Hurwitz space since the main contribution to the right hand side of this equation is given by the {\it  Bergman 
projective connection}.
In the framework of conformal field theory $\tau_B(\CC,df)$ has the
meaning of chiral partition function of  free bosons on a Riemann surface with flat metric of infinite volume $|df|^2$ \cite{Kniz87,Sonoda}.
 An explicit formula for $\tau_B$ was found in \cite{IMRN2}.

Another class of  explicitly solvable Riemann-Hilbert problems
corresponds to monodromy groups with quasi-permutation monodromies; such RH problems were solved in \cite{Annalen} in terms of 
Szeg\"o kernel 
on  branched coverings of the Riemann sphere (the $2\times 2$ case was treated earlier in \cite{KitKor98,DIKZ}).
The corresponding Jimbo-Miwa tau-function is a product of three 
factors:
\be
\tau_{JM}=\tau_0\;\tau_B^{-1/2}\, \Thpq(Q)
\la{tauJMint}\ee
 each of which has an independent meaning. The last factor is the theta-function with characteristics $\pb,\qb\in \C^g$; the Malgrange divisor is defined by vanishing of this theta-function.  The factor $\tau_0$ satisfies a system of equations
$$
\f{\p\log\tau_0}{\p\ z_m}=\f{1}{2}\res|_{z=
z_m}\sum_{j=1}^n\f{W^2(z^{(j)})}{dz};\hskip0.8cm W(x)=  \sum_{m=1}^M\sum_{j=1}^n r_m^{(j)}d_x\log E(x,z_m^{(j)})
$$
(here $n$ is the degree of the function $f$, $r_m^{(j)}$ are constants which, together with $\pb,\qb$ define monodromy matrices;
$z_m^{(j)}$ is the point on the $j$th sheet projecting to the branch point $z_m\in\C$)
 which  resembles equations from
Seiberg-Witten theory (this analogy  was recently observed in  \cite{GavMar}); for a special class of coverings this factor can be expressed via a product of prime-forms \cite{GavMar} as follows:
$$
\tau_0^2=\prod_{\lmj\neq\lki} E(\lmj,\lki)^{\rmj\rki}\;.
$$
 We give an alternative proof of this
formula for another characteristic special case of Riemann-Hilbert problems in this paper. The vector $Q$ from (\ref{tauJMint}) is given by $Q=\sum_{j,m} r_m^{(j)}\Acal_{x_0} (z_m^{(j)})$ where $\Acal_{x_0}$ is the Abel map with basepoint $x_0$.
The second factor in (\ref{tauJMint}) is the $-1/2$ power of the Bergman tau-function (\ref{tauint}).   

The first applications of tau-functions (\ref{tauint}), (\ref{tauJMint}) we consider here are in the area of explicit solutions of Einstein's equations. The first 
case is self-dual Einstein manifolds with $SU(2)$ symmetry which are equivalent to a special case of $2\times 2$ fuchsian Schlesinger system with four singularities. The metric on such manifolds is written in the form
$$
g=F\left\{d\mu^2 +\f{\si_1^2}{W_1^2} +\f{\si_2^2}{W_2^2}+\f{\si_3^2}{W_3^2}\right\}
$$
where the 1-forms $\si_i$ satisfy $d\si_i=\si_j\wedge \si_k$ where $(i,j,k)$ is any cyclic permutation of $(1,2,3)$; $\mu$ is the "euclidean time" variable and functions 
$W_j$ and $F$ depend only on $\mu$.
In the most non-trivial case of such manifolds the self-duality conditions are equivalent to the following system of equations:
\be
\f{dW_j}{d\mu}= - W_k W_l+ 2 W_j\f{d}{d\mu}\log(\th_{k+1}\th_{l+1})\;, 
\la{Wjint}\ee
(where $(i,j,k)$ is an arbitrary permutation of $(1,23)$ and $\theta_j,\;\; j=2,3,4$ are theta-constants with module $i\mu$) which is equivalent to a 4-point $2\times 2$ Schlesinger system. The conformal factor $F$ can be chosen such that the Einstein's equations (with cosmological constant $\Lambda$) hold for the metric $g$  when an integral of motion of the system has a special value corresponding to an explicitly solvable case of
(\ref{Wjint}). The remarkably simple explicit formulas for $W_j$ can be found from the  tau-function 
 $\tau_{JM}=\theta[p,q])/(\theta_2\theta_4)$ (where $p,q\in\C$ satisfy appropriate reality conditions) \cite{KitKor98}:
$$
W_1=-\f{i}{2}\th_3\th_4\f{\f{d}{d q}\th[p,q+\f{1}{2}]}{e^{\pi i p}\th[p,q]}\;,\hskip1.0cm
W_2=\f{i}{2}\th_2\th_4\f{\f{d}{d q}\th[p+\f{1}{2},q+\f{1}{2}]}{e^{\pi i p}\th[p,q]}\;,
$$
$$
W_3=-\f{1}{2}\th_2\th_3\f{\f{d}{d q}\th[p+\f{1}{2},q]}{\th[p,q]}\;.
$$
The corresponding conformal factor $F$ is given by the formula
$$
F= \f{2}{\pi\Lambda} \f{W_1 W_2 W_3}{\Big(\f{d}{dq}\log\th[p,q]\Big)^2}\;.
$$
Such metrics have remarkable modular properties which were  exploited  in various contexts (see \cite{ManMar15,FaFaMa15}).

 Another appearance of the tau-function (\ref{tauJMint}) is in the theory of stationary axially symmetric Einstein's equation (in this case the spacetime has physical signature $3+1$ and there is no cosmological constant).  The non-trivial part of Einstein's equations is then encoded in the Ernst equation
 $$
(\Ecal+\overline{\Ecal})(\Ecal_{xx}+\rho^{-1}\Ecal_{\rho}+\Ecal_{zz})=2(\Ecal_x^2+\Ecal_\rho^2)
$$
where $\Ecal$ is a complex-valued function of variables $z$ and $\rho$. The Ernst equation admits a class of solutions associated to hyperelliptic Riemann 
surfaces of the form 
$$
y^2=(w-\xi)(w-\xb)\prod_{j=1}^{2g} (w-w_j)
$$
where two branch points $\xi=z+i\rho$ and $\xb$ depend on space time variables. The corresponding solution of the Ernst equation depending on two constant vectors $\pb,\qb\in \C^g$ (which satisfy certain reality conditions) has the form \cite{TMF1989}
$$
\Ecal(\xi,\xb)=\f{\Theta\left[^\pb_\qb\right](\Acal(\infty^1)-\Acal(\xi))}{\Theta\left[^\pb_\qb\right](\Acal(\infty^2)-\Acal(\xi))}\;.
$$
Such solutions can be applied to solve various physically meaningful boundary value problems for Ernst equation \cite{Klein_book}.

For a given Ernst potential coefficients of the corresponding metric can be found in quadratures. However, having an explicit formula for the tau-function (\ref{tauJMint})
in the hyperelliptic case one can find these coefficients explicitly, in particular, the so-called {\it conformal factor} is given by the formula \cite{KorMatFUNKAN}
$$
e^{2k}=\f{\Theta\left[^\pb_\qb\right](0)\Theta\left[^\pb_\qb\right](\f{1}{2}{\bf e})}{\Theta(0)\Theta(\f{1}{2}{\bf e})}
$$
where
${\bf e}=(1,\dots,1)$

The key property of the Bergman tau-function which makes it useful in geometry of moduli spaces is its transformation law under the change of Torelli marking of $C$:
 \be
\tau_B (\CC,v)\to \epsilon {\rm det} (C\Omega+D)\tau_B (\CC,v)
\la{transint}\ee
where $\Omega$ is the period matrix of $\CC$ and $\left(\ba{cc} C & D\\ B & A\ea\right)$ is an $Sp(2g,\Z)$ transformation of the 
canonical basis of cycles on $\CC$. and
where $\e$ is a root of unity of degree which depends on multiplicities of zeros of the differential  $v$.
The property (\ref{transint}) allows to interpret  the  Bergman tau-function$\tau_B$ as a section of the determinant of the Hodge vector bundle
(or  a higher genus version of the Dedekind eta function) and can be used to study
 the geometry of moduli spaces. 

 The list of moduli spaces where this strategy was successfully applied is quite long.
 Studying analytical properties of Bergman tau-function on  Hurwitz spaces and their compactifications, called spaces of admissible covers   
gives an  expression for the  Hodge class $\lambda$ on the space of admissible covers of degree $n$ and genus $g$ \cite{Advances}.
 An alternative algebro-geometric proof  of this relation was obtained in \cite{GK1}
 and further used to compute  classes of certain divisors within $\Mcal_g$ in \cite{GK2}.

Applying a similar strategy to the moduli space $\Hb_g$ of stable abelian differentials on  Deligne-Mumford stable Riemann surfaces of genus $g$ 
we get the following relation 
in the rational Picard group   ${\rm Pic}(P\Hb_g)\otimes\Q$: 
$$
\lambda=\f{g-1}{4}\phi+\f{1}{24}\d_\deg+\f{1}{12}\d_0+\f{1}{8}\sum_{j=1}^{[g/2]}\d_j\;
$$
where $\lambda$ is the  Hodge class,  $\phi$ is the first Chern class of
the line bundle associated to the projection $\Hb_g\to P\Hb_g$,
 $\d_\deg$ is the class of the divisor of abelian differentials with multiple zeros and $\d_j$ are classes of   Deligne-Mumford boundary divisors.
 
 Further generalization  to moduli spaces $\Qcal_g$ of holomorphic  quadratic differentials looks as follows. For a pair $(\CC,Q)$ where $\CC$ is a Riemann surface of genus $g$ and $Q$ is a holomorphic quadratic differential with simple zeros on $\CC$ we define the canonical cover $\Ch$ by $v^2=Q$; the genus of $\Ch$ equals $4g-3$. Then we define two  vector bundles over $\Qcal_g$: one is the Hodge vector bundle whose fiber has rank $g$; the fiber is the 
 space of abelian differentials on $\Ch$ invariant under the natural involution (this fiber can be identified with the space of abelian differentials on $\CC$). The second is the Prym vector bundle whose fiber has rank $3g-3$; the fiber is the space of abelian differentials on $\Ch$ which are skew symmetric under the natural involution. The fiber of the Prym vector bundle turns out to be isomorphic
 to the space of holomorphic quadratic differentials on $\CC$. One can define then two tau-functions: the {\it Hodge} tau-function 
 $\tau_+(\CC,Q)$ and the {\it Prym} tau-function $\tau_-(\CC,Q)$; analysis of their analytical properties allows to express the
 Hodge class and the class $\lambda_2$ (the first Chern class of the vector bundle of quadratic differentials) via  $\psi$-classes, the class 
 $\delta_{\deg}$ of quadratic differentials with multiple zeros and the  Deligne-Mumford boundary divisor classes. Further elimination of the classes $\psi$ and $\delta_{\deg}$ from these expressions leads to the famous Mumford's formula in the rational Picard group of $\ol{\Mcal}_g$
$$
\l_2-13\lambda =\delta_{DM}\;.
$$
Generalizing this scheme to spaces of meromorphic quadratic differentials with $n$ second order poles the formalism of  
Prym and Hodge tau-functions reproduces another
classical relation which holds in the Picard group of $\ol{\Mcal}_{g,n}$:
\be
\lambda_2^{(n)}-13\lambda+\sum_{j=1}^n \psi_j=-\delta_{DM}\;.
\la{l2lint}\ee
Here $\lambda_2^{(n)}$ is the first Chern class of the   vector bundle of quadratic differentials with first order 
poles at the punctures; $\psi_j$ is the  class of line bundle whose fiber is the cotangent space to $C$ at the $j$th marked point.

The tau-functions turn out to  be a useful tool also in the real analytic context of combinatorial model of $\Mcal_{g,n}$ based
on Strebel differentials. This combinatorial model, denoted by  $\Mcal_{g,n}[\pb]$,  can be considered as a real slice of the moduli space  $\Qcal_{g,n}[\pb]$
of meromorphic quadratic differentials on $\CC$ with second order poles at $n$ marked points and biresidues given by
$-p_j^2/4\pi^2$. The real slice is defined by the condition that all periods of $v$ on the canonical cover $v^2=Q$ are real.
Such combinatorial model is a union of strata labelled by the set of multiplicities of zeros of $Q$; each stratum consists 
of several topologically trivial cells while the facets between the cells belong to strata of lower dimension.

Strata corresponding to odd multiplicities of all zeros of $Q$ turn out to be cycles; they are named after Witten and Kontsevich.
Combinations of these strata are known to be Poincar\'e dual to  tautological classes  \cite{ArbCor,Mondello2}.
In real codimension 2 there are two such cycles: the Witten cycle $W_5$ whose largest stratum corresponds to differentials $Q$ with one triple zero while other zeros are simple, and Kontsevich boundary $W_{1,1}$ whose largest stratum corresponds to quadratic differentials with two simple poles at 
points obtained by resolution of the node of a stable  Riemann surface with one node. Arguments of Hodge and Prym tau-functions $\tau_\pm$ then give sections of 
certain circle bundles over cells in the largest stratum of $\Mcal_{g,n}[\pb]$ which can be continuously propagated from cell to cell but have monodromies around $W_5$ and $W_{1,1}$. Calclulation of these monodromies leads to the following two relations \cite{BK2}:
$$
12\kappa_1=W_5+W_{1,1}
$$
where $\kappa_1$ is the first kappa-class (this relation is an analog of the relation derived by Penner in the framework of combinatorial model based on hyperbolic geometry \cite{Penner}) and
$$
\lambda_2^{(n)}- 13 \lambda- \sum_{i=1}^n \psi_i= - \Wbdr
$$
which is a combinatorial version of the complex analytic formula (\ref{l2lint}).

Tau-functions on  moduli spaces of $N$-differentials were used in \cite{KSZ} to compute  classes of determinants of  Prym-Tyurin 
vector bundles over these moduli spaces. In \cite{Faddeev} these results were applied to find the class of the universal Hitchin's discriminant 
in the Picard group of the universal moduli space of Hitchin's spectral covers.

The equation of a general $GL(n)$ spectral cover $\Ch$ over a Riemann surface $\CC$ of genus $g$ is given by
\be
P_n(v)=v^n+\dif_{n-1}v^{n-1}+\dots+\dif_{1}v+\dif_0=0
\la{Spcov}\ee
where $Q_j$ is a holomorphic $(n-j)$ - differential over $\CC$. The branch points of $\Ch$ are zeros of discriminant $W$ of equation (\ref{Spcov})
which is a holomorphic $n(n-1)$-differential; thus the total number of branch points  equals $n(n-1)(2g-2)$ and the genus of $\Ch$ equals $\gh=n^2(g-1)+1$.
Denote the moduli space of covers (\ref{Spcov}) when both the base $\CC$ and coefficients $Q_j$ are allowed to vary by $\Mcal_H$; this is the universal 
space of Hitchin's spectral covers.  We denote by $\Mcal_H^\CC$ the space of spectral covers for fixed base $\CC$ . The universal discriminant $D_H\subset \Mcal_H$ consists of covers with coinciding branch points; ($D_H$ consists of three components corresponding to different structures
of ramification points corresponding to the double branch point). The class of divisor $D_H$ in the space $P\Mcal_H$ can be computed 
using the formula for the Hodge class on the moduli space of holomorphic $n(n-1)$-differentials \cite{KSZ}:
$$
\f{1}{n(n-1)}[D_H]=(n^2-n+1)(12\lambda -\delta)- 2(g-1)(2n^2-2n+1) \phi 
$$
where $\l$ is (the pullback of) the Hodge class on $\overline{\Mcal}_g$, $\phi$ is the tautological class of the line bundle arising from  projectivization 
$\overline{\Mcal}_H\to P\overline{\Mcal}_H$, and $\delta$ is the (pullback to $P\Mcal_H$) of the class of DM boundary of $\overline{\Mcal}_g$.

We mention also an  interesting fact about the space of spectral covers  with fixed base $\Mcal_H^\CC$. The set of natural coordinates on 
$\Mcal_H^\CC$ is given by $a$-periods $A_j$ of the abelian differential $v$ on $\Ch$. The variation of the period matrix $\Oh$ of $\Ch$ with 
respect to these coordinates can be deduced from variational formulas (\ref{varO2}) on the moduli space of abelian differentials on Riemann surfaces of genus $\gh$. Applying the elementary chain rule to the formulas (\ref{varO2}) one gets certain sum of residues over ramification points $\hat{x}_r$ of $\Ch$ corresponding to branch points $x_r\in \CC$ \cite{BK3}
(assuming that all of these branch points are simple)
\be
\f{d \Oh_{lk}}{d A_j}=-2\pi i \sum_{ramification \;\;points\;\;\hat{x}_r}  {\rm res}|_{\hat{x}_r}\f{v_l v_k v_r}{d\xi \,d(v/d\xi)}
\la{DMint}\ee
where $\xi$ is an arbitrary local parameter on $\CC$ near branch point $x_r$; $\{v_j\}_{j=1}^{\gh}$ are normalized abelian differentials on $\Ch$.
The formula (\ref{DMint}) provides an explicit coordinate realization of the Donagi-Markman cubic \cite{DonagiMarkman}

The Bergman tau-functions have also an application to the spectral theory of Riemann surfaces with flat singular metrics and special holonomy: they allow to calculate the determinant of Laplace operator in such metrics. An example of such formula
is given by \cite{JDG}
$$
\det\Delta^{C,|v|^2}= const\, (\det \Im\Omega)\,{\rm Area}(\CC,|v|^2)\,|\tau(\CC,v)|^2 
$$
where $v$ is a holomorphic abelian differential on $\CC$. This formula as well as its generalization to spaces of quadratic differentials turn out to be useful in the theory of the Teichm\"uller flow \cite{EKZ}.

We did not cover all aspects of the theory and applications of the Bergman tau-function. 
Among other applications of this subject are  the theory of random matrices \cite{EKK}, the Chern-Simons theory \cite{McIntyre} and the theory of topological recursion \cite{Baraglia}.

The goal of this paper is to to summarize these links and establish a few new facts along the way. We start from describing the role of Bergman tau-function on Hurwitz spaces in the theory of Dubrovin-Frobenius manifolds in section \ref{DubrFrob}. This section is based on \cite{Dubr2,KokKor03,IMRN2}. Here we introduce the Schlesinger system and Jimbo-Miwa tau-function and present an explicit formula for the Bergman tau-function on Hurwitz spaces. In section \ref{RHqp}, following \cite{Annalen}, we describe solutions of an arbitrary Riemann-Hilbert problem with quasi-permutation monodromies and give a detailed calculation of its tau-function.
In section \ref{SDEsec}, following \cite{BabKor98}, we show how to use the most elementary version of this tau-function to get elementary expressions for metric coefficients of $SU(2)$ invariant self-dual Einstein manifolds. In section \ref{Ernstsec} we show how to use the solutions of $2\times 2$ Riemann-Hilbert problems with quasi-permutation monodromies to get 
solutions of Einstein equations with two Killing vectors in terms of hyperelliptic theta-functions and express  metric coefficients in terms of corresponding theta-functions \cite{TMF1989,KorNicPRL,KorMatFUNKAN}. In section \ref{Bergmod} we describe applications of Bergman tau-function to geometry of moduli spaces of abelian, quadratic and $N$-differentials. In section \ref{Hodgeadm}, following \cite{Advances} we compute the Hodge class
on the space of admissible covers in terms of boundary divisors. In section \ref{spacehol} which is based on \cite{MRL} we apply the formalism to compute the Hodge class on spaces of abelian differentials. In section \ref{spaceQHP} we extend the formalism to spaces of quadratic differentials 
by defining the Hodge and Prym tau-functions \cite{contemp}. We
show in particular how Mumford's relations between Hodge class and the class of the determinant of the vector bundle of quadratic differentials can be derived by analyzing the analytical properties of tau-functions. In section \ref{highQ} we introduce the Hodge and Prym tau-functions on an arbitrary stratum of the space of holomorphic quadratic differentials.  In section \ref{clMgn} we extend this 
formalism to spaces of meromorphic quadratic differentials with second order poles and get the analog of Mumford's relations
for $\ol{\Mcal}_{g,n}$. In section \ref{flatco} (following \cite{BK2}) we consider the flat combinatorial model of $\Mcal_{g,n}$ based on Jenkins-Strebel differentials.
Arguments of Hodge and Prym tau-functions give sections of circle line bundles which are combinations of tautological classes; computation of monodromy of these arguments  around Witten's cycle and Kontsevich's boundary of the combinatorial model gives combinatorial analogs of Mumford's relations in the Picard group of $\ol{\Mcal}_{g,n}$. In section \ref{nspin} we briefly describe the application of tau-functions to spaces of holomorphic $N$-differentials and spin moduli spaces following
\cite{KSZ,Basok}. In particular we show how the analytical properties of tau-functions on spin moduli spaces imply the Farkas-Verra formula for the divisor of degenerate 
odd spinors. In section \ref{secHit} we use tau-functions to study moduli spaces of Hitchin's spectral covers following \cite{Faddeev,BK3}. 
In section \ref{secdiscr} we determine the class of the universal discriminant locus in the space of all spectral curves i.e. the locus where not all branch points of the spectral cover are simple. In section \ref{varperH} we, following \cite{BK3}, derive variational formulas for the period matrix on the moduli space of spectral covers with fixed base from variational formulas on moduli spaces of abelian differentials \cite{JDG}, establishing the link with the Donagi-Markman cubic.
In section \ref{detteich} we summarize the links between tau-functions, determinants of Laplace operator on Riemann surfaces with flat singular metrics and
the sum of Lyapunov exponents of the Teichm\"uller flow on spaces of abelian and quadratic differentials. In Appendix (section \ref{appsec}) we describe canonical objects associated to Riemann surfaces which are used in the main text as well as variational formulas on Hurwitz spaces and spaces of abelian and $N$-differentials.

{\bf Acknowledgements.}
The author thanks Marco Bertola, Alexey Kokotov and Peter Zograf 
for  numerous illuminating discussions.
This work was supported in part by the Natural Sciences and Engineering Research Council of Canada grant
RGPIN/3827-2015.

 \section{Tau-function of  Dubrovin-Frobenius manifold structure on Hurwitz spaces}
\la{DubrFrob}
\subsection{Schlesinger system and Jimbo-Miwa tau-function}

Consider the system of linear differential equations with the initial condition:
\be
\frac{d\Psi}{dz} = A(z)\Psi \;,\hs \Psi(z_0)=I
\la{ls}
\ee
where $A(z)$ is an $N \times N$ matrix whose entries are meromorphic functions on $\CP1$.
The system (\ref{ls}) is
called Fuchsian if the matrix $A(z)$ has only simple poles, i.e.
$A(z)=C+\sum_{i=1}^M \frac{A_i}{z-z_i}$. We assume that $z=\infty$ is not a singular point of (\ref{ls}) i.e.
$\sum_{i=1}^M A_i=0$.
The solution    $\Psi(z)$    of   (\ref{ls})  is single-valued on
the universal cover of $\CP1\setminus\{z_i\}_{i=1}^M$.
If one starts at a point $z_0$ on some sheet of the universal cover, and analytically
continues $\Psi$   along a loop $\gamma\in \pi_1( \CP1\setminus\{z_i\}_{i=1}^M )$ one gets a new solution, $\Psi_\gamma$, of the same system; therefore, $\Psi_\gamma$ is related to $\Psi$ by a right multiplier, $M_\gamma$, which is called the monodromy matrix: $\Psi_\gamma=\Psi M_\gamma$. In this
way, one gets an anti-homomorphism from the fundamental group $\pi_1( \CP1\setminus\{z_i\}_{i=1}^M )$ to $GL(N)$. The image of this anti-homomorphism
is called the monodromy group of the system (\ref{ls}).  The monodromy matrices and the 
positions of singularities form the set of monodromy data of the equation (\ref{ls}). The
Riemann-Hilbert (or inverse monodromy) problem is the problem of finding a
matrix-valued function $\Psi$ [and, therefore, also the coefficients $A_j$] knowing the
monodromy data.
One can vary the positions of singularities $z_j$  such that the
 monodromy matrices remain unchanged. Such a deformation
(called the isomonodromic deformation) generically implies the following set of non-linear differential {\it Schlesinger equations }
for $A_j$ as functions of $\{z_k\}$:
\be
\f{\p A_j}{\p z_k}= \f{[A_j,\,A_k]}{z_j-z_k} - \f{[A_j,\,A_k]}{z_0-z_k}\;,\hskip0.6cm
j\neq k\; ;
\la{Schl1}
\ee
\be 
\f{\p A_k}{\p z_k}= -\sum_{j\neq k}\left(\f{[A_j,\,A_k]}{z_j-z_k} - 
\f{[A_j,\,A_k]}{z_j-z_0}\right)\;.
\la{Schl2}\ee

In terms of a solution of the Schlesinger system  the  Jimbo-Miwa tau-function is defined 
by the system of equations \cite{JimboMiwa}
\be
\f{\p}{\p z_j}\log\tau = H_j:= \f{1}{2}{\rm res}|_{z=z_j}\f{\tr\left(d\Psi\,\Psi^{-1}\right)^2}{ dz}\;;
\la{taudefJM}\ee

According to Jimbo-Miwa and  Malgrange \cite{Malg83}, the isomonodromic tau-function is a holomorphic section of a holomorphic line bundle
 on the universal covering of the space 
$\C^n\setminus \{z_k=z_j\}$. The divisor $(\tau)$ of its zeros (the {\it Malgrange divisor}) has an important meaning: if $\{z_j\}_{j=1}^n\in (\tau)$ then
the Riemann-Hilbert problem with the given  set of monodromy data does not have a solution. Moreover, the solution $\{A_j\}_{j=1}^n$ of the Schlesinger system is singular on $(\tau)$.

\subsection{Bergman tau-function on Hurwitz space}

Here we discuss the appearance of the Bergman tau-function in the theory of Dubrovin-Frobenius manifolds \cite{Dubr1,Dubr2}. The class of Dubrovin-Frobenius manifolds which admits a complete analytical description is provided by
 Hurwitz spaces. Restricting ourselves to the  semi-simple  manifolds consider a Hurwitz space $H_{g,n}$; a point of   $H_{g,n}$ is a pair $(\CC,f)$ where $\CC$ is a Riemann surface of genus $g$ and $f$ is a function of degree $n$ on $\CC$ with  simple poles and simple critical points. To each Dubrovin-Frobenius manifold one can associate a natural Riemann-Hilbert problem
 whose solution, together with associated solutions of the Schlesinger system,  encodes all essential ingredients of the  manifold - from flat coordinates to the prepotential. Matrix entries of  the solution $\{A_j\}$ of  the Schlesinger system are given    by the rotation coefficients
 of a flat metric associated to each Dubrovin-Frobenius manifold, see eq (3.70) of \cite{Dubr1}. Therefore, the rotation coefficients
 define also the Jimbo-Miwa isomonodromic tau-function  which in turn gives the genus one contribution (the $G$-function) to the free energy
 (see Th.3.2 of \cite{Dubr2}). Referring for details to \cite{Dubr1,Dubr2} we will only present formulas which are
 necessary to make this section reasonably  self-contained.
 
 The Jimbo-Miwa tau-function of a Dubrovin-Frobenius manifold,  which in the present context is nothing but  the Bergman tau-function on the Hurwitz space, is defined  by the system of equations (see Eq. (3.40) of \cite{Dubr2})
\be
\f{\p\log \tau_B}{\p z_j}=\frac{1}{2}\sum_{i\neq j} \f{\beta_{ij}^2}{z_j-z_i}
\la{tauFr}
\ee
where $\beta_{ij}$ are the rotation coefficients of the Darboux-Egoroff metric associated to the Frobenius manifold and $z_j$ are
the canonical coordinates on the manifold.

For semi-simple Frobenius structures on the Hurwits space $H_{g,n}$ the canonical coordinates are given by the  critical values
of the function $f$:  $z_i=f(x_i)$ where $x_i$ are  critical points of the function $f$ ($df(x_i)=0$). Equivalently, the critical values $z_i$ are called the
branch points of the corresponding $n$-sheeted branched cover while for the critical points $x_i$ we reserve the term "ramification points". The natural local coordinates (also called ``distinguished'') on $\CC$
near $x_j$  are given by $\zeta_j(x)=[f(x)-z_j]^{1/2}$. The rotation coefficients for the Frobenius structures on $H_{g,n}$ are expressed in terms of the canonical bidifferential $B(x,y)$ as follows \cite{IMRN1}:
\be
\beta_{jk}=\frac{1}{2}\f{B(x,y)}{d\zeta_j(x)d\zeta_k(y)}\Big|_{x=x_j,\;y=x_k}\;.
\ee
Then the equations (\ref{tauFr}) can be alternatively written as follows:
\be
\f{\p\log \tau_B}{\p z_j}=-{\rm res}\Big|_{x_k}\frac{B_{reg}(x,x)}{df(x)}
\la{tauFr1}
\ee
where $B_{reg}(x,x)$ is the meromorphic  quadratic differential which equals to  the  constant term of $B(x,y)$ on the diagonal:
\be
B_{reg}(x,x)=\left(B(x,y)-\f{df(x)df(y)}{(f(x-f(y))^2}\right)\Big|_{x=y}\;.
\la{Bregf}
\ee
The differential $B_{reg}(x,x)$ can also be represented in terms of the Bergman projective connection $S_B$ and the projective connection $S_{df}(\cdot)=\{f,\cdot\}$ as follows:
$$
B_{reg}(x,x)=\f{1}{6}(S_B-S_{df})\;.
$$

Let us introduce the divisor of the differential $(df)$:
\be
(df)=\sum k_j x_j
\ee
where $k_j=1$ if $x_j$ is a (simple) ramification point. If $x_j$ is a pole of $f$ then  the order of the pole of $f$ at $x_j$ equals $-(k_j+1)$.
Near pole $x_j$  the distinguished local parameters are given by
\be
\zeta_j(x)={f(x)}^{1/(k_j+1)}\;.
\la{xijint}
\ee
 Near zeros  $x_j$ of $df$   the distinguished local coordinates are defined by
\be
\zeta_j(x)=(f(x)-f(x_j))^{1/(k_j+1)}\;.
\la{lcoHur}
\ee

Introduce also the following notations:
\be
E(x,x_j)=\lim_{y\to x_j} E(x,y) \sqrt{d\zeta_i(y)}
\la{defEpi}
\ee
and
\be
 E(x_i,x_j)=\lim_{x\to x_j, y\to x_i} E(x,y) \sqrt{d\zeta_i(y)} \sqrt{d\zeta_j(x)}.
\la{defEppi}
\ee

The solution of the system (\ref{tauFr1}) is given by the following theorem:
 
\begin{proposition} \cite{IMRN2} Let $(df)=\sum k_i q_i$ be the divisor of the differential $df$.
Then the solution $\tau_B$ of the system (\ref{tauFr1}) is given by the following expression:
\be
\tau_B=
\Ccal^{2/3}(x)\left(\f{df(x)}{\prod_{i}E^{k_i}(x,q_i)}\right)^{(g-1)/3} 
\left(\prod_{i<j} E(q_i,q_j)^{\f{k_i k_j}{6}}\right)\;e^{\f{\pi i}{6}\langle {\bf s},\Omega{\bf s}\rangle-\f{2\pi i}{3}\langle {\bf s},K^x\rangle}\, .
\la{tauBerg}
\ee
where $K^x$ is the vector of Riemann constants.  Here the vector ${\bf s}$ is defined by relation
\be
\Acal_x((df))+2K^x+\Omega {\bf s}+ {\bf r}=0
\la{defsrHur}
\ee
where $ {\bf s}, {\bf r}\in \Z^g$.
\end{proposition}

For the space of hyperelliptic coverings given by $w^2=\prod_{i=1}^{2g+2} (z-z_i)$  the following simpler expression for the  Bergman tau-function was found earlier in \cite{KitKor98}:
\be
\tau_B(\{z_m\})=\det{A}
\prod\limits_{m<n}(z_m-z_n)^{\frac 14}
\la{Berghyp}
\ee
where $A_{ij}=\int_{a_i} z^{j-1}dz/w$ is the matrix of $a$-periods of the non-normalized holomorphic differentials on $\CC$.
The expression (\ref{Berghyp})  appeared first as  the correlation function of the Ashkin-Teller model
\cite{Zamo86}.

The function $\tau_B$ is simultaneously the Jimbo-Miwa tau-function of two Riemann-Hilbert problems whose solutions are related by the Laplace transform. One of them if non-Fuchsian; it's solution 
was implicitly given  in \cite{Dubr1} and  later given in the complete  form in \cite{Schram}.
Another Riemann-Hilbert problem is Fuchsian; its monodromy group and solution were 
described in \cite{KorSch}. Monodromy matrices of such  monodromy group always have integer values; therefore, they do not contain any parameters
and represent isolated points in the  monodromy manifold.
The  function $\tau_B$ does not vanish unless two  critical values of the function $f$ coincide; therefore, the Malgrange divisor of the Dubrovin's Riemann-Hilbert problems is empty.

There exists another class of monodromy groups which can be associated to the
branch coverings: these are the monodromy groups with quasi-permutation monodromies \cite{Annalen}.  
These monodromy groups depend on  parameters, and the Bergman tau-function gives only one of the contributions to the Jimbo-Miwa tau-function.

\section{Riemann-Hilbert problem  for quasi-permutation monodromy groups }
\la{RHqp}


Riemann-Hilbert problems with the quasi-permutation monodromy groups can be solved as follows \cite{Annalen}. Suppose we are given a
$GL(n)$ representation of $\pi_1(\CP1\setminus \{z_j=z_k\},z_0)$ such that all monodromy matrices have exactly one non-vanishing 
entry in each column and in each row. Then for any monodromy matrix $M_\gamma$ ($\gamma\in \pi_1(\CP1\setminus \{z_j=z_k\},z_0)$ one can construct an associate permutation
matrix $M_\gamma^0$ by replacing all non-vanishing entries of $M_\g$ by $1$. In this way we get a {\it permutation} representation of $\pi_1(\CP1\setminus \{z_j=z_k\},z_0)$,
which, according to Riemann's theorem, defines an $n$-sheeted branched covering  of $\CP1$. This  gives the pair $(\CC,f)$ where $\CC$ is the  Riemann surface of a genus
$g$ (the genus is determined by the monodromy group) and $f$ is the meromorphic function on $\CC$ of degree $n$. 

Assume that the monodromy representations can not be decomposed into direct sum of  two other 
 quasi-permutation representations.
Then the Riemann surface $\CC$ is connected.
Denote by $z^{(j)}$ the point on $j$th sheet (under some enumeration of sheets) of  $(\CC,f)$ having projection $z$ to the base;  $\{z^{(j)}\}_{j=1}^n=f^{-1}(z)$ if $z$ is not a branch point.

The solution of an arbitrary RH problem with quasi-permutation monodromies constructed in \cite{Annalen} depends on the following set of parameters:
\begin{itemize}
\item
Two vectors $\pb,\qb\in \C^g$.
\item
Constants 
$r_m^{(j)}\in \C$ assigned to each point $z_m^{(j)}$; we assume that the 
constants $r_m^{(j)}$ and $r_m^{(j')}$ coincide if $z_m^{(j)}=z_m^{(j')}$ i.e. if $z_m^{(j)}$ is
a ramification point. These constants are assumed to satisfy the relation
\be
\sum_{m=1}^M\sum_{j=1}^n r_m^{(j)} =0\;\;.
\la{sumr}\ee
Therefore, there are
$Mn-2g-2n+1$
independent parameters among the  constants $r_m^{(j)}$.  
\end{itemize}

Altogether we have $Mn-2n+1$ independent constants
$\pb,\qb$ and $r_m^{(j)}$; 
this number coincides with the number of non-trivial parameters
carried by the  quasi-permutation 
monodromy matrices $M_1,\dots, M_M$.

Denote by  $\Sh(x,y)$  the modified Szeg\"o kernel  given by the following
formula (we assume that $x$ and $y$ lie inside of the fundamental polygon $\tilde{\CC}$ of $\CC$):
\be
\Sh(x,y):=\f{\Th\left[^\pb_\qb\right]\left(\Acal(x)-\Acal(y)+Q\right)}
{\Th\left[^\pb_\qb\right](Q)E(x,y)}\prod_{m=1}^M \prod_{l=1}^n \left[\f{E(x,z_m^{(l)})}
{E(y,z_m^{(l)})}\right]^{r_m^{(l)}}\;
\la{phinew}
\ee
where
$$
Q:= \sum_{m=1}^M\sum_{j=1}^n r_m^{(j)} \Acal_{x_0}(z_m^{(j)})\;.
$$
The vector $Q$ does not depend on the choice of initial point $x_0$ of the Abel map due to assumption
(\ref{sumr}). The kernel (\ref{psinew}) is well-defined
 if $\Th\left[^\pb_\qb\right](Q)\neq 0$.

Define 
\be
\psi(x,y)=
\Sh (x,y)E_0(f(x),f(y))
\la{psismall}
\ee
where $$ E_0(z,w)=
\frac{z-w}{\sqrt{dz\,dw}}
$$
is the prime form on $\CP1$.

According to \cite{Annalen} the solution of the RH problem is given by the analytical continuation (in $z$) of the  matrix function 
\be
\Psi_{kj}(z_0,z)=\psi(z^{(j)},z_0^{(k)})
\la{psinew}\ee
from a neighbourhood of the normalization point $z\sim z_0$.

The function $\Psi$ satisfies the condition $\Psi(z_0,z_0)=I$. Moreover, the Fay identity (\ref{ident}) allows to compute its determinant:
\ben
\det\Psi = \prod_{m=1}^M\prod_{j,k=1}^N \left[\f{E (z^{(j)},z_m^{(k)})}
{E(z_0^{(j)},z_m^{(k)})}\right]^{r_m^{(k)}}\;.
\een
In the proof of this expression for ${\rm det}\Psi$  one uses the elementary  fact  that  
$\sum_{p\in f^{-1}(z)} \frac{v}{df}(p)=0\;$ for any holomorphic differential $v$ on $\CC$,
see \cite{Annalen}.

\subsubsection{Jimbo-Miwa tau-function}

\vskip0.2cm

The computation of the Jimbo-Miwa tau-function   (\ref{taudefJM}) corresponding to the solution (\ref{psinew}) starts from transforming 
$\tr\left(\Psi_z\Psi^{-1}\right)^2$ to a suitable form. Since this expression 
is independent of the choice of the 
normalization point $z_0$ one can take the
 limit $z_0\to z$ in the formulas (\ref{psinew}), (\ref{phinew}): 
\be
\Psi_{kj}(z,z_0)= (z_0-z)\f{\Sh(z^{(j)},z^{(k)})} {d z} + O((z_0-z)^2)\;,\hskip0.6cm k\neq j
\ee
\be
\Psi_{jj}(z,z_0)= 1 + (z_0-z)\f{W_1(z^{(j)}) - W_2(z^{(j)})}{dz}
\ee  
where $W_1(x)$ is the  linear combination of the basic   holomorphic differentials on $\CC$:
\be
W_1(x) = \f{1}{\Thpq(Q)} \sum_{\a=1}^g \p_{z_\a}\{\Thpq(Q)\} v_\a (x)
\la{W1}\ee
and $W_2(x)$ is the following meromorphic differential  with simple poles at the points $z_m^{(j)}$ and
residues $r_m^{(j)}$:
\be
W_2(x)=  \sum_{m=1}^M\sum_{j=1}^n r_m^{(j)}d_x
\log E(x,z_m^{(j)})\;.
\la{W2}
\ee

Then
\ben
\tr\left(\Psi_z\Psi^{-1}\right)^2(dz)^2=
2\sum_{j<k}\Sh(z^{(j)},z^{(k)})\Sh(z^{(k)},z^{(j)}) + \sum_{j=1}^n \left(W_1(z^{(j)})-
W_2(z^{(j)})\right)^2\;.
\een
Due  to (\ref{SSB}) we have
\ben
\Sh(x,y)\Sh(y,x)= -B(x,y)-\sum_{\a,\b=1}^g\p^2_{\a \b}\{\log\Thpq(Q)\}v_\a (x) v_\b(y)\;.
\een
Furthermore, since $W_1(x)$ is  holomorphic  on $\CC$, we have
$\sum_{j=1}^N W_1(z^{(j)})=0$ and
\ben
\sum_{j=1}^n \{W_1(z^{(j)})\}^2 
= -2\sum_{\stackrel{j,k=1}{j<k}}^N \sum_{\a,\b=1}^g\p_{\a}\{\log\Thpq(Q)\}
\p_{\b}\{\log\Thpq(Q)\} v_\a(z^{(j)})v_\b(z^{(k)})\;.
\een

Therefore, 
\be
\f{1}{2}\tr\left(\Psi_z\Psi^{-1}\right)^2(dz)^2=
-\sum_{j<k} B(z^{(j)},z^{(k)})+
\f{1}{2}\sum_{j=1}^n W^2_2(z^{(j)})
\la{trace}
\ee
\ben
-\f{1}{\Thpq(Q)}\sum_{j< k}\sum_{\a,\b} \p^2_{\a \b}\{\Thpq(Q)\} v_\a(z^{(j)})
v_\b(z^{(k)})
\een
\ben
-\f{1}{\Thpq(Q)}\sum_{\a}\p_{\a}\{\Thpq(Q)\}\sum_{m}\sum_{j} r_m^{(j)}v_\a(z^{(j)})
d_x \log E(x,z_m^{(j)})\;.
\een

Using the heat equation for the theta-function (\ref{heat}) and the variational
 formula (\ref{varBxy3}),
we see that the contribution of the last two terms in (\ref{trace}) to the residue at $z_m$ is given by $\p_{z_m}\Thpq(Q)$ and, therefore,
$$
\f{\p\log\tau_{JM}}{\p\ z_m}=\frac{1}{2}{\rm res}|_{z=z_m} \left\{\tr\left(\Psi_z\Psi^{-1}\right)^2 dz\right\} 
$$
\be
= -\res|_{z=
z_m}\sum_{j<k} \f{B(z^{(j)},z^{(k)})}{dz} + \f{1}{2}\res|_{z=
z_m}\sum_{j=1}^n \f{W^2_2(z^{(j)})}{dz}
+\p_{z_m}\Thpq(Q)\;.
\la{Hamil}\ee


We get the following proposition
 \cite{Annalen}:
\begin{proposition}
The tau-function $\tau_{JM}$ is given by
\be
\tau_{JM}=\tau_0\;\tau_B^{-1/2}\, \Thpq(Q)
\la{tauJM1}
\ee
where $\tau_B$ is the Bergman tau-function defined by the system (\ref{tauFr1}) and given by the formula
(\ref{tauBerg}).

The factor $\tau_0$ satisfies the system of equations
\be
\f{\p\log\tau_0}{\p\ z_m}=\f{1}{2}\res|_{z=
z_m}\sum_{j=1}^n\f{W^2_2(z^{(j)})}{dz}
\la{deftau0}
\ee
where the Abelian  differential of third kind  $W_2$ is given by (\ref{W2}).
\end{proposition}
{\it Proof.}
The theorem follows from (\ref{Hamil}) and  the identity proven in  \cite{IMRN1}:
\be
2\res|_{z=
z_m}\f{1}{dz}\left\{\sum_{j<k} B(z^{(j)},z^{(k)}) \right\}=\f{1}{6} {\rm res}|_{z=z_m}\sum_{p\in f^{-1}(z)}\frac{S_B-S_{df}}{df}
\la{defBalt}
\ee
where $S_B$ is the Bergman projective connection and $f(x)=z$ is the meromorphic function defining the covering $\CC$. The right hand side of (\ref{defBalt}) coincides
 with the logarithmic derivative of the Bergman tau-function (\ref{tauBerg}).
 $\Box$

\subsubsection{Function $\tau_0$.}

The factor $\tau_0$ in (\ref{tauJM1}) is given by the following theorem

\begin{proposition}
The solution of the system of equations  
\be
\f{\p\log\tau_0}{\p z_m}=\f{1}{2}\res|_{z=
z_m}\sum_{j=1}^n\f{W^2_2(z^{(j)})}{dz}\;.
\la{sys}
\ee
with 
\be
W_2(x)=  \sum_{m=1}^M\sum_{j=1}^n r_m^{(j)}d_x
\log E(x,z_m^{(j)})
\la{defW2}
\ee
is given by 
\be
\tau_0^2=\prod_{\lmj\neq\lki} E(\lmj,\lki)^{\rmj\rki}
\la{sol}
\ee
where the prime-forms whose arguments coincide with $z_m^{(j)}$ are computed with respect to the distinguished local parameters as in (\ref{defEppi})
(notice that all the  terms in the r.h.s. of (\ref{sol}) enter twice; we use this convention  since there is no natural ordering of the points $\lmj$).
\end{proposition}
{\it Proof.}
The proof is based on the variational formula for the prime-form (\ref{varExy3});
to avoid unnecessary technicalities we give the proof in the special situation which, however, reflects all the characteristic features of the general case.
Namely, assume that
$z_1$ is not a critical value (this means that the corresponding monodromy matrix $M_1$ is diagonal) while all residues corresponding to other $z_m$'s equal zero.
In this case  (\ref{sol}) gives 
\be
\tau_0^2= \prod_{i\neq j} E(z_1^{(i)},z_1^{(j)})^{r_1^{(i)} r_1^{(j)}}
\la{tau0sp}\ee
  and
\be
W_2=\sum_{i=1}^n r_1^{(i)} d\log E(\l,z_1^{(i)})
\la{W2sp}\ee
with $\sum_{i=1}^n r_1^{(i)}=0$. 

Let us first  differentiate (\ref{tau0sp})  with respect to some branch point $z_2$ using  (\ref{varExy3}):
\be
\f{\p}{\p z_2}\log \prod_{i\neq j} E(z_1^{(i)},z_1^{(j)})^{r_1^{(i)} r_1^{(j)}}
\la{v1}\ee
\be
=-\f{1}{2} \sum_{p\in f^{-1}(z_2)} \res\Big|_{t=p}(z_2)\f{1}{d f(t)}\sum_{i\neq j} r_1^{(i)} r_1^{(j)}
(d\log E(t,z_1^{(i)})- d\log E(t,z_1^{(j)}))^2
\ee
$$
=-\f{1}{2} \sum_{p\in \pi^{-1}(z_2)} \res\Big|_{t=p}\f{1}{d f(t)}\sum_{i\neq j} r_1^{(i)} r_1^{(j)}
$$
$$
\times
\left[(d\log E(t,z_1^{(i)}))^2+ (d\log E(t,z_1^{(j)}))^2 -2 d\log E(t,z_1^{(i)}))d\log E(t,z_1^{(j)}))\right]\;.
$$
In the first term of the last expression we  perform the summation over $j$ using $\sum_{i=1}^n r_1^{(i)}=0$ and in the second term the summation is performed 
over $i$. The results equal to each other which gives an extra factor of $-2$. Summing only over $i<j$ in the last sum gives the following expression for (\ref{v1}):
\be
\sum_{p\in \pi^{-1}(z_2)} \res\Big|_{t=p}\f{1}{d f(t)}\left[\sum_{i} (r_1^{(i)})^2 
(d\log E(t,z_1^{(i)}))^2
+2 \sum_{i<j} r_1^{(i)} r_1^{(j)}d\log E(t,z_1^{(i)}))\,d\log E(t,z_1^{(j)}))\right]
\ee
$$
=\sum_{p\in \pi^{-1}(z_2)} \res\Big|_{t=z_2^{(k)}}\f{1}{df(x)}\left[\sum_{i}  r_1^{(i)} d\log E(t,z_1^{(i)}\right]^2
$$
which coincides with the r.h.s.  of (\ref{sys}).
\vskip0.2cm
Consider now the derivative with respect to $z_1$. The moduli of the Riemann surface $\CC$ remain unchanged
but each prime form has to be differentiated with respect to its  arguments.
Then
\be
\f{\p}{\p z_1}\log \prod_{i\neq j} E(z_1^{(i)},z_1^{(j)})^{r_1^{(i)} r_1^{(j)}}
=2\sum_{i\neq j} r_1^{(i)} r_1^{(j)} \f{d_x\log E(x,y)}{df(x)}\Big|_{x=z_1^{(i)}, y=z_1^{(j)}}\;.
\la{v2}\ee
On the other hand, the r.h.s. of (\ref{sys}) gives
$$
\sum_{k=1}^n \res\Big|_{x=z_1^{(k)}}\f{1}{df(x)}\left[\sum_{i}  r_1^{(i)} d\log E(x,z_1^{(i)})\right]^2
$$
$$
=\res\Big|_{z=z_1}\f{1}{d z}\sum_{k}\sum_i (r_1^{(i)})^2(d\log E(z^{(k)},z_1^{(i)}))^2
$$
\be
+\res\Big|_{z=z_1}\f{1}{dz}\sum_{k}\sum_{i\neq j}r_1^{(i)}r_1^{(j)} d\log E(z^{(k)},z_1^{(i)}) d\log E(z^{(k)},z_1^{(j)})\;.
\la{rhs2}\ee

In the first sum in (\ref{rhs2}) we can interchange the order of summation and sum over $k$ first; this gives the meromorphic differential on the base:
\be
\f{1}{dz}\sum_{k}(d\log E(z^{(k)},z_1^{(i)}))^2 =\f{dz}{(z-z_1)^2}
\ee
which does not have residue at $z_1$.

The second sum contributes when either $k=i$ or $k=j$; due to the  symmetry we get
$$
2\sum_{i\neq j}r_1^{(i)}r_1^{(j)} d\log E(z_1^{(j)},z_1^{(i)})
$$
computed in the local parameter $z-z_1$, which coincides with (\ref{v2}).

$\Box$

Factors $\tau_0$ and $\tau_B^{-1/2}$ in  the formula  (\ref{tauJM1}) for the tau-function never vanish in $\C^M\setminus\{z_j=z_k\}$.
Therefore, the Malgrange divisor of the Riemann-Hilbert problems with quasi-permutation monodromies is defined by the equation
$$
\Thpq\left(\sum_{m=1}^M\sum_{j=1}^n r_m^{(j)} \Acal_{x_0}(z_m^{(j)})\right)=0\;.
$$

\begin{remark} \rm 
The formula for $\tau_0$ proposed in  \cite{Annalen} was incorrect. The expression (\ref{sol}) was first given 
in \cite{GavMar} for the case of coverings with special  monodromy groups (the monodromy matrices from    \cite{GavMar}  were split in pairs
such that $M_{2k} M_{2k-1}=I$); the tau-function $\tau_0$ was named in  \cite{GavMar} the "Seiberg-Witten" tau-function although 
we are not aware of the logic behind this terminology. The full expression (\ref{tauJM1}) 
was interpreted in \cite{GavMar}  as  conformal block of a conformal field theory in the spirit of  recent works
on the link between conformal blocks and isomonodromic tau-functions (see for example \cite{Lisov}).
\end{remark}

\section{Self-dual $SU(2)$ invariant Einstein manifolds}
\la{SDEsec}

The complete 
description of $SU(2)$ invariant self-dual Einstein metrics metrics
of this type was given by Hitchin \cite{Hitc94} in terms of special solutions of Painlev\'e 6 equation. The analysis presented 
in \cite{Hitc94} essentially relied on the previous works \cite{Tod91,Tod94}. The final formulas derived in  \cite{Hitc94} were 
rather complicated; the elementary formulas for these metrics were obtained later in \cite{BabKor98} using the expression for the  Bergman tau-function on the space of elliptic curves derived in \cite{KitKor98}. In this section we summarize the approach of  \cite{BabKor98}.

The $SU(2)$-invariant self-dual
Einstein metric can be written in the following form \cite{Tod91,Tod94}
\be
g=F\left\{d\mu^2 +\f{\si_1^2}{W_1^2} +\f{\si_2^2}{W_2^2}+\f{\si_3^2}{W_3^2}\right\}
\la{g}\ee
where the 1-forms $\si_j$ satisfy
\be
d\si_1 =\si_2\wedge \si_3\;, \hskip0.5cm
d\si_2 =\si_3\wedge \si_1\;, \hskip0.5cm
d\si_3 =\si_1\wedge \si_2\; ,
\la{ds}\ee
and the functions $W_j$ depend only on the  "euclidean time" $\mu$.
In terms of the  new variables $A_j(\mu)$  (the connection coefficients) defined  by  equations
\be
\f{d W_j}{d\mu} = - W_k W_l + W_j (A_k+ A_k)\;,
\la{motion}
\ee
where  $(j,k,l)$ is an arbitrary permutation of the indexes $(1,2,3)$ the
condition of the self-duality of the metric  is given by  the classical Halphen system:
\be
\f{d A_j}{d\mu} = - A_k A_l + A_j (A_k+ A_k)\;.
\la{Halphen}\ee

Any solution of (\ref{Halphen}) can be substituted into the system (\ref{motion}), which defines then 
metric coefficients $W_j$.

The full system (\ref{motion}), (\ref{Halphen}) is invariant 
with respect to an $SL(2,\R)$ M\"{o}bius transformations:
\be
\mu\to \tilde{\mu} \f{a\mu + b}{c\mu + d}\;,\hskip1.0cm ad-bc=1 \;,\hskip0.6cm
\la{conmu}\ee
\be
W_j\to  \tilde{W}_j = (c\mu+d)^2 W_j\;,\hskip2.0cm
\la{conW}\ee
\be
A_j\to  \tilde{A}_j=  (c\mu+d)^2 A_j +c (c\mu+d)^2\;.
\la{conA}\ee

The system (\ref{Halphen}) has several interesting special cases. When 
one chooses the trivial solution $A_j=0$ of the system (\ref{Halphen}), the  corresponding  system  (\ref{motion}) 
reduces to the equations of the Euler top. 
If two of the  functions $A_j$ vanish, the remaining one must be  constant. 
Another special case is  when for all $j$ one chooses $W_j=A_j$.  We refer to \cite{Hitc94} for the list of references where these special cases were discussed.

The system (\ref{motion}), corresponding to the  general solution of the
system  (\ref{Halphen}), was related in the papers \cite{Tod94,Hitc94} to the four-point 
Schlesinger system. Moreover, it turned out that the conformal factor $F$ 
can be chosen to make the metric (\ref{g}) satisfy the  Einstein equation  exactly in the case when the
system (\ref{motion}) can be solved in terms of elliptic functions.

It is well-known that the functions
\be
A_1= 2\f{d}{d\mu}\log\th_2\;,\hskip0.5cm 
A_2= 2\f{d}{d\mu}\log\th_3\;,\hskip0.5cm A_3= 2\f{d}{d\mu}\log\th_4\;
\la{Wtheta}\ee 
where
$$
\th_2 \equiv \th\Big[\f{1}{2},0\Big](0,i\mu)\;, \hskip0.5cm
\th_3 \equiv \th[0,0](0,i\mu)\;, \hskip0.5cm
\th_4 \equiv \th\Big[0,\f{1}{2}\Big](0,i\mu)\;
$$
are standard theta-constants, solve the Halphen system (\ref{Halphen}). 
 The general solution of (\ref{Halphen}) may be obtained applying the M\"{o}bius transformations 
(\ref{conmu}), (\ref{conA}) to solution (\ref{Wtheta}). Therefore, it is sufficient to 
solve the system  (\ref{motion}), where the functions  $A_j$ are given by  (\ref{Wtheta}). 
Then the general solution of the system (\ref{motion}) can be obtained by the M\"obius transformation from the 
general solution of the system
\be
\f{dW_j}{d\mu}= - W_k W_l+ 2 W_j\f{d}{d\mu}\log(\th_{k+1}\th_{l+1})\; .
\la{eqOt}\ee

In general, the system (\ref{eqOt}) can not be solved in elementary or special functions. 
However, such solution is possible if $W_i$ satisfy the following additional condition (which is nothing but the fixing of the value of 
an integral of motion of the system (\ref{eqOt}):
\be
\th_2^4\Ot_1^2-\th_3^4\Ot_2^2+\th_4^4\Ot_3^2=\f{\pi^2}{4}\th_2^4\th_3^4\th_4^4\; .
\la{intt}\ee

The relationship of the system (\ref{eqOt}) to isomonodromic deformations  and the tau-function is most naturally seen in terms of variables
\be
\Om_1 =  -\f{W_2}{\pi \th_2^2 \th_4^2} \; ,\hskip0.5cm
\Om_2 =  -\f{W_3}{\pi \th_2^2 \th_3^2} \; ,\hskip0.5cm
\Om_3 =  -\f{W_1}{\pi \th_3^2 \th_4^2} \;.
\la{Om}\ee
which were used in \cite{Tod94,Hitc94}. Then the relation (\ref{intt}) takes the simple form
\be
-\Om_1^2+\Om_2^2+\Om_3^2=\f{1}{4}\;.
\la{int}\ee

To rewrite the system (\ref{eqOt}) in terms of $\Om_j$'s one also makes the change of the independent   variable from $\mu$ to $x$ such that the period 
of the elliptic curve
\be
\nu^2= \l(\l-1)(\l-x)\;.
\la{L}\ee
equals $i\mu$. It is assumed that $0<x<1$, the $a$-cycles goes around $[0,x]$ and $b$-cycle goes around $[x,1]$.
Then the system (\ref{eqOt}) is equivalent to
\be
\f{d\Om_1}{d x}= -\f{\Om_2\Om_3}{x(1-x)}\;, \hskip0.5cm
\f{d\Om_2}{d x}= -\f{\Om_3\Om_1}{x}\;, \hskip0.5cm
\f{d\Om_3}{d x}= -\f{\Om_1\Om_2}{1-x}\;. \hskip0.5cm 
\la{eqOm}\ee
To derive equations (\ref{eqOm}) from (\ref{eqOt}) one needs to use the elementary version of the variational formula (\ref{varO3}):
\be
\f{d\mu}{d x}=\f{\pi }{4K^2 x(x-1)}\;.
\la{sx}\ee
where $K$ is the full elliptic integral
\be
K=\f{1}{2}\int_0^x\f{d\l}{\sqrt{\l(\l-1)(\l-x)}}\;
\la{w}\ee
and the standard expressions for theta-constants are given by
\be
\th_2^4 =  \f{4}{\pi^2} K^2 x\;, \hskip0.8cm
\th_3^4 =  \f{4}{\pi^2} K^2 \;, \hskip0.8cm
\th_4^4 =  \f{4}{\pi^2} K^2 (1-x)\;.
\la{thomae}\ee

The metric (\ref{g}) in terms of $\Om_j$'s can be written as follows
\be
g=\tilde{F}\left\{\f{dx^2}{x(1-x)} +\f{\si_1^2}{\Om_1^2} +\f{(1-x)\si_2^2}{\Om_2^2}+
\f{x\si_3^2}{\Om_3^2}\right\}\;,
\la{g1}\ee 
where the conformal factors $\tilde{F}$ and $F$ are related by
$
\tilde{F}=\f{1}{\th_2^4\th_4^4} F\;.
$

Let variables $\Om_j$ solve the system (\ref{eqOm}),
and satisfy the  condition (\ref{int}).  Then the 
metric (\ref{g1}) satisfies the  Einstein equations with the cosmological constant  $\Lambda$
 if the conformal factor $\tilde{F}$ is given by the complicated explicit expression found in
 \cite{Tod94}.

The system (\ref{eqOm}) arises in the context of isomonodromic deformations
of the  equation 
\be
\f{d\Psi}{d\l} = \Big(\f{A^0}{\l}+\f{A^1}{\l-1}+\f{A^x}{\l-x}\Big)\Psi\;;
\la{lsSc}\ee
the isomonodromy condition (assuming the normalization $\Psi(\infty)=I$)
 implies the equation
\be
\f{d\Psi}{d x}= - \f{A^x}{\l-x}\Psi\;.
\la{lsx}\ee
The compatibility condition of equations (\ref{lsSc}) and (\ref{lsx}) is
given by the  Schlesinger system 
\be
\f{dA^0}{dx}=\f{[A^x,A^0]}{x}\;, \hskip0.5cm
\f{dA^1}{dx}=\f{[A^x,A^1]}{x-1}\;, \hskip0.5cm
\f{dA^x}{dx}=-\f{[A^x,A^0]}{x}-\f{[A^x,A^1]}{x-1}\;.
\la{Sch}\ee

Eigenvalues of the matrices $A^0$, $A^x$ and $A^1$ are integrals of motion of the Schlesinger system.
If  one chooses
\be
 \tr (A^0)^2=\tr (A^1)^2=\tr (A^x)^2=\f{1}{8}\;,
\la{trA}\ee
then the formulas
\be
\Om_1^2=-(\f{1}{8} +\tr A^0 A^1)\;, \hskip0.5cm
\Om_2^2=\f{1}{8} +\tr A^1 A^x\;, \hskip0.5cm
\Om_3^2=\f{1}{8} +\tr A^0 A^x
\la{defOm}\ee
give the solution of  the system (\ref{eqOm}), (\ref{int}) (see \cite{Hitc94}).

As a corollary of the conditions  (\ref{trA}), the eigenvalues of all
monodromies $M^0$, $M^1$ and $M^x$ equal to $\pm i$.
Such sets of monodromy matrices can be completely classified   \cite{Hitc94}: up to a  simultaneous constant
similarity transformation they are either given by the triple
$$                                                 {}
M_0=\left(\ba{cc}0 & -ie^{-2\pi i q} \\
                 -ie^{2\pi i q} & 0      \ea\right)\;; \hskip0.8cm
M_1=\left(\ba{cc}0 & ie^{-2\pi i (p+q)} \\
                 i e^{2\pi i (p+q)} & 0      \ea\right)\;;
$$
\be
M_x=\left(\ba{cc}0 & -ie^{-2\pi i p} \\
                 -i e^{2\pi ip} & 0      \ea\right)\; \hskip0.8cm
p,q\in \C\;,
\la{genM}\ee
or the triple
\be
M_0=\left(\ba{cc}  -i & q_0 \\
                    0 & i     \ea\right)\;; \hskip0.5cm
M_1=\left(\ba{cc} -i  & -i+ q_0   \\
                   0  & i      \ea\right)\;; \hskip0.5cm
M_x=\left(\ba{cc} -i  & -i \\
                   0  & i     \ea\right)\;, \hskip0.5cm  q_0\in \C
\,.                                      {}
\la{degM}\ee

In \cite{Hitc94} the system (\ref{eqOm}), (\ref{int})  was solved in terms
of elliptic functions
(independently in the cases (\ref{genM}) and  (\ref{degM}))
using the link between $\{\Om_j\}$ and the solution $y(x)$ of the Painlev\'{e} 6
equation with coefficients
$(\f{1}{8},\,-\f{1}{8},\,\f{1}{8},\,\f{3}{8})$, which is known to be
equivalent to the system (\ref{Sch}).

If one  then  directly expresses $\{\Om_j\}$ in terms of elliptic
functions then the resulting expressions turn out to be very
cumbersome (see \cite{Hitc94}).

Simple 
formulas for variables $\Om_j$ and $W_j$ were derived in \cite{BabKor98} using the expression for the Jimbo-Miwa tau-function 
 of the Schlesinger system
(\ref{Sch}) obtained in \cite{KitKor98}. This tau-function is   defined by equation
\be
\f{d}{dx}\log\tau (x)= \f{\tr A^0 A^x}{x} + \f{\tr A^1 A^x}{x-1}\;.
\la{taudef11}\ee

The solution of the system
(\ref{eqOm}) 
can be expressed in terms of the
$\tau$-function (\ref{taudef11}) as follows:

\be
\Om_1^2=\f{d}{dx}\{x(x-1)\f{d}{dx}\log\tau(x)\}\;,
\la{Om1}\ee
\be
\Om_2^2=(1-x)\f{d}{dx}\{x(x-1)\f{d}{dx}\log\tau(x)\}+x(x-1)\f{d}{dx}\log\tau(x)+\f{1}{8}\;,
{}
\la{Om2}\ee
\be
\Om_3^2=x\f{d}{dx}\{x(x-1)\f{d}{dx}\log\tau(x)\}-x(x-1)\f{d}{dx}\log\tau(x)+\f{1}{8}
\la{Om3}\ee
solve   (\ref{eqOm}), (\ref{int}).

 The Jimbo-Miwa $\tau$-function of the Schlesinger system (\ref{Sch}), corresponding
to monodromy matrices (\ref{genM}), is given by  the formulas (\ref{Berghyp}) and (\ref{tauJM1})
where $\tau_0=1$ and
$$
\tau_B=K[x(1-x)]^{1/4}\;.
$$
(In present case 
${\rm det}\,A$ from (\ref{Berghyp}) equals 4 times the elliptic integral $K$ (\ref{w})).  Therefore, 
\be
\tau(x)=\f{\th[p,q]}{K^{1/2}[x(1-x)]^{1/8}}
\la{tau0}
\ee
If, moreover, we make use of the Thomae formulas (\ref{thomae}) 
we get (up to an unessential constant)
\be
\tau(x)=\f{\th[p,q]}{\sqrt{\th_2\th_4}}\;.
\la{tau}
\ee

Now a  long but straightforward computation leads to the following theorem.

\begin{theorem}\cite{BabKor98}
The general two-parametric family of solutions of the system 
(\ref{eqOt}),  satisfying condition
(\ref{intt}), is given by the following formulas:
$$
W_1=-\f{i}{2}\th_3\th_4\f{\f{d}{d q}\th[p,q+\f{1}{2}]}{e^{\pi i p}\th[p,q]}\;,\hskip1.0cm
W_2=\f{i}{2}\th_2\th_4\f{\f{d}{d q}\th[p+\f{1}{2},q+\f{1}{2}]}{e^{\pi i p}\th[p,q]}\;,$$
\be
W_3=-\f{1}{2}\th_2\th_3\f{\f{d}{d q}\th[p+\f{1}{2},q]}{\th[p,q]}
\la{WWW}\ee
where $\th[p,q]$ denotes the  theta-function with characteristics
of vanishing first argument $\th[p,q](0,i\mu)$; $p,q\in\C$. 

The corresponding metric (\ref{g}) is real and satisfies  Einstein's equations with  the negative cosmological constant $\Lambda$
if 
\be
p\in\R\;, \hskip0.8cm \Re q=\f{1}{2}
\la{rc2}\ee
and the conformal factor $F$ is given by the following formula:
\be
F= \f{2}{\pi\Lambda} \f{W_1 W_2 W_3}{\Big(\f{d}{dq}\log\th[p,q]\Big)^2}\;.
\la{Fgen}\ee
The metric (\ref{g}) is real and satisfies  Einstein's equations with the positive cosmological constant $\Lambda$
if 
\be
q\in\R\;, \hskip0.8cm \Re p=\f{1}{2}\;,
\la{rc4}
\ee
and the conformal factor is given by the same formula (\ref{Fgen}).

There exists also the additional one-parametric family of solutions of the system  (\ref{eqOt}):
\be
W_1=\f{1}{\mu+ q_0}+ 2\f{d}{d\mu}\log\th_2 \;,\hskip0.7cm
W_2=\f{1}{\mu+ q_0}+ 2\f{d}{d\mu}\log\th_3 \;,\hskip0.7cm
W_3=\f{1}{\mu+ q_0}+ 2\f{d}{d\mu}\log\th_4 \;
\la{WWWdeg}\ee
 ($q_0\in\R$), which defines  manifolds with the vanishing cosmological constant $\Lambda=0$
if the  conformal factor $F$ is defined by the formula
\be
F= C (\mu+q_0)^2 W_1 W_2 W_3\;,
\la{Fdeg}\ee
where $C>0$ is an arbitrary constant.
\end{theorem}

The metric coefficients (\ref{WWW}) and (\ref{WWWdeg}) have nice modular properties which in turn translate into modular properties 
of objects arising in the spectral geometry of such Einstein's manifolds \cite{FaFaMa15}. These modular properties were analyzed from cosmological perspective in
\cite{ManMar15}.

\section{Einstein equations with two Killing symmetries}
\la{Ernstsec}

Here we give another example of appearance of the Schlesinger equations and the Jimbo-Miwa tau-function in the theory of Einstein equations. Our presentation   is based on the papers \cite{TMF1989,KorNicPRL,KorMatFUNKAN}. Unlike the previous example here we consider the 
Einstein equations without cosmological constant in the physical $3+1$ signature.
The physical context depends on the choice of two commuting Killing vectors: if both Killing vectors are space-like one gets either the 
Einstein-Rosen waves with two polarizations or the Gowdy model. If one of these Killing vectors is space-like and another
one is time-like, the resulting spacetimes are stationary and axially symmetric. It is the latter  case which was considered in the original paper \cite{TMF1989} although the solutions found there can be easily extended to two other cases. 

The metric on the  stationary axially symmetric manifold can be written in the standard form
\be
ds^2 = f^{-1}[e^{2k}(dz^2 + d\rho^2) + \rho^2 d\phi^2] -f(dt + F d\phi)^2
\la{metricSA}
\ee
The metric coefficients $k, A $ and $f$ are functions of two coordinates $(\rho,z)$ only.
The most non-trivial part of the Einstein equations (called the {\it Ernst equation}) can be written in terms of only one complex-valued function 
$\Ecal(\rho,z)$ (the {\it Ernst potential}): 
\be
(\Ecal+\overline{\Ecal})(\Ecal_{xx}+\rho^{-1}\Ecal_{\rho}+\Ecal_{zz})=2(\Ecal_x^2+\Ecal_\rho^2)\;.
\la{Ernst}
\ee

The metric coefficients of (\ref{metricSA}) can be restored from the Ernst potential via equations
\be
f=\Re \Ecal\;,\hskip0.7cm \f{\p F}{\p \xi}=2\rho\f{(\Ecal-\overline{\Ecal})_\xi}{(\Ecal+\overline{\Ecal})^2}\hskip0.7cm 
\f{\p k}{\p \xi}=2i \rho\f{\Ecal_\xi\overline{\Ecal}\xi}{(\Ecal+\overline{\Ecal})^2}
\la{eqfAk}
\ee
where $\xi=x+i\rho$.

The Ernst equation is integrable in the sense of the  theory of integrable systems. Being nonlinear itself, it can be represented as the compatibility condition of the
associated  linear system found in \cite{Maison,BelZak}:
\be
\Psi_\xi=\f{G_\xi G^{-1}}{1-\gamma}\Psi\;,\hskip0.7cm 
\Psi_\xi=\f{G_\xi G^{-1}}{1-\gamma}\Psi
\ee
where
\be
G=\left(\ba{cc}2 & i(\Ecal-\Eb)\\ i(\Ecal-\Eb) & 2\Ecal\Eb\ea\right)\;;
\la{GE}
\ee
\be
\gamma=\f{2}{\xi-\xb}\left\{w-\f{\xi+\xb}{2}+\sqrt{(w-\xi)(w-\xb)}\right\}
\la{gammavar}
\ee
is the "variable spectral parameter" while
 $w\in \C$ is the constant spectral parameter; $\Psi(x,\rho,w)$ is the $2\times 2$ matrix-valued function.
 
 The link between the Fuchsian Schlesinger system and the Ernst equation was established in \cite{KorNicPRL}. Namely, let a solution $\Psi$ of the system (\ref{ls}) with simple poles
 (and $z$ replaced by $\gamma$)
 satisfy two additional conditions:
 \be
 \Psi^t (1/\gamma)\Psi^{-1}(0)\Psi(\gamma)=I\;,
 \ee
 \be
 \Psi(-\bar{\gamma})=\overline{\Psi}(\gamma)\;.
 \ee

 Let matrices $\{A_j\}$ be the corresponding solution of  the Schlesinger system (\ref{Schl1}),
  (\ref{Schl2}) with $z_0=\infty$ and let $z_j=\gamma(w_j,\xi,\xb)$ where the function $\gamma$ is given by
  (\ref{gammavar}). Then the matrix $G=\Psi(\gamma=0,\xi,\xb)$ has the structure (\ref{GE}) and 
  the corresponding function $\Ecal$ satisfies the Ernst equation (\ref{Ernst}).
  
  Moreover, the metric coefficient $k$ is related to the Jimbo-Miwa tau-function of the Schlesinger system by the following formula:
  \be
  e^{2k}=C \prod_{j} \left\{\f{\p\g_j}{\p w_j}\right\}^{\tr A_j^2/2} \tau_{JM}
  \la{contau}
  \ee
where $C$ is a constant. We notice that $\tr A_j^2$ are also constants of motion of the Schlesinger system which can be found directly from the monodromy matrices.
  
  This observation means that knowing a solution of Riemann-Hilbert problem with 
  some set of monodromy matrices one can always find related solution of the Ernst equation.
  In particular, solutions of RH problem with quasi-permutation monodromies described above give rise to
  algebro-geometric solutions of Ernst equation found in  \cite{TMF1989}.
  
Define the hyperelliptic curve $\CC$ of genus $g$ by
\be
y^2=(w-\xi)(w-\xb)\prod_{j=1}^{2g} (w-w_j)
\la{Ernstcurve}
\ee
where $w_j\in \C$. The curve (\ref{Ernstcurve}) is assumed to be real i.e. the branch points
 $w_j$ are either real or form conjugated pairs. 
 The branch points $\xi$ and $\xb$ of the curve $\CC$ depend on the space variables $(x,\rho)$.
 Denote the Abel map on $\CC$ by $\Acal$ and introduce two constant vectors $\pb,\qb\in \C^g$
which satisfy appropriate reality conditions \cite{TMF1989,KorMatFUNKAN}. 
These vectors define the monodromy group with  off-diagonal $2\times 2$ monodromy matrices,
see \cite{KitKor98}.

The the solution of (\ref{Ernstcurve}) is given by 
\be
\Ecal(\xi,\xb)=\f{\Theta\left[^\pb_\qb\right](\Acal(\infty^1)-\Acal(\xi))}{\Theta\left[^\pb_\qb\right](\Acal(\infty^2)-\Acal(\xi))}
\la{Ernstag}
\ee
where $\infty^1$ and $\infty^2$ are points at infinity on different sheets of $\CC$.

The expression (\ref{tauJM1}) for the tau-function of $2\times 2$ Riemann-Hilbert problem with quasi-permutation monodromies, combined with the link (\ref{contau}) between the conformal factor and the 
Jimbo-Miwa tau-function, leads then to the following simple formula for  the conformal factor $e^{2k}$ defined by  equations (\ref{eqfAk}):
\be
e^{2k}=\f{\Theta\left[^\pb_\qb\right](0)\Theta\left[^\pb_\qb\right](\f{1}{2}{\bf e})}{\Theta(0)\Theta(\f{1}{2}{\bf e})}
\la{conag}
\ee
where ${\bf e}=(1,\dots,1)$.

The theta-functional solutions (\ref{Ernstag}) in genus 2 turned out to have a physical significance:
they arise in the problem of analytical description of rigidly rotating infinitely thin dust disk (see \cite{Klein_book} and references therein for details).

\section{Bergman tau-function and moduli spaces }
\la{Bergmod}
\subsection{Hodge class on spaces of admissible covers}
\la{Hodgeadm}

The Bergman tau-function $\tau_B$ (\ref{tauBerg}) turns out to be an efficient tool in the study of geometry of moduli spaces. In particular, the close look at the 
properties of $\tau_B$ allows to interpret it as the natural higher genus analog of the Dedekind's eta-function. 

Consider the Hurwitz space $\Hcal_{g,n}$ of pairs $(\CC,f)$ where $\CC$ is a Riemann surface of genus $g$ and $f$ is  a meromorphic function of degree
$n$ on $\CC$ such that all poles and critical points of $f$ are simple. Then the number of critical points equals to $M=2g+2n-2$. 
Consider the quotient $\Hcal_{g,n}/\sim$ where $\sim$ is the equivalence relation between pairs $(\CC,f)$ and $(\CC,(af+b)/(cf+d))$
for an arbitrary $GL(2,\C)$  matrix $\left(\ba{cc} a & b \\ c & d\ea\right)$.  There exists the compactification of the space $\Hcal_{g,n}/\sim$ , called the 
space of {\it admissible covers} \cite{HMu}, which we denote by $\Hcal^{adm}_{g,n}$. Study of analytical properties of the Bergman tau-function (\ref{tauBerg})
near various boundary components of the space $\Hcal^{adm}_{g,n}$ allows to get a new relation in the Picard group of this space \cite{Advances}. \footnote{The tau-function (2.9) used in \cite{Advances} equals to 24th power of the Bergman tau-function on a Hurwitz space used in this paper}

The first key fact about $\tau_B$ is its transformation law under a change of Torelli marking of $\CC$ which is given by the following
proposition.
\begin{proposition}\la{sympltr}
Under the symplectic change of canonical basis of cycles $(a_i,b_i)$ on $\CC$ given by 
\be
\left(\ba{c} \tilde{b}_i \\ \tilde{a}_i  \ea\right)=\left(\ba{cc} A & B \\ C & D\ea\right)\left(\ba{c} b_i \\ a_i\ea\right)
\la{sympltrans}
\ee
the function $\tau_B$ transforms as follows:
\be
\tau_B (\CC,f)\to \epsilon\, {\rm det} (C\Omega+D)\tau_B (\CC,f)
\la{transT}
\ee
where $\epsilon^{24}=1$.
\end{proposition}
This proposition can be easily deduced  from the explicit formula (\ref{tauBerg}) (equations (\ref{tauFr1}) give this relation up to a multiplicative constant;
to prove that this constant is a $24$th root of unity one needs to use  the explicit formula (\ref{tauBerg})).

Another important property of $\tau_B$ is given by the following

\begin{proposition}\la{Mobtr}\cite{Advances}
Let $V(z_1,\dots, z_M)= \prod_{i<j} (z_i-z_j)$ be the Vandermonde determinant of the critical 
values $z_1,\dots, z_M$ of $f$. Then 
\be
\eta= \tau_B^{24(M-1)}V^{-6}
\la{eta}
\ee
is invariant under any M\"obius transformation $f\to (af+b)/(cf+d)$.
\end{proposition}

Propositions \ref{sympltr} and \ref{Mobtr} imply that $\eta$ is the section of the determinant of Hodge vector bundle over $\Hcal_{g,n}^{adm}$.
This section is singular on boundary divisors $\Delta_\mu^{(k)}$ of $\Hcal_{g,n}^{adm}$ which are parametrized by the following data \cite{HMu}:
$\mu=[n_1,\dots,n_r]$ is a partition of $1,\dots,n$ and  $(k,M-k)$ is a partition of the set of critical values. A generic nodal curve from $\Delta_\mu^{(k)}$ is obtained as follows. Consider the degeneration of the base $\CP1$ into the union of two Riemann spheres such that one of them carries $k$ critical values and another one carries the remaining $M-k$ critical values (such stable curves  form the divisor $D_k$ in $\Mcal_{0,M}$). Then $\mu$ is the partition corresponding to $S_n$  monodromy of the covering around the nodal point formed on the base.   The number $r$ of cycles in the permutation corresponding to the nodal point on the base equals to the number of nodes of the coverings forming $\Delta_\mu^{(k)}$.
The union of  $\Delta_\mu^{(k)}$ over all $\mu$ for a given $k$ is the  pre-image of the divisor $D_k$ under the inverse of the {\it branching map } from the Hurwitz space to the set of critical values of the function $f$ i.e. to $\Mcal_{0,M}$.

By studying the asymptotics of  $\tau_B$ and $\eta$ near $\Delta_\mu^{(k)}$ and using  (\ref{sympltr}) we get the following
\begin{theorem}\cite{Advances}
The Hodge class $\lambda\in {\rm Pic}(\Hb_{g,n})\otimes \Q$ can be expressed as follows in terms of the classes of boundary divisors:
\be\la{lamhur}
\lambda=\sum_{k=2}^{g+n-1}\sum_{\mu=[n_1,\dots, n_r]} \prod_{i=1}^r n_i
\left(\f{k(M-k)}{8(M-1)}-\f{1}{12}\sum_{i=1}^r\f{n_i^2-1}{n_i}\right)\delta_\mu^{(k)}\,.
\ee
where $\delta_\mu^{(k)}$ is the class of divisor $\Delta_\mu^{(k)}$.
\end{theorem}

In the case $n=2$, i.e. for the moduli spaces of hyperelliptic curves of given genus $(M-2)/2$ the formula (\ref{lamhur})  was found earlier by Cornalba and Harris in \cite{CH}: 
\be
\lambda=\sum_{i=1}^{[(g+1)/2]}\f{i(g+1-i)}{4g+2}\delta_{[1^2]}^{(2i)}+\sum_{j=1}^{[g/2]}\f{j(g-j)}{2g+1}\delta_{[2]}^{(2j+1)}\;.
\label{CorHar}
\ee
The formula (\ref{lamhur}) got later an alternative proof using algebro-geometric techniques and then used to 
compute the classes of the  Mumford divisors in $\Mcal_g$  by van der Geer and Kouvidakis in \cite{GK1,GK2,GK3}.

The theory of the Bergman tau-function on Hurwitz spaces can be naturally extended to other moduli spaces: 
spaces of holomorphic Abelian, quadratic and $n$-differential, as well as to moduli spaces of spin curves.
In all of these cases it allows to derive old and new facts about the geometry of these spaces.

\subsection{Spaces of holomorphic abelian  differentials}
\la{spacehol}

Denote by $\Hcal_g$ the Hodge bundle over $\Mcal_g$ and by $\ol{\Hcal}_g$ its extension to $\ol{\Mcal}_g$.

The space $\Hcal_g$ is the union of strata $\Hcal_g(k_1,\dots, k_n)$ such that $k_1+\dots+k_n=2g-2$;
an element of   $\Hcal_g(k_1,\dots, k_n)$ is the pair $(\CC,v)$, where $\CC$ is a Riemann surface of genus $g$ and $v$ is a  holomorphic abelian differential on $\CC$ with zeros $x_1,\dots,x_n$ of multiplicities $(k_1,\dots,k_n)$. Then the tau-function $\tau_B(\CC,v)$ on
$\Hcal_g(k_1,\dots, k_N)$ is defined by the formula which looks exactly as (\ref{tauBerg}) but where the divisor of the exact meromorphic differential $df$ is replaced by the divisor of the holomorphic differential $v$: $(v)=\sum_{j=1}^n k_j q_j=\sum_{j=1}^n k_j x_j$. 

To write down the system of equations for $\tau_B(\CC,v)$ similar to (\ref{tauFr1}) we introduce the
 system of local {\it homological} coordinates on 
$\Hcal_g(k_1,\dots, k_n)$  given by integrals of $v$ over a basis of the relative homology group $H_1(\CC,\{x_j\}_{j=1}^n)$; the number $2g+n-1$ of
cycles in this basis coincides with the dimension of $\Hcal_g(k_1,\dots, k_n)$. The set of independent cycles in this homology group can be chosen as follows (this is a special case of the basis (\ref{defsmerom}) in the case of meromorphic differentials):
\be
\{s_j\}_{j=1}^{2g+n-1} =(a_i,b_i,l_j)\;,\hskip0.7cm
i=1,\dots,g,\;\;\; j=1,\dots,n-1
\la{relbas}
\ee
where $l_j$ is a contour connecting $x_1$ with $x_{j+1}$. The homology group $H_1(\CC\setminus\{x_j\}_{j=1}^n$ is dual  to $H_1(\CC,\{x_j\}_{j=1}^n)$.
 The basis in the dual group which is dual to the basis (\ref{relbas}) in $H_1(\CC,\{x_j\}_{j=1}^n)$ is given by (this is the special case of 
(\ref{defsstarmer})):
\be
\{s_j^*\}_{j=1}^{2g+n-1}=(-b_i,a_i,\tilde{c}_j)
\la{dualba}
\ee
where $\tilde{c}_j$ is a small counter-clockwise contour around $x_{j+1}$; $s_j^*\circ s_k=\delta_{jk}$.

The equations for $\tau_B$ on the stratum $\Hcal(k_1,\dots, k_n)$ are given by the following analog of (\ref{tauFr1}):
\be
\f{\p \log \tau_B}{\p (\int_{s_j} v)}=-\f{1}{2\pi i} \int_{s_j^*} \f{B_{reg}(x,x)}{v(x)}\;.
\la{egtauab}
\ee

The main properties of $\tau_B$ on $\Hcal_g(k_1,\dots, k_n)$ are the following. First, $\tau_B$ is  holomorphic and non-vanishing
as long as the curve $\CC$ is non-degenerate i.e. $\CC$ does not approach the boundary of $\Mcal_g$ and the zeros of $v$ do not change their multiplicities (i.e. do not merge). Second, as well as in the case of Hurwitz spaces, $\tau_B$ transforms as follows under the change of the Torelli marking of $\CC$ given by (\ref{sympltrans}):
\be
\tau_B (\CC,f)\to \epsilon {\rm det} (C\Omega+D)\tau_B (\CC,f)
\la{transT1}
\ee
where $\e$ is a root of unity of degree $12 N$ and  $N$ is the least common multiple of $k_1+1,\dots,k_n+1$
(see (4.27) of \cite{CMP}).

Third, $\tau_B$ transforms as follows under the rescaling of $v$:
 
\be
\tau_B(\CC,\kappa v)=\tau_B(\CC,v)\kappa^{\f{1}{12}
\sum_{i=1}^n \f{k_i(k_i+2)}{k_i+1}}\;.
\la{Hom1}
\ee
These  properties can not be used to get relations in the Picard group of compactification of any stratum  
$\Hcal(k_1,\dots, k_n)$ since the geometric structure of such compactification is not well understood yet. However,
when all zeros of $v$ are simple such compactification  coincides with $P\ol{\Hcal}_g$. In this case one can prove the following theorem. 

\begin{theorem}\cite{MRL}
In the rational Picard group of  ${\rm Pic}(P\Hb_g)\otimes\Q$  the following relation holds:
\be
\lambda=\f{g-1}{4}\phi+\f{1}{24}\d_\deg+\f{1}{12}\d_0+\f{1}{8}\sum_{j=1}^{[g/2]}\d_j\;.
\label{mf}
\ee
Here $\lambda$ is the pullback of the Hodge class from $\Mb_g$ to $\Hb_g$, $\phi$ is the first Chern class of
the line bundle associated to the projection $\Hb_g\to P\Hb_g$,
 $\d_\deg$ is the class of the divisor of abelian differentials with multiple zeros, and $\d_j,\;j=0,\dots,[g/2],$ are the pullbacks of the classes of the Deligne-Mumford boundary divisors from $\ol\Mcal_g$ to $P\Hb_g$.
\label{mt}
\end{theorem}

\subsection{Spaces of quadratic differentials: Hodge and Prym tau-functions}
\la{spaceQHP}

Denote by $\Qcal_g$ the moduli space of pairs $(\CC,Q)$ where $\CC$ is a Riemann surface of genus $g$ and $Q$ is a holomorphic quadratic differential on $\CC$.
The canonical covering  $\Ch$ of $\CC$ is defined by the equation
\be
v^2=Q
\la{cc1}
\ee
in $T^*\CC$. Introduce a subspace $\Qcal_g^0$ of $\Qcal_g$ corresponding to differentials $Q$ with $4g-4$ simple zeros at $x_1,\dots,x_{4g-4}$; we have ${\rm dim} \Qcal_g^0= {\rm dim} \Qcal_g=6g-6$. For a point $(\CC,Q)\in \Qcal_g^0$ denote zeros of
$Q$ by $x_1,\dots,x_{4g-4}$; the 
 genus of the canonical cover $\Ch$  equals $4g-3$. The zeros $x_j$ are branch points of the
covering $\pi:\Ch\to\CC$. Denote the involution on $\Ch$ interchanging the sheets by $\mu$. The linear operator $\mu_*$ acting in $H_1(\Ch,\R)$ 
has eigenvalues $\pm 1$; denote the corresponding eigenspaces by $H_+$ and $H_-$. We have ${\rm dim} H_+=2g$ ($H_+$ can be identified with $H_1(\CC,\R)$) and ${\rm dim} H_-=2g_-$ with $g_-=3g-3$. The system of homological coordinates on  $\Qcal_g^0$ is given by choosing a symplectic basis $\{s_j\}_{j=1}^{2g_-}=\{a_i^-,b_i^-\}_{i=1}^{g_-}$ in $H_-$ and integrating $v$ over this basis:
\be
\Pcal_{s_i}=\int_{s_i} v
\la{homolcoord}
\ee

Similarly, we decompose the space $H^{(1,0)}(\Ch)$ of holomorphic differentials on $\Ch$ into $\pm 1$ eigenspaces of $\mu^*$:  $H^{(1,0)}(\Ch)=H^+\oplus H^-$ where elements of $H^-$
are called {\it Prym differentials}.

Denote the canonical bidifferentials on $\CC$ and $\Ch$ by $B$ and $\Bh$,
respectively and define
\be
B_{reg}(x,x)=\left(B(x,y)-\f{v(x)v(y)}{(\int_x^y v)^2}\right)\Big|_{y=x}
\la{BregC}
\ee
($B_{reg}(x,x)$ is the meromorphic quadratic differential on $\CC$ depending on the choice of  the Torelli marking on $\CC$; we can also pull it back   to $\Ch$ and denote by the same letter, slightly abusing the notations) and
\be
\Bh_{reg}(x,x)=\left(\hat{B}(x,y)-\f{v(x)v(y)}{(\int_x^y v)^2}\right)\Big|_{y=x}
\la{Bhreg}
\ee
which is the   meromorphic quadratic differential on $\Ch$ depending on the choice of the Torelli marking on $\Ch$.

Equivalently, 
\be
B_{reg}(x,x)=\f{1}{6}(S_B-S_v)\;,\hskip0.7cm
\Bh_{reg}(x,x)=\f{1}{6}(\widehat{S}_B-S_v)
\ee
where $S_B$ and $\widehat{S}_B$ are the Bergman projective connections on $\CC$ and $\Ch$, respectively and $S_v=\{\int^x v, \cdot\}$ is the
projective connection (defined on $\Ch$ and invariant under $\mu$, thus also giving the projective connection on $\CC$) defined by differential $v$.

Now we define two Bergman tau-functions, $\tau(\CC,Q)$ and 
$\tauh(\CC,Q)$, by equations
\be
\f{\p \log\tau}{\p\int_{s_i}v}=-\f{1}{4\pi i }\int_{s_i^*} \f{B_{reg}(x,x)}{v(x)}\;,
\la{tauqu}
\ee
\be
\f{\p \log\tauh}{\p\int_{s_i}v}=-\f{1}{4\pi i }\int_{s_i^*} \f{\hat{B}_{reg}(x,x)}{v(x)}\;.
\la{tauquh}
\ee
Let us assume now that the Torelli markings used to define $B$ and $\Bh$ are compatible i.e. we choose the canonical basis in
$H_1(\Ch,\Z)$ as follows:
\be
\{a_j,a_j^\mu,\tilde{a}_k,b_j,b_j^\mu,\tilde{b}_k\}
\la{canbCh}
\ee
such that
\be
\mu_* a_j=a_j^{\mu},\hskip0.7cm \mu_* b_j=b_j^{\mu},\hskip0.7cm
\mu_*\tilde{a}_k+\tilde{a}_k=\mu_*\tilde{b}_k+\tilde{b}_k=0
\ee
where $(a_i,b_i,a_i^\mu,b_i^\mu)$ is the pullback of the canonical basis of cycles $(a_i,b_i)$ from  $\CC$ to $\Ch$.

Denote by
$$
(\hat{v}_j,\hat{v}_j^\mu,\wh_k)
$$
the corresponding basis of normalized abelian differentials on $\Ch$. The differentials $v_j^+=\vh_j+\vh_j^\mu$ for $j=1,\dots,g$ form
a basis in the space $H^+$; these differentials are naturally identified with the normalized holomorphic differentials $v_j$ on $\CC$.
The basis in $H_-$ is given by $g_-=3g-3$ Prym differentials $v_l^-$:
\be
\{\vh_l-\vh_l^\mu, \wh_k\}\;,\hskip0.7cm l=1,\dots,g,\;\;\; k=1,\dots,2g-3
\la{Prymdif}
\ee

The classes in $H_1(\Ch,\R)$ given by
\be
a_j^+=\f{1}{2}(a_j+a_j^\mu)\;,\hskip0.7cm
b_j^+=b_j+b_j^\mu
\la{apbp}
\ee
with the intersection index $a_j^+\circ b_k^+=\f{1}{2}\delta_{jk}$,
form the basis in $H_+$ whereas the classes
\be
a_l^-=\f{1}{2}(a_l-a_l^\mu),\hskip0.7cm b_l^-=b_l-b_l^\mu\;\hskip0.7cm l=1,\dots,g
\la{prc1}
\ee
\be
a_l^-=\tilde{a}_{l-g},\hskip0.7cm b_l^-=\tilde{b}_{l-g}  \;\hskip0.7cm l=g+1,\dots,g_-
\la{prc2}
\ee
form the basis in $H_-$ with the  intersection indeces $a_j^-\circ b_k^-=\delta_{jk}$.

The period matrix $\Omega^+$ of differentials $v_j^+$ is equal to twice  the period matrix of $\CC$:
\be
\Omega_{jk}^+=\int_{b_k^+} v_j^+=2\Omega_{jk}\;.
\ee
Interating the Prym differentials (\ref{Prymdif}) over the cycles (\ref{prc1}),  (\ref{prc2}) we get the matrix $\Omega^-$ which is equal to $2\Pi$ where $\Pi$ is the Prym matrix defined in \cite{Fay73}, p. 86:
\be
\Omega_{jk}^-=\int_{b_k^-} v_j^-=2\Pi_{jk}\;.
\ee
Then the period matrix $\Oh$ of $\Ch$ is given by
\be
\widehat{\Omega}=
T^{-1}
 \left(\ba{cc} \Omega^+ & 0\\
                0  & \Omega^-  \ea\right) T^{-1}=2 T^{-1}
 \left(\ba{cc} \Omega & 0\\
                0  & \Pi  \ea\right) T^{-1} 
 \la{shs}
\ee
where
\be
T=  \left(\ba{ccc} I_g & I_g& 0     \\
                     I_g & -I_g  & 0 \\
                    0& 0 & I_{\gm-g}  \ea\right)\;.
\ee

Under this choice of Torelli marking of $\CC$ and $\Ch$ we introduce  (following \cite{Leipzig,BK2})
the {\it Hodge tau-function}
\be
\tau_+(\CC,Q)=\tau(\CC,Q)
\la{Hodgetau}
\ee
and {\it Prym tau-function}
\be
\tau_-(\CC,Q)=\frac{\tauh(\CC,Q)}{\tau(\CC,Q)}\;.
\la{Prymtau}
\ee

The explicit formula for $\tau_+(\CC,Q)$ is given by  (\ref{tauBerg}) with only a slight modification:
the divisor $\sum k_i q_i$ should be formally  replaced by  $\sum_{i=1}^{4g-4}\f{1}{2} x_i$; the distnguished local parameters near
$x_i$ should be chosen as $[\int_{x_i}^x v]^{2/3}$ and the vectors ${\bf s},{\bf r}\in \Z^g/2 $ should be  defined by  the following analog
of (\ref{defsrHur}):
\be
\f{1}{2}\Acal_x((Q))+2K^x+\Omega {\bf s}+ {\bf r}=0\;.
\ee

The tau-function   $\tauh(\CC,Q)$ is also  given by  (\ref{tauBerg}), but this time applied to the pair $(\Ch,v)$ i.e. it coincides with the
tau-function on the space of holomorphic Abelian differentials $\tau(\Ch,v)$ as defined by (\ref{tauBerg}) with $\CC$ replaced by $\Ch$ and the differential $df$ replaced by $v$;   the divisor $(v)$ on $\Ch$ is given by
$\sum_{i=1}^{4g-4} 2 x_i$).

The key property of the tau-functions $\tau_\pm$ is their transformation under the change of the symplectic bases
in $H_+$ and $H_-$. Namely, for a symplectic transformation $\sigma$ in $H_1(\Ch)$ which is commuting with $\mu_*$ and acting by matrices
\be
\s_\pm=\left(\ba{cc} A_\pm & B_\pm  \\ C_\pm & D_\pm \ea\right)
\la{spm}
\ee
in $H_\pm$ the tau-functions $\tau_\pm$ transform as follows:
\be
\f{\tau_\pm^\s}{\tau_\pm}=\gamma_\pm (\s) \det(C_\pm \O_\pm + D_\pm)
\la{tausigma}
\ee
where $\gamma_\pm^{48}(\s)=1$.

Another important property is the behaviour of $\tau_\pm$ under the rescaling of $Q$:

\be
\tau_+(\CC,\kappa Q)=\kappa^{\f{5(g-1)}{{36}}} \tau_+(\CC,\kappa Q)
\;,\hskip0.7cm
\tau_-(\CC,\kappa Q)=\kappa^{\f{11(g-1)}{{36}}} \tau_-(\CC,\kappa Q)\;.
\la{rescsz}
\ee

Both tau-functions, $\tau_\pm$, are holomorphic and non-vanishing on $\Qcal_g^0$. Denote by $D_{\deg}$ the divisor in $\ol{\Hcal}_g$
containing differentials $Q$ with multiple zeros. The asymptotics of $\tau_\pm$ near  $D_{\deg}$ was computed in \cite{contemp}.
Namely, let two zeros of $Q$, say $x_1$ and $x_2$, merge.  It is easy to show (Lemma 8 of \cite{contemp})  that the transversal local coordinate
on $\ol{\Hcal}_g$ in a neighbourhood of  $D_{\deg}$ can be chosen to be $t_{\deg}=\int_{x_1}^{x_2}v$, and near $D_{\deg}$ the tau-functions have the following asymptotics:
\be
\tau_+\sim t_{\deg}^{1/72} (const + o(1))\;,\hskip0.7cm
\tau_-\sim t_{\deg}^{13/72} (const + o(1))\;.
\ee

The transversal local coordinate $t$ on $\Hcal_g$ near components of Deligne-Mumford boundary can be expressed in terms of the periods of $v$
along the corresponding vanishing cycle $a$ and its dual cycle $b$ as $t=\exp\{2\pi i \int_b v/\int_a v\}$. The asymptotics of $\tau_\pm$ in terms of $t$ can also be computed (Sec 5.2 and 5.3 of
\cite{contemp}) to give $\tau_\pm\sim t^{1/12}(1+o(1))$.

These analytical facts about $\tau_\pm$  allow to relate different classes in $Pic(P\ol{\Qcal}_g,\Q)$. Namely,
denote by $\lambda=\lambda_H$ the first Chern class of the  Hodge vector bundle and by $\Lambda_P$ the first Chern class of the Prym vector bundle with the fiber $H^-$. Denote also by $\phi$ the first Chern class of the tautological line bundle $L$
associated to the projection $P\ol{\Qcal}_g\to \ol{\Qcal}_g$.

Then the analytical properties of $\tau_\pm$ imply the following expressions for the classes $\l$ and $\l_P$ in  $Pic(P\ol{\Qcal}_g,\Q)$ (Th.1 of \cite{contemp}):
\be
\lambda=\f{5(g-1)}{36}\,\phi +\f{1}{72}\, \delta_{\deg}+\f{1}{12}\delta_{DM}
\la{Hodgecl}
\ee
\be
\lambda_P=\f{11(g-1)}{36}\,\phi +\f{13}{72}\, \delta_{\deg}+\f{1}{12}\delta_{DM}
\la{Prymcl}
\ee
where $\delta_{DM}$ is the class of pullback of the Deligne-Mumford boundary to $P\ol{\Qcal}_g$.

Excluding the class $\delta_{\deg}$ from (\ref{Hodgecl}), (\ref{Prymcl}), we get

\be
\lambda_P - 13 \l= -\sum_{i=0}^{[g/2]}\delta_i -\f{3g-3}{2} \phi \;.
\la{pt10}
\ee

For each pair $(\CC,Q)$ the vector space $H^-$ of Prym differentials is closely related to the 
space $\Lambda^{(2)}$ of holomorphic quadratic differentials on $\CC$. 
Namely, for each quadratic differential $\tilde{Q}\in \Lambda^{(2)} $ the ratio $\pi_*\tilde{Q}/v$
in a Prym differential. This relation implies the following equality between the determinant classes: $\l_P=p^*\l_2-\f{3g-3}{2}\psi$ where $\l_2$  is $c_1$ of the determinant line bundle of the vector bundle of quadratic differentials and $p$ is the natural projection of 
$P\ol{\Qcal}_g$ to $\Mcal_g$. In this formula $3g-3$ is the dimension of the fiber and $1/2$ appears since multiplication of $Q$ with $\kappa$ 
corresponds to transformation $v\to\kappa^{1/2}v$ of $v$.
Now (\ref{pt10}) leads to the celebrated Mumford formula
\be
\l_2-13\lambda =\delta_{DM}
\label{M}
\ee

\subsection{Tau-functions on  higher strata of $\Qcal_g$ }
\la{highQ}

The construction of the previous section can be extended to an arbitrary stratum of the space $\Qcal_g$ with fixed multiplicities of zeros of
$\qd$. 
Namely, consider a stratum  $\Qcal_{g}^{\bk,\bl}$ of the space $\Qcal_g$ such that the differential $Q$ has $\mo$  zeros of odd multiplicity and  $\me$   zeros of even multiplicity. Then the  divisor $(Q)$ has the  form
\be
(\qd)=\sum_{i=1}^{m} d_i q_i\equiv\sum_{i=1}^\mo (2k_i+1)x_i+\sum_{i=\mo+1}^{m} 2l _i x_i
\la{divqd1}
\ee
where $k_i,l_i\geq 0$,  $m= \mo + \me$.

The canonical covering $v^2=\qd$
(\ref{cancov}) is branched at  zeros of odd multiplicity $\{x_i\}_{i=1}^\mo$.
Note that $\mo$ is always even because  $\deg (\qd) = 4g-4$ is even.
Therefore the  genus of $\Ch$ equals
 \be
\hat{g}=2g+\f{\mo}{2}-1.
\la{genush}
\ee

Each  zero of even multiplicity $\{x_i\}_{i=\mo+1}^m$ has  two  pre-images on $\Ch$  which we denote by $\xh_i$ and 
$\xh_i^\mu$.
 The pre-images  of poles and zeros
of odd multiplicity $\{x_i\}$ on  $\Ch$ are branch points  (as before,  we continue to denote them by the same letters, omitting the hat).

Decomposing the space of holomorphic differentials into invariant subspaces of $\mu^*$, $H^{(1,0)}(\Ch) = H^+ \oplus H^-$, and computing their dimensions, we get $\dim \, H^+ = g$ and 
$ \dim \, H^{-} =\gm= g+\f{\mo}{2}-1$. The space $H_+$ is the fiber of the  Hodge vector bundle $\Lambda_H$ over 
$\Qcal_{g}^{\bk,\bl}$ while the space  $H^-$  is the fiber of the  Prym vector bundle  $\Lambda_P$ over $\Qcal_{g}^{\bk,\bl}$.

\be
H^{(1,0)}(\Ch) = H^+ \oplus H^- \ ,\qquad \dim \, H^+ = g\ ,\qquad \dim \, H^{-} =\gm 
\ee 
where
\be
\gm= g+\f{\mo}{2}-1\;.
\la{gmd}
\ee

To introduce homological coordinates on the stratum  $\Qcal_{g}^{\bk,\bl}$ decompose the relative homology group
\be
H_1(\Ch\,,
\{\hat{x}_j,\hat{x}_j^\mu\}_{j=\mo+1}^{m})\hskip0.5cm {\rm as}\hskip0.5cm H_+\oplus H_-
\la{homsplit}
\ee
where $H_+$ can be identified with $H_1(\CC,\{{x}_j\}_{j=\mo+1}^{m})$.
We have for $\me\geq 1$
$$
\dim H_1(\Ch\,,
\{\hat{x}_j,\hat{x}_j^\mu\}_{j=\mo+1}^{m}) = 2\gh+2 m_{even}-1\;,
 $$
$$
\dim H_+=2g+m_{even}-1\;, \hskip0.7cm
\dim H_-=2g_-+m_{even}\;.
$$
(for $\me=0$ we have $\dim H_+=2g$, $\dim H_-=2g_-$ and $g_-=g+m/2-1$).

Similarly to the case of all simple zeros we have the equality
$\dim \Qcal_{g}^{{\bk,\bl}}=\dim H_-\;,$
and, therefore, choosing some set of independent cycles $\{s_i\}_{i=1}^{\dim H_-}$ in $H_-$ one can use the integrals $\int_{s_i}v$
as local coordinates on  the stratum $\Qcal_{g}^{{\bk,\bl}}$.

The homology group dual to (\ref{homsplit}) is 
\be
H_1(\Ch\setminus
\{\hat{x}_j,\hat{x}_j^\mu\}_{j=\mo+1}^{m})=H_+^*\otimes H_-^*\;;
\la{dualq}\ee
its symmetric part $H_+^*$ can be identified with $H_1(\CC\setminus\{{x}_j\}_{j=\mo+1}^{m})$.

The skew-symmetric part $H_-^*$ of the group (\ref{dualq}) is dual to $H_-$, and we denote by $\{s_i^*\}$ the basis in  $H_-^*$ which is  dual to the
basis  $\{s_i\}$ in $H_-$.

Now the  tau-functions $\tau_+=\tau(\CC,Q)$ and $\tauh(\CC,Q)$ on the space  $\Qcal_{g}^{{\bk,\bl}}$ formally coincide with (\ref{tauqu}) and
 (\ref{tauquh}) assuming that the Torelli markings of $\CC$ and $\Ch$ agree.
 
 As well as in the case of simple zeros, we introduce  the Prym tau-function via the ratio $\tau_-(\CC,Q)=\tauh(\CC,Q)/\tau(\CC,Q)$.
 
 The explicit formulas for $\tau$ and $\tauh$ can formally be written in the same form as (\ref{tauBerg}) under an appropriate identification of 
 the corresponding divisors. 
 Namely, 
 $\tau_+=\tau(\CC,Q)$  is then expressed in terms of the divisor $(\qd)=\sum d_i q_i$ as follows:
\be
\tau(\CC,Q)=
\Ccal^{2/3}(x)\left(\f{\qd(x)}{\prod_{i=1}^{\nz}E^{d_i}(x,q_i)}\right)^{(g-1)/6} 
\prod_{i<j} E(q_i,q_j)^{\f{d_i d_j}{24}}e^{\f{\pi i}{6}\langle {\bf s},\Omega{\bf s}\rangle-\f{2\pi i}{3}\langle {\bf s},K^x\rangle}\, .
\la{taupint}
\ee
where the vectors   $ {\bf s}, {\bf r}\in \Z^g/2$ are defined by the equations
$$
\f{1}{2}\Acal_x((\qd))+2K^x+\Omega {\bf s}+ {\bf r}=0\;.
$$
Near zeros  $x_j$ of odd multiplicities   the distinguished local coordinates are given by 
\be
\zeta_j(x)=\left[\int_{x_j}^x v\right]^{2/(2k_j+3)}\;,\hskip0.4cm j=1,\dots, \mo
\la{distC1}
\ee
and near zeros of even multiplicity by
\be
\zeta_j(x)=\left[\int_{x_j}^x v\right]^{1/(l_j+1)}\;,\hskip0.4cm j=\mo+1,\dots,m\;.
\la{distC2}
\ee

To give the  explicit expression for $\tauh(\CC,Q)$ we introduce the following notation for the divisor $(v)$ on $\Ch$:
\be
(v)=\sum\dh_i \qh_i = \sum_{i=1}^\mo (2k_i+2)x_i+\sum_{i=\mo+1}^{m} l_i (\xh_i+\xh_i^\mu)\;.
\la{divv}
\ee
Let $\Eh$ and $\Ccalh$  be the prime-form and the multidifferential (\ref{Ccal}) associated to $\Ch$. Then
\be
\tauh(\CC,\qd)=\Ccalh^{2/3}(x)\left(\f{v(x)}{\prod_{i=1}^{\nzh}\Eh^{\dh_i}(x,\qh_i)}\right)^{(\gh-1)/3} 
\left(\prod_{i<j} \Eh(\qh_i,\qh_j)^{\dh_i \dh_j}\right)^{1/6}\,e^{-\f{\pi}{6} \langle\Oh \hat{{\rb}}, \hat{{\sb}}\rangle-\f{2\pi i}{3}\langle\hat{{\rb}},\Kh^x\rangle }
\la{tauhfor}
\ee
where the vectors $\hat{\rb},\hat{\sb}\in\Z^{{\hat g}}$
are defined by the relation 
\be
\Acalh_x((v))+2\Kh^x+\Oh\hat{\rb}+\hat{\sb}=0
\la{defZh1}
\ee
and $\Acalh_x$ is the Abel map on $\Ch$ with the basepoint $x$.

The distinguished local coordinates on $\Ch$ near zeros of $v$ are given by
\be
\zetah_i=\zeta_i^{1/2}=\left[\int_{x_i}^x v\right]^{1/(2k_i+3)}
\la{distCh1}\ee
for $i=1,\dots,\mo$ and
\be
\zetah_i=\zeta_i=\left[\int_{x_i}^x v\right]^{1/(l_i+1)}
\la{distCh2}\ee
for $i=\mo+1,\dots,m$.

The tau-functions $\tau_\pm$ on  $\Qcal_{g}^{\bk,\bl}$   transform according to (\ref{tausigma}) with $\gamma_\pm^{48\alpha}=1$ where
$$
\a=LCM(d_1+2,\dots,d_m+2)\;.
$$

Under a rescaling of $\qd$ the functions $\tau_\pm$ transform as $\tau_\pm(\e Q,\CC)=\e^{\kappa_\pm}\tau_\pm(Q,\CC)$ with 
the homogeneity coefficients given by
\be
\ka_+= \f{1}{48}\sum_{i=1}^{\nz}\f{d_i(d_i+4)}{d_i+2}   \;,
\la{kapl}
\ee
\be
\kappa_-= \ka_+
+\f{1}{8}\sum_{i=1}^\mo \f{1}{d_i+2}\;.
\la{kam}
\ee

The direct use of $\tau_\pm$ to study relations between various classes in the  Picard group of  compactification of $\Qcal_{g}^{\bk,\bl}$ for arbitrary $\{\bk,\bl\}$ is problematic since the structure of such compactification is not well understood until now. 

The explicit formulas for $\tau_\pm$ as well as their homogeneity coefficients turn out to be useful in the problems of the holomorphic factorization of 
the determinant of Laplacian in flat metrics over Riemann surfaces and also in the theory of Teichm\"uller flow on moduli spaces, as we discuss in  section \ref{detteich} below.

\subsection{Quadratic differentials with poles of second order and classes on $\ol{\Mcal}_{g,n}$}
\la{clMgn}

Here we show that the tau-functions $\tau_\pm$ on spaces of quadratic differentials with second order poles can be used to get the generalization of the Mumford relation (\ref{M}) to moduli spaces $\Mcal_{g,n}$.

Let $\qd$ be a meromorphic quadratic differential on a curve $\CC$ of genus $g$ such that all poles $z_1,\dots,z_n$ of $\qd$
  are of second order. We now   write the divisor $(Q)$  in the form
\be
(\qd)=\sum_{i=1}^{m+n} d_i q_i\equiv\sum_{i=1}^\mo (2k_i+1)x_i+\sum_{i=\mo+1}^{m} 2l _i x_i -\sum_{i=1}^{n} 2 z_i
\la{divqd2}
\ee
where $k_i\in \Z$, $l_i\in \mathbb N$, $m= \mo + \me$.

Let us fix $n$ non-vanishing numbers $p_i\in\C$ and assume that the singular part of
$\qd$ at $z_i$ looks as follows:
\be
\qd(\zeta)=-\f{p_i^2}{4\pi^2}\f{(d\zeta)^2}{\zeta^2}\;.
\la{saingpart}
\ee

The moduli space of pairs $(\CC,\qd)$ where the divisor of $\qd$ has form 
(\ref{divqd2})  will be denoted by
$\Qcal_{g,n}^{\bk,\bl}$.
The subspace of $\Qcal_{g,n}^{\bk,\bl}$ such that   singular parts of $\qd$ near $x_i$ are given by (\ref{saingpart}) will be denoted by
$\Qcal_{g,n}^{\bk,\bl}[\pb]$.

The divisor of Abelian differential of third kind $v$ on the canonical cover $v^2=\qd$ now has the form
\be
(v)=\sum_{i=1}^{\nzh}
\dh_i\qh_i\equiv\sum_{i=1}^\mo (2k_i+2)x_i+\sum_{i=\mo+1}^{m} l_i (\xh_i+\xh_i^\mu) -\sum_{i=1}^{n} (\zh_i+\zh_i^\mu)\;.
\la{divv1}
\ee
where $\nzh=\mo+2\me+2n$.

The genus of $\Ch$ is agian given by $\gh=g+g_-$ where $g_-=g+\mo/2-1$. The distinguished local coordinates on $\CC$ and $\Ch$
near zeros $x_i$ and $\xh_i$ are given by the same formulas (\ref{distC1}), (\ref{distC2}),  (\ref{distCh1}), (\ref{distCh2})
as in the case $n=0$. To define distinguished local coordinates near $z_i$ (both on $\CC$ and $\Ch$) we fix some zero, say, $x_i$
and choose a set of cuts  $\g_1,\dots,\g_n$ within the fundamental polygon $\CC_0$ of $\CC$ connecting $x_1$ with $z_1,\dots,z_n$.
Now we define the local coordinates near $z_i$ and $z_i^\mu$ by
\be
\hat{\xi}_i^\pm(x)=\exp\left\{\f{\pm 2\pi i}{p_i}\int_{x_1}^x v\right\}
\la{dist3}
\ee
where the integration paths are chosen not to cross the cuts $\gamma_i$.

Now tau-functions $\tau$ and $\tauh$ are defined by formally the same expressions (\ref{taupint}), (\ref{tauhfor}) on the whole space
$\Qcal_{g,n}^{\bk,\bl}$ as well as  on each subspace $\Qcal_{g,n}^{\bk,\bl}[\pb]$.

However, besides Torelli markings of $\CC$ and $\Ch$ they now depend on the choice of contours $\gamma_i$ through the
distinguished coordinates near $z_i$ and $\hat{z}_i$. On the other hand, the
 ($48\a$)th power of the expressions
\be
\tau_\pm (\CC,\qd)\left(\prod_{i=1}^n d \xi_i(z_i)\right)^{1/12}\;.
\la{defTcalp}
\ee
where
\be
\alpha=LCM(d_1+2,\dots,d_{\nz}+2)\;.
\la{LCD}\ee
is independent of the choice of local parameters $\xi_i$ near $z_i$.

Transformations of $\tau_\pm$ under a change of canonical bases in $H_+$ and $H_-$, respectively, are again given by (\ref{tausigma}).

Therefore, on each space  $\Qcal_{g,n}^{\bk,\bl}[\pb]$  the powers $\tau_\pm^{48\a}$ are sections of the line bundles $(\det \Lambda_\pm)\prod_{j=1}^n L_j^{1/12}$ where $L_i$ is the line bundle associated to the marked point $z_i$ (the fiber of $L_i$ is the cotangent space to $\CC$ at $z_i$).

Consider now the largest stratum $\Qcal_{g,n}^{0}[\pb]$ which corresponds to differentials with all simple zeros $x_1,\dots,x_{4g-4+2n}$.
We have $\dim \Qcal_{g,n}^{0}[\pb]=\dim \Qcal_{g,n}[\pb]=6g-6+2n$ which coincides with dimension of $T^*\Mcal_{g,n}$. Actually, the space
$ \Qcal_{g,n}[\pb]$ can be identified with  $T^*\Mcal_{g,n}$ if on each Riemann surface one chooses a quadratic differential $\qd_0$ with given
singular part which holomorphically depends on moduli. Than any other differential with the same singular parts can be obtained by adding
to it a quadratic differential with simple poles at $z_i$ i.e. a cotangent vector to $\Mcal_{g,n}$. Thus $ \Qcal_{g,n}[\pb]$ is the affine space modelled on the vector space $T^*\Mcal_{g,n}$.

For a point $(\CC,\qd)\in  \Qcal_{g,n}^{0}[\pb]$ we have $\gh=4g-3+n$. Let $D_{\deg}= \Qcal_{g,n}[\pb]\setminus  \Qcal_{g,n}^{0}[\pb]$ be
the divisor corresponding to differentials with multiple zeros. Consider the partial compactification
$\tilde{\Qcal}_{g,n}[\pb]$ of  $\Qcal_{g,n}[\pb]$  obtained by the natural extension of  fibers of the affine bundle $\Qcal_{g,n}[\pb]$
to the Deligne-Mumford boundary of $\Mcal_{g,n}$; denote this divisor in $\tilde{\Qcal}_{g,n}[\pb]$ by $D_{DM}$.
Introduce the classes $\psi_i=c_1(L_i)$, the  Hodge class $\l=c_1(\det \Lambda_+)$ and the Prym class  $\l_P=c_1(\det \Lambda_-)$
in the Picard group of  $\Qcal_{g,n}^{\bk,\bl}[\pb]$.

Asymptotics of $\tau_\pm$ near $D_{\deg}$ and  $D_{DM}$ are the same as in the case of holomorphic quadratic differentials.
This implies the following formulas relating the Hodge and Prym classes with classes $\psi_i$ and classes of divisors
 $D_{\deg}$ and  $D_{DM}$ in $Pic(\tilde{\Qcal}_{g,n}[\pb],\Q)$:
\be
\lambda+\f{1}{12}\sum_{i=1}^n \psi_i=\f{1}{72}\, \delta_{\deg}+\f{1}{12}\delta_{DM}\;,
\la{Hodgeclm}
\ee
\be
\lambda_P+\f{1}{12}\sum_{i=1}^n\psi_i= \f{13}{72}\, \delta_{\deg}+\f{1}{12}\delta_{DM}\;.
\la{Prymclm}
\ee
One difference between these formulas and the formulas (\ref{Hodgecl}), (\ref{Prymcl}) is  the absence of the class $\phi$
which appears in (\ref{Hodgecl}), (\ref{Prymcl})  due to projectivization of the underlying moduli space. 
Another difference is  the presence of the  classes $\psi_i$ in  (\ref{Hodgeclm}), (\ref{Prymclm}).

Linear combinations of   (\ref{Hodgeclm}) and (\ref{Prymclm}) gives  the following relations between the 
classes in $Pic(\tilde{\Qcal}_{g,n}[\pb],\Q)$:
\be
\lambda_P-13\lambda=\sum_{i=1}^n \psi_i-\delta_{DM}
\la{rel1}
\ee
and
\be
\l_P-\l=\f{1}{6}\delta_{\deg}
\la{rel2}
\ee

Observe now that for each point of ${\Qcal}^0_{g,n}[\pb]$
the fiber $H^-$ of the Prym bundle can be identified with the space of quadratic differentials on $\CC$ with at most simple poles at
the marked points. The correspondence is given by the relation
\be
\pi^*\qd_1= u v
\ee
where $u\in H^-$ is a holomorpic Prym differential on $\Ch$ and $\qd_1$ is a quadratic differentials on $\CC$ with at most simple poles at $z_1,\dots,z_n$. Since both $u$ and $v$  change their sign under the interchange of the sheets of $\Ch$ the product $uv$ can be identified with a quadratic differential on $\CC$. Such correspondence naturally extends to  the space  $\tilde{\Qcal}_{g,n}[\pb]$  (see \cite{KSZ}).

Denoting the (pullback from $\ol{\Mcal}_{g,n}$) of the first Chern class of the vector bundle of quadratic differentials with simple poles
by $\l_2^{(n)}$ we get  the equality $\l_P=\l_2^{(n)}$ and the relations (\ref{rel1}), (\ref{rel2}) imply 
can be rewritten as
\be
\lambda_2^{(n)}-13\lambda+\sum_{i=1}^n \psi_i=-\delta_{DM}
\la{rel11}
\ee
and
\be
\kappa_1=\f{1}{6}\delta_{\deg}
\la{rel22}
\ee
where $\kappa_1=\l_2^{(n)}-\l$ is the pullback of the first kappa-class from $\ol{\Mcal}_{g,n}$ to $\tilde{\Qcal}_{g,n}[\pb]$.

The relation (\ref{rel11}) on the affine bundle over $\ol{\Mcal}_{g,n}$ implies the identical relation for the corresponding classes on
$\ol{\Mcal}_{g,n}$. This reproduces  the  classical extension of the Mumford relation (\ref{M}) to $\ol{\Mcal}_{g,n}$ (see the formula (7.8) of \cite{ArbCorGri}).

The relation (\ref{rel22}) is the  analog of the Penner  expression for the class $\kappa_1$ in the hyperbolic
combinatorial model \cite{Penner}; however, unlike the Penner relation, our formula (\ref{rel22}) does not have any contribution from hte Deligne-Mumford boundary. We notice that the divisor $\delta_{\deg}$ in   (\ref{rel22}) essentially depends on $[\pb]$;
this situation is parallel to the analog of the Penner formula which is written in the framework of the 
combinatorial model of $\Mcal_{g,n}$ based on Jenkins-Strebel differentials. 

\section{Flat combinatorial model of ${\mathcal M}_{g,n}$: tau-functions and Witten's classes}
\la{flatco}

The formalism of the previous  section can be further applied to the  combinatorial  model   of $\Mcal_{g,n}$ which is based on  the Jenkins-Strebel (JS) differentials \cite{BK2}. This combinatorial model is based on the  fact  that for a given 
Riemann surface $\CC$ with $n$ marked points $z_1,\dots,z_n$ and for a given   vector $\pb\in \R_+^n$ there exists the
unique quadratic differential $\qd$  with the singular part at $z_i$ of the form (\ref{saingpart}) which has all real periods
of $v=\sqrt{\qd}$ on $\Ch$ (this means that all periods of $v$ over cycles from $H_-$ are real, since all integrals of $v$ over cycles from $H_+$ vanish).  Riemann surfaces with $n$ marked points which are equipped with  Jenkins-Strebel differentials form  the {\it flat} combinatorial model  $\Mcal_{g,n}[{\bf p}]$ of  $\Mcal_{g,n}$. 
 This model  was developed starting from   ideas of  Harer, Mumford and Thurston; the modern exposition can be found in \cite{Mondello2}). 
 The flat combinatorial model of $\Mcal_{g,n}$ was used in  Kontsevich's proof \cite{Kontsevich} of the 
 Witten conjecture \cite{Witten}.  Another combinatorial model of $\Mcal_{g,n}$ which is based on   hyperbolic geometry 
 of $\CC$ was proposed in  \cite{Penner}.

The cells of $\Mcal_{g,n}[{\bf p}]$ with ${\bf p} = (p_1,\dots, p_n)\in \R^n_+$   are labeled by  ribbon graphs  of given topology on a Riemann surface $\CC$ 
of genus $g$ punctured at $n$ points. The vertices of the ribbon graph are  zeros of $\qd$ while each face 
contains one pole of $\qd$.

The quadratic residues at the poles are given by $-p_i^2/4\pi^2$, where $p_1,\dots,p_n$ are the perimeters of the faces.
The lengths of the edges in the metric $|\qd|$ are used as coordinates on a given  cell; these lengths
are equal to  (half-integer combinations of) the  integrals of $v$ over cycles in $H_-$.
 The union of  cells with given multiplicities $k_1,\dots,k_m$ of  zeros of  $\qd$ forms the stratum  of
$\Mcal_{g,n}[{\bf p}]$  labeled by the vector ${\bf k}$. The valencies of vertices of the ribbon graph  are equal to $k_j+2$.
 The special role is played by the strata corresponding to all odd $k_i$; these strata turn out to be cycles, called 
 {\it Kontsevich-Witten cycles} \cite{Mondello2}.
The compactification $\overline{\Mcal}_{g,n}[\pb]$  is obtained by adding  the
 {\it Kontsevich boundary} $W_{1,1}$ to $\Mcal_{g,n}[{\bf p}]$. Cells of $W_{1,1}$ of the highest dimension correspond to JS differentials with two simple poles;
 corresponding ribbon graphs have two univalent vertices while all other vertices are three-valent.

In the main stratum $W$ of $\Mcal_{g,n}[{\bf p}]$  all    zeros of $\qd$ are simple, so that the ribbon graph has only three-valent vertices.  The boundaries of real co-dimension $1$  of the cells of $W$ (the {\it facets})  correspond to JS differentials $\qd$ with one
double zero, while all other zeros remain simple;  the corresponding ribbon graphs have one 4-valent vertex while all other vertices are three-valent.
The union of cells of $W$ and their facets will be denoted $\tilde{W}$. The complement, $\Mcomb\setminus \tilde{W}$,
contains cells of (real) co-dimension 2 and higher.

In the (real)  co-dimension 2 there exist two special sub-complexes of $\Mcomb$ which are also cycles.
 The first one is the "Witten's cycle" $\Wfive $ which corresponds to ribbon graphs with at least one  
 5-valent vertex; in the cells of highest dimension of $W_5$ the Strebel differential  $\qd$ has one zero of order 3
 while all other zeros are simple. Ribbon graphs corresponding to cells of  $W_5$ are obtained from three-valent ribbon graphs by
 degeneration of two edges having one common vertex.
 The second one  is the  Kontsevich's boundary $\Wbdr$; ribbon graphs corresponding to cells of $\Wbdr$ are obtained from three-valent graphs by
 degeneration of two edges having two common vertices.
The cycle $\Wbdr $ is the sum of several sub-cycles that correspond to  different topological types of the boundary of
$\Mcal_{g,n}$. 
While $\Mcal_{g,n}[\pb]$ is in one-to-one correspondence with $\Mcal_{g,n}$,
this isomorphism does not extend to the boundary: some components of the Deligne-Mumford boundary  $\overline{\Mcal}_{g,n}\setminus \Mcal_{g,n}$ (namely, the components corresponding 
to homologically trivial vanishing cycles which leave all punctures in  one of the connected components of the stable curve) are not represented by cells of highest dimension of $\Wbdr $.

The orientation on  $\Mcal_{g,n}[\pb]$ is induced by the symplectic form $\sum_{i=1}^n p_i^2 \psi_1$, where $\psi_i$ is the 2-form representing the corresponding tautological class and which can be written in terms of lengths of the edges \cite{Kontsevich}. This symplectic form on  each cell of a Kontsevich-Witten
cycle can be written as $\sum_{k=1}^{g_-}dA_k\wedge dB_k$ where $A_k$ and $B_k$ are periods of $v$ over a set of
cycles $(a_k^-,b_k^-)$ in $H_-$ satisfying $a_j^-\circ b_k^-=\delta_{jk}$ \cite{BK2} ( $(a_k^-,b_k^-)$ do not form a basis in $H_-$ since $H_-$ also contains combinations of small circles around $z_i$ and $z_i^\mu$, but the integrals of $v$ along these circles are combinations of $p_i$,
which are constant in a given combinatorial model).

The real-analytic complex  $\Mcal_{g,n}[\pb]$ is nothing but the real slice  (i.e. all homological coordinates including the perimeters $p_j$ are real) of the complex-analytic  space $\Qcal_{g,n}[{\bf p}]$ considered in previous section. In particular, the stratum $W$ is the real slice  of
the space $\Qcal^0_{g,n}[{\bf p}]$.

Therefore, the Hodge and Prym vector bundle, as well as tau-functions $\tau_\pm$ can be defined over
  $\Mcal_{g,n}[\pb]$ and  $\overline{\Mcal}_{g,n}[\pb]$ by restriction of constructions of the previous section to the real slice.
However, then  $\tau_\pm$ become only real-analytic on each cell of  $\Mcal_{g,n}[\pb]$. Defining $\tau_\pm$ on each cell of $W$ one can
verify that $|\tau_\pm|$ diverge at the facets betwen cells, but $\phi_\pm={\rm arg}\tau_\pm$ vary continuously and define sections of
$U(1)$ (circle) bundles which are combinations of $U(1)$ bundles related to $\det\Lambda_P$, $\det\Lambda_H$ and classes $\psi_i$.

Further (rather technically non-trivial) computation of the increments of $\arg\tau_\pm$ around Witten's cycle $W_5$ and the
Kontsevich's boundary $W_{1,1}$ within $\tilde{W}$ 
leads to the following  relations which should be understood as equivalences between classes and their Poincare dual cycles
\cite{BK2}:
\be
\lambda+\f{1}{12}\sum\psi_i= \f{1}{144}\Wfive  +\f{13}{144} \Wbdr \la{112}\;\;\;,
\ee
\be
\lambda_2^{(n)}+\f{1}{12}\sum\psi_i= \f{13}{144}\Wfive  +\f{25}{144}\Wbdr \la{113}\;\;\;.
\ee
The derivation of these formulas also uses  the isomorphism between the Prym vector bundle $\Lambda_P$ and the vector bundle  $\Lambda_2^{(n)}$ of meromorphic quadratic differentials with
simple poles at the marked points over  $\Mcomb$.

Taking into account that $\kappa_1 =\lambda_2^{(n)}-\lambda$ 
one can express the class $\ka_1$ via $\Wfive $ and $\Wbdr$:
\be
12\kappa_1=W_5+W_{1,1}
\la{kapcl}
\ee
reproducing the Arbarello-Cornalba formula \cite{ArbCor}.


 Eliminating 
the Witten cycle $\Wfive $ from \eqref{112}, \eqref{113} we get to the following "combinatorial" version of the Mumford relation
 (\ref{rel11}):
\be
\lambda_2^{(n)}- 13 \lambda- \sum_{i=1}^n \psi_i= - \Wbdr \;.
\la{Mumcomb}
\ee

\section{Other examples:  moduli of  $N$-differentials and spin moduli spaces}
\la{nspin}

Here we briefly outline the applications of the formalism of the Bergman tau-function to two other problems: the computation of the class of degenerate $n$-differentials in the vector bundle of holomorphic $N
$-differentials over $\Mcal_g$ \cite{KSZ}
and the computation of the class  of degenerate odd spinors in the spin moduli space \cite{Basok}.

\subsubsection{$N$-differentials}

Denote by $\Mgn$ the moduli space of pairs $(\CC,W)$ where $\CC$  is a Riemann surface of genus $g$ and
$W$ is a holomorphic $N$-differential on $\CC$; ${\rm dim} \Mgn=  (2N+2)(g-1)$.  Projectivisation of this  space admits the natural compactification
$P\Mgn$, see \cite{KSZ}. The pair $(\CC,W)$ defines an $N$-sheeted   covering $\Ch$ of $\CC$ via equation
\be
v^N=W
\la{covN}\ee
The covering (\ref{covN}) possesses $\Z_N$ symmetry; denote the natural  automorphism of order $N$ on $\Ch$ by $\mu$. Then the homology space $H_1(\Ch,\C)$ can be represented as the direct sum of $N$ invariant subspaces of $\mu_*$:
$\Hcal_0$ (which can be identified with $H_1(\CC)$, so $\dim\Hcal_0=2g$), $\Hcal_1,\dots,\Hcal_{N-1}$.
If all zeros of $W$ are simple then the  genus of $\Ch$ equals $\gh=n^2(g-1)+1$.
Then also $\dim \Hcal_k=(2N+2)(g-1)$ for $k=1,\dots,N-1$  which coincides with $\dim\Mgn$. Integrals of $v$ over a basis in $\Hcal_1$ can be used
as local homological coordinates on each stratum of $\Mgn$ (integrals of $v$ over cycles from other $\Hcal_j$ vanish).

The Bergman tau-function on the stratum of generic $N$-differentials i.e. differentials with simple zeros can formally be defined by the formula (\ref{tauBerg}) where the multiplicities $k_i$ are formally replaced by $1/N$ and points $q_i$ are replaced by zeros
$x_1,\dots,x_{N(2g-2)}$ of $W$. The same substitution should be done in expressions for distinguished local coordinates (\ref{lcoHur}).

In analogy to the $N=2$ case, computing the asymptotics of the tau-function near Deligne-Mumford boundary and the subspace $D_{\deg}$ of differentials with multiple zeros we get the following  relation  in
$Pic(P\Mgn,\Q)$ \cite{KSZ}: 
\be
\l=\f{(g-1)(2N+1)}{6N(N+1)}\phi+\f{1}{12N(N+1)} \delta_{\deg}+\f{1}{12}\delta_{DM}
\la{Hodgeint}
\ee
where, as before,  $\phi$ is the class of the line bundle arising from projectivization,
$\lambda$ is the (pullback from $\Mcal_g$ of) the Hodge class, $\delta_{\deg}$ is the class of $D_{\deg}$.

The formula (\ref{Hodgeint}) will be used below to compute the class of the universal Hitchin discriminant in the 
universal moduli space of Hitchin spectral covers.

\subsubsection{Spin moduli spaces and Farkas-Verra formula}

The spin moduli space $\Scal_g$ is the space of  pairs $(\CC,\chi)$ where $\CC$ is a Riemann surface of genus $g$ and
$\chi$ is a spin line bundle over $\CC$. This space has two connected components, $\Scal_g^+$ and $\Scal_g^-$,
 corresponding to even and odd $\chi$, respectively, which are finite-sheeted coverings of $\Mcal_g$. We  refer to
\cite{Farkas} for more detailed description of these spaces and references. The formalism of the Bergman tau-function,
being  applied to the space $\ol{\Scal}_g^-$  \cite{Basok}, allows to reproduce the formula for the class of degenerate odd spinors
originally derived by Farkas and Verra in Th.0.5 of \cite{Farkas} (this theorem was the main tool in determining the
type of spaces $\ol{\Scal}_g^-$ for $g>11$).

Here we briefly explain the construction of \cite{Basok}. An odd spin line bundle for the generic curve $\CC$  has
exactly one holomorphic section (the co-dimension of the subspace of $\Mcal_g$ where there exist 3 or more holomorphic sections equals two), and this section (for $\CC\in \Mcal_g$) is nothing but the spinor $h$  (\ref{hp}) which enters the definition of
the prime form.

The divisor $\Zcal_g$ of \cite{Farkas} corresponds to the spinors $h$ with one zero of
order 2 while other zeros remain simple.
To compute the class of this divisor via the tau-function formalism consider the Bergman  tau-function on the stratum
$\Hcal_g(2,\dots,2)$ of Abelian differentials (the Abelian differential $v$ is taken to be $h^2$).
Then  $\Zcal_g$ is a subspace of the boundary component of $\Hcal_g(2,\dots,2)$ where two zeros merge, i.e.
$\Zcal_g\subset \Hcal_g(4,2,\dots,2)$. The computation of the asymptotics of $\tau_B(\CC,h^2)$ near  $\Zcal_g$ and
boundary components of ${\Scal}_g^-$ which correspond to DM boundary of $\Mcal_{g}$ gives, as shown in \cite{Basok}, the following expression
of the class of $\Zcal_g$ in terms of the Hodge class $\lambda$ and boundary divisors.  This expression reproduces the
original Farkas-Verra formula
of \cite{Farkas}:
\be
[{\mathcal Z}_g]= (g+8)\lambda-\f{g+2}{4}[A_0]-2[B_0]-\sum_{j=1}^{[g/2]} 2(g-j)[A_j]-\sum_{j=1}^{[g/2]} 2j[B_j]
\la{FV}
\ee
where 
for each $j$ the union of divisors $A_j\cup B_j$ in the formula (\ref{FV}) is the pullback of  the  component $D_j$ of the Deligne-Mumford boundary. For $j=1,\dots,[g/2]$ the divisor $A_j$ corresponds to a reducible curve such that on the component of
genus $j$ the spin line bundle is  odd. The divisor $B_j$ corresponds to reducible curve such that on component of
genus $j$ the spin line bundle is even.
The definition of the divisors $A_0$ and $B_0$ whose union forms the pullback of the irreducible Deligne-Mumford boundary components $\Delta_0$ is slightly more subtle, and we refer to \cite{Farkas,Basok} for more details.

\section{Moduli spaces of Hitchin's spectral covers}
\la{secHit}

Hitchin in \cite{Hitchin1,Hitchin2} proposed the dimensional reduction 
of the self-dual Yang-Mills equations by
splitting the 4-dimensional space into the product of a Riemann surface $\CC$ and the real plane $\R^2$; the gauge fields are assumed to 
be independent of coordinates on $\R^2$. As a result of such dimensional reduction, one arrives at the class of  the finite-dimensional 
completely integrable systems, called {\it Hitchin's systems}. Here we refer to  Atiyah's book \cite{Atiyah} (Sect. 6.3) for an introduction to the topic and to the original papers of Hitchin \cite{Hitchin1,Hitchin2}.  In the  paper by van Greemen and Previato \cite{Previato}  the detailed treatment of genus 2 case was given; \cite{DonagiMarkman} contains the detailed review of the subject.
Hitchin  systems  provide the most general class of integrable systems associated to Riemann surfaces of an
arbitrary genus. 

The Hamiltonians of the Hitchin system are encoded  in the so-called {\it spectral cover}
$\Ch$ which is the $n$-sheeted cover of $\CC$ defined by the following equation in $T^*\CC$:
\be
\Ch=\{(x,v)\in\CC\times T^*_x\CC\;|\; P_n(v)=0\}
\la{cancov}
\ee
where 
$$P_n(v)=v^n+\dif_{n-1}v^{n-1}+\dots+\dif_{1}v+\dif_0,$$
$\dif_k$ is a holomorphic $n-k$-differential on $\CC$, and $v$ is  a holomorphic 1-form on 
 $\Ch$.  In the framework of \cite{Hitchin1} the equation (\ref{cancov}) is given by the characteristic polynomial
 $P_n(V)={\rm det}(\Phi-vI)$ 
  of the  {\it Higgs field} $\Phi$ on $\CC$ ($\Phi$ is a section of $ad_\chi\otimes K$ where $\chi$ is a vector bundle and $K$ is the canonical line bundle over $\CC$).

For the case of $GL(n)$ Hitchin's systems all differentials $\dif_k$ from  (\ref{cancov}) are arbitrary;
in the case of $SL(n)$ systems  $\dif_1=0$.

The branch points of the cover $\Ch$ are  zeros of the discriminant $W$ of $P_n(v)$ which is the holomorphic $n(n-1)$-differential on $\CC$. Thus, the number $m$ of zeros of $W$, counted with multiplicities, equals 
\be
m=n(n-1)(2g-2)\;,
\ee
and the Riemann-Hurwitz formula 
gives  the genus $\gh$ of $\Ch$:
\be
\gh=n^2(g-1)+1\;
\la{gh}
\ee
so that $m=2(\gh-1-n(g-1))$.
When all zeros of $W$ are simple, all branch points of the covering $\pi:\Ch\to \CC$ are also simple.

Let $\CC$ be a smooth curve of genus $g$ and denote by  $\Mcal_H^{\CC}$ the moduli space of $GL(n)$ Hitchin spectral covers of the form  (\ref{cancov}). Then
\be
\Mcal_H^{\CC}=\bigoplus_{j=1}^n H^0(\CC,K_{\CC}^{\otimes j})
\ee
where $K_{\CC}=T^*\CC$ is the canonical line bundle on $\CC$,
and
\be
\dim\,\M_H^{\CC}=\gh=n^2(g-1)+1
\la{dimMC}
\ee
(recall that $\dim H^{0}(\CC,K)=g$ and $ \dim H^{0}(\CC,K^j)=(2j-1)(g-1)$  for $j\geq 2$).

There is the natural coordinate system on $\Mcal^{\CC}$ given by the $a$-periods of $v$  on $\Ch$
\be
A_j=\oint_{\ah_j}v
\la{coordM0}
\ee
where  
$\{\ah_j,\bh_j\}_{j=1}^{\gh}$ is a canonical symplectic basis in $H_1(\Ch,\Z)$.

\subsection{Class of universal Hitchin's discriminant locus}
\la{secdiscr}

 The {\it Hitchin's discriminant locus}  $D_H^{\CC}\subset \Mcal^{\CC} $ is defined by the condition that  the discriminant
 $W$ of $P_n(v)$ has multiple zeroes i.e. $\Ch$ has non-simple branch points.

 Denote by $\Mcal_H$ the space of all $GL(n)$ spectral covers for a fixed genus of $\CC$ (the base $\CC$ is allowed to vary);
\be
\dim \Mcal_H=\dim\Mcal_H^{\CC}+3g-3=(n^2+3)(g-1)+1\;.
\la{dimMcal}
\ee
 
 For the precise definition of the space  $\Mcal_H$ and the  compactification of its projectivization (with respect to the
 equivalence under rescaling $\dif_j\mapsto\epsilon^j\dif_j,\;\epsilon\in\C^*,\;j=1,\ldots,n$.)
 $P\ol{\Mcal}_H$
 we refer to \cite{Faddeev}. The divisor $D_H\subset P\ol{\Mcal}_H$ is called the {\it universal Hitchin discriminant locus};
 it's the union of (projectivizations) of $D_H^{\CC}$ over all stable curves $\CC\in \ol{\Mcal}_g$.

 Two zeros of $W$ can coalesce in three different ways. First, the double branch point can form on $\Ch$; this component of $D_H$  is called the {\it caustic} and denoted by $D_H^{(c)}$. Second, two simple branch points with the same projection to $\CC$ can form;  this component of $D_H$  is called the {\it Maxwell stratum} and denoted by $D_H^{(m)}$.
 Finally, the node can form  i.e. $\Ch$ gets to   the boundary of $\Mcal_{\gh}$. This component of $D_H$ is called {\it the boundary} and denoted by  $D_H^{(b)}$.
 The simple local analysis \cite{Faddeev} shows that the divisor class of $D_H$ equals
 \be
[D_H]= [D_H^{(b)}]+2[D_H^{(m)}]+3[D_H^{(c)}]\;.
\la{DWdec}
\ee

Consider the following natural bundles on the moduli space $P\overline{\Mcal_H}$:
\begin{itemize}
\item
The Hodge vector bundle $\Lambda\to\overline{M}_g$ (the fiber of $\Lambda$ over a smooth $\CC$ is the $g$-dimensional vector space of holomorphic 1-forms on $\CC$).
This bundle naturally lifts to $P\overline{\Mcal}_H$, and we put $\lambda=c_1(\Lambda)$. 

\item
The Hodge vector bundle $\Lh\to\overline{\Mcal}_{\gh}$ (the fiber of $\Lh$ over a smooth spectral cover $\Ch$ is the $\gh$-dimensional vector space of holomorphic 1-forms on $\Ch$.
This bundle also lifts to $P\overline{\Mcal}_H$, and, similarly, we put $\lh=c_1(\Lh)$.

\item There is the natural action of $\C^*$ on $\Mcal_H$ by
\be
\dif_k\mapsto \mu^k \dif_k \;,\qquad \mu\in \C^*
\la{rescale}
\ee
Denote by $P\Mcal_H$ the projectivization of $\Mcal_H$
 with respect to the action (\ref{rescale}). Let $L$ be the tautological line bundle of the natural projection $\Mcal_H\to P\Mcal_H$. The bundle $L$ extends to $P\overline{\Mcal}_H$, and we put $\phi=c_1(L)$.
\end{itemize}

The space $\Mcal_H$ can be naturally embedded into the moduli space of holomorphic $n(n-1)$ differentials on
Riemann surfaces of genus $g$:  the $n(n-1)$ - differential corresponding to a point of $\Mcal_H$ is the 
discriminant $W$. This allows to define the tau-function $\tau_B$ on $\Mcal_H$ by the pullback of the tau-function 
$\tau(\CC,W)$ from the moduli space of $N$-differentials for $N=n(n-1)$ (notice that the dimension of $\Mcal_H$ 
which equals to $3g-3+n^2(g-1)-1=(n^2+3)(g-1)+1$ is  smaller than the dimension of the moduli space of $n(n-1)$ differentials which equals to $(2n(n-1)-1)(g-1)+3g-3=(2n^2-2n+3)(g-1)$).

As before, computing the asymptotics of the tau-function near the different components of $D_H$ we conclude that
the class $D_H$ of the  universal Hitchin discriminant      (\ref{DWdec}) is expressed in terms of the
standard generators of ${\rm Pic}(P\overline{\Mcal}_H)\otimes\Q$ as follows:
\be
\f{1}{n(n-1)}[D_H]=(n^2-n+1)(12\lambda -\delta)- 2(g-1)(2n^2-2n+1) \phi \;.
\la{formMW}
\ee
Here
$\l$ is (the pullback of) the Hodge class on $\overline{M}_g$, $\phi$ is the tautological class of the projectivization 
$\overline{\Mcal}_H\to P\overline{\Mcal}_H$, and $\delta$ is the pullback of the class of the
 $\ol{\Mcal}_g\setminus \Mcal_g$ to $P\overline{\Mcal}_H$.

\subsection{Variation of the period matrix}
\la{varperH}
To conclude this section we show how the variational formulas on the space $\Mcal_\CC$ can be deduced 
from the  variational formulas (Th.\ref{thvarfo}) on strata of the moduli space of Abelian differentials found in
\cite{JDG}. This section is based on the paper  \cite{BK3} to which  we refer for more details.

We shall discuss here only variations of the period matrix $\Oh$ of $\Ch$ on the moduli space $\Mcal_H^\CC$
with a fixed base $\CC$. We get these formulas by the pullback of the variational formulas on the space
$H_{\gh}(1,\dots,1)$ of abelian differentials on Riemann surfaces of genus $\gh$.
The map of $\Mcal_H^\CC$ to $H_{\gh}(1,\dots,1)$ is defined by assigning to a point of $\Mcal_H^\CC$ the
pair $(\Ch,v)$; for a generic point of $ \Mcal_H^\CC$ all zeros of $v$  are simple. Assume that the branch points of $\Ch$ i.e. the zeros of $W$ are also simple.
The total number of zeros of $v$ is $2\gh-2=2n^2(g-1)$; in a general position $(v)=D_{br}+D_0$
where $D_{br}$ is the divisor of branch points of $\Ch$; the projection of $D_{br}$ on $\CC$ coincides with the divisor of discriminant $W$: $\pi(D_{br})=(W)$; ${\rm deg} D_{br}=n(n-1)(2g-2)$. The projection of $D_0$ on $\CC$ coincides with the divisor of the
$n$-differential $Q_0$:
$\pi(D_0)=(Q_0)$; ${\rm deg} D_0=n(2g-2)$ i.e. ${\rm deg} D_{br}+{\rm deg} D_0={\rm deg} (v)$ as expected.
Let
$$D_{Br}=\{x_i\}_{i=1}^{{\rm deg} D_{br}}\;,\hskip0.7cm D_0=\{x_i\}_{{\rm deg} D_{br}+1}^{{\rm deg}(v)}\;.$$

The homological coordinates on $H_{\gh}(1,\dots,1)$ are given by   (\ref{homcoormain}):
$$
A_j=\int_{\ah_j}v\;,\hskip0.7cm
B_j=\int_{\bh_j}v\;,\hskip0.7cm j=1,\dots,\gh\;,
$$
$$
z_k=\int_{x_{2\gh-2}}^{x_k}\;,\hskip0.7cm k=1,\dots, 2\gh-3
$$

On the submanifold $\Mcal_H^\CC$ one can choose $A_j$ to be the set of independent coordinates; then
all remaining coordinates $B_j$ and $z_j$ become $A_j$-dependent. Their dependence on $A_j$'s can be computed taking into account that
(see \cite{BK3} for the proof)
\be
\f{\p v(x)}{\p A_j}= v_j(x)
\la{varofv}
\ee
where $v_j$ is a  holomorphic differential on $\Ch$ normalized by $\int_{\hat{a}_k}v_j=\delta_{jk}$ and  the projection of $x$ on the base $\CC$ is assumed to be fixed under differentiation.
Then we have on $\Mcal_H^\CC$
\be
\f{\p B_k}{\p A_j}=\Oh_{jk}\hskip0.7cm j,k=1,\dots,\gh
\la{ABjk}
\ee
and, since $v$ vanishes at the endpoints of contours $[x_{2\gh-2},x_k]$ for $k={\rm deg} D_{br}+1,\dots,2\gh-3$,
\be
\f{\p z_k}{\p A_j}=\int_{x_{2\gh-2}}^{x_k} v_j\;, \hskip0.7cm k={\rm deg} D_{br}+1,\dots,2\gh-3\;.
\la{Azjk}
\ee
The variational formulas for  $z_k$, $k=1,\dots,\deg D_{br}$  are more subtle.

Let $x_k\in\Ch$ be a ramification point of $\Ch$ and $\xi_k=\xi(\pi(x_k))\in\CC$ be the corresponding critical value
in some local coordinate $\xi$ on $\CC$. We assume that $\xi$ remains fixed under a deformation of $\Ch$;
let $\zeta=\xi-\xi_k$ be a coordinate on $\CC$ vanishing at $\pi(x_k)$ (the coordinate $\zeta$ deforms when $\Ch$ varies). The local coordinate on $\Ch$ near
$x_k$ can then be chosen to be $\zetah(x)=\zeta^{1/2}$.

Then the differentiation of the endpoint of the integration contour with respect to $A_j$ also gives the contribution to $\p z_k/\p A_j$ and we get \cite{BK3}
\be
\f{\p z_k}{\p A_j}=\int_{x_{2\gh-2}}^{x_k} v_j+ \f{\p\xi_k}{\p A_j}\f{v}{d\xi}(\xi_k)
\la{zkaj1}
\ee
for  $k=1,\dots,{\rm deg} D_{br}$.

To compute the derivative  ${\p\xi_k}/{\p A_j}$ we write $v(\xi)$ near $\xi_k$ in the form
\be
v=(a+b\sqrt{\xi-\xi_k}+\dots)d\xi
\ee
(recall that $v$ has simple zero in the local parameter $\sqrt{\xi-\xi_k}$).
Then
\be
b=\f{d(v/d\xi)}{d\sqrt{\xi-\xi_k}}\Big|_{\xi=\xi_k};
\ee
and
\be
v_j=\f{\p v}{\p A_j}=(a_{A_j}-\f{({\xi_k})_{A_j}}{2\sqrt{\xi-\xi_k}}b+\dots)d\xi\;.
\ee
Therefore,
\be
-\f{v_j}{d\sqrt{\xi-\xi_k}}\Big|_{\xi=\xi_k}=b\f{\p\xi_k}{\p A_j}
\ee
and,
\be
\f{\p\xi_k}{\p A_j}=-\f{v_j/d\zetah_k}{(v/d\xi)_{\zetah_k}}(x_k)
\ee
Now (\ref{zkaj1}) takes the form
\be
\f{\p z_k}{\p A_j}=\int_{x_{2\gh-2}}^{x_k} v_j-\f{v_j/d\zetah_k}{[\log(v/d\xi)]_{\zetah_k}}(x_k)
\la{zkaj2}
\ee
for  $k=1,\dots,{\rm deg} D_{br}$.

Now, applying the chain rule we get on $\Mcal_H^\CC$
\be
\f{d \Oh_{lk}}{d A_j}=\f{\p \Oh_{lk}}{\p A_j}+\sum_{s=1}^{\gh}\f{\p \Oh_{lk}}{\p B_s}\f{\p B_s}{\p A_j} +\sum_{r=1}^{2\gh-3}\;.
\f{\p \Oh_{lk}}{\p z_r}\f{\p z_r}{\p A_j}\;.
\la{varOA}
\ee
Using the variational formulas (\ref{varO2})  
\be
\f{\p \Oh_{lk}}{\p A_j} =-\int_{\bh_j}\f{v_l v_k}{v}\;,\hskip0.7cm
\f{\p \Oh_{lk}}{\p B_j} =\int_{\ah_j}\f{v_l v_k}{v}\;,\hskip0.7cm
\f{\p \Oh_{lk}}{\p z_r} =2\pi i\; {\rm res}|_{x_r}\f{v_l v_k}{v}
\ee
for $r=1\dots,2\gh-3$,
we can transform (\ref{varOA}) by applying the Riemann bilinear identity to the pair
$(v_j,\f{v_lv_k}{v})$ to get the sum of residues only at the branch points of $\Ch$. The result looks as follows (see \cite{BK3} for details):
\be
\f{d \Oh_{lk}}{d A_j}=-2\pi i \sum_{branch \;\;points\;\;x_r} \f{v_j}{d\log(v/d\xi)}(x_r)\; {\rm res}|_{x_r}\f{v_\a v_\b}{v}
\la{Oh1}
\ee
or, equivalently, since the differential under the residue has the first order poles at $x_r$,  we get the more familiar form of this variational formula which makes the symmetry between $(l,k,r)$ manifest:
\begin{proposition}
\be
\f{d \Oh_{lk}}{d A_j}=-2\pi i \sum_{branch \;\;points\;\;x_r}  {\rm res}|_{x_r}\f{v_l v_k v_r}{d\xi \,d(v/d\xi)}
\la{Oh2}
\ee
where $\xi$ is a local coordinate on $\CC$ near $x_r$. 
\end{proposition}

Notice that the right-hand side of (\ref{Oh2}) is independent of the choice of the local coordinates coordinates near $x_r$.
The formula (\ref{Oh2}) is compatible with variational formula for $B(x,y)$ derived in \cite{Baraglia}. It provides the explicit 
version of the variation of moduli based on the Donagi-Markman cubic \cite{DonagiMarkman}.

\section{Tau-functions, determinants of flat Laplacians and Teichm\"uller flow}
\la{detteich}

\subsection{Determinant of Laplace operator in flat metrics with trivial and $\Z_2$ holonomy}


Any metric with trivial holonomy and finite volume on a Riemann surface $\CC$ can be represented in the 
form $|v|^2$ where $v$ is a holomorphic abelian differential on $\CC$. 
Denote zeros of $v$ by $x_1,\dots,x_N$ and their multiplicities by $m_1,\dots,m_N$.
Since the metric has conical singularities, 
the definition of the Laplace operator $\Delta$ is ambiguous; the most natural is the  Friedrich's 
self-adjoint extension (see \cite{JDG}). The spectrum of $\Delta$ can then be shown to be discrete and the determinant of Laplacian can then be defined via the 
usual $\zeta$-function regularization: 
$$
\det\Delta=e^{-\zeta'(0)}\;,\hskip0.7cm 
\zeta(s)=\sum_{n=1}^\infty \lambda_n^s
$$
where $\lambda_1,\lambda_2,\dots$ are  eigenvalues of $\Delta$.

Then the determinant of Laplacian can be expressed as follows \cite{JDG}
\be
\det\Delta^{C,|v|^2}= const\, (\det \Im\Omega)\,{\rm Area}(\CC,|v|^2)\,|\tau(\CC,v)|^2 
\la{DetA}
\ee
where $\tau(\CC,v)$ is the tau-function on the connected component 
of the moduli space $\Hcal(m_1,\dots,m_N)$;
 $\Omega$ is the period matrix of $\CC$ and 
\be
{\rm Area}(\CC,|v|^2)=\Im\sum_{j=1}^g A_j\overline{B}_j
\la{Areaab}
\ee
The constant in (\ref{DetA}) depends on the partition $(m_1,\dots,m_N)$ and may depend also on the  connected component of the moduli space $\Hcal(m_1,\dots,m_N)$.

The formula (\ref{DetA}) is the natural generalization of the first Kronecker limit formula which holds in genus one.
Namely, on the torus with the flat coordinate $z$ and the periods $(A,B)$ 
$$
\det\Delta^{|dz|^2} =const\,\Im(B/A)\Im(A\bar{B})|\eta(B/A)|^4  
$$
where the constant  is  explicitly computable.

Similarly, an arbitrary metric on $\CC$ with $\Z_2$ holonomy and finite volume has the form $|Q|$ where $Q$ is a holomorphic quadratic differential on $\CC$. 
Denote multiplicities of zeros of $Q$ by $k_1,\dots,k_M$ and define the canonical cover $\Ch$
by the  equation
$$
v^2=Q 
$$
whose genus equals $\gh=2g+m_{odd}/2+1$ where  $m_{odd}$ is the number of zeros of $Q$ of odd multiplicity.
Then $v$ is a holomorphic abelian differential on $\Ch$; multiplicities of zeros of $v$ are determined by
$k_1,\dots,k_M$.

Introduce the Hodge tau-function $\tau(\CC,Q)$ and the tau-function 
$\tauh(\Ch,v)$; define also the Prym tau-function via the ratio (\ref{Prymtau})
\be
\tau_-(\CC,Q)=\f{\tau(\Ch,v)}{\tau(\CC,Q)}\;.
\la{taum}
\ee

The homogeneity coefficient of $\tau_-$ is the difference between the homogeneity coefficients of $\tau(\Ch,v)$ and $\tau(\CC,Q)$:

\be
\kappa_-=\kappa_+ + \f{1}{8}\sum_{k_j\;\; {\rm even}} \f{1}{k_j+2}
\la{kappammin}
\ee

The formula for the determinant of Laplacian on $\CC$ in the metric $|Q|$ is given by the formula which looks like a direct analog of  (\ref{DetA}):
\be
\det\Delta^{\CC,|Q|}=const\,(\det \Im\Omega)\, {\rm Area}(\CC,|Q|)|\tau(\CC,Q)|^2 
\la{detQ}
\ee
where the area is expressed in terms of the periods $(A_i,B_i)_{i=1}^{g_-}$ of $v$:
\be
{\rm Area} (\CC)= \Im \sum_{j=1}^{g_-} A_i \bar{B}_i\;.
\ee
The constant in the r.h.s. of (\ref{detQ})  depends on the connected component of the space $\Qcal(k_1,\dots,k_M)$.

On the other hand, applying the formula (\ref{DetA}) to the canonical cover $\Ch$ equipped with the abelian differential $v=\sqrt{Q}$ we have the following formula for the determinant of Laplacian on $\Ch$:
\be
\det\Delta^{\Ch,|Q|}=const\,(\det \Im\widehat{\Omega}) \, {\rm Area}(\Ch,|Q|) \,|\tau(\Ch,v)|^2
\la{detQ1}
\ee
where
\be
{\rm Area}(\Ch,|Q|)=2{\rm Area}(\CC,|Q|)\;;
\la{Area1}
\ee
and the constant in (\ref{detQ1}) may be different from the constant in (\ref{detQ}), but  may depend on the connected component in the moduli space $\Qcal(k_1,\dots,k_M)$ as well.

Take the ratio of (\ref{detQ1}) and (\ref{detQ}) and define the {\it Prym Laplacian} $\Delta_-^{\Ch,Q}$ which is the Laplacian on $\Ch$ with metric $|Q|$, restricted to the subspace of functions which are skew-symmetric under the involution interchanging the sheets of $\Ch$. Then the
determinant of $\Delta_-^{\Ch,Q}$ is naturally defined by 
\be
\det\Delta_-^{\Ch,|Q|}=\frac{\det\Delta^{\Ch,|Q|}}{\det\Delta^{\CC,|Q|}}\;.
\la{detPrym}
\ee

The  determinant of Prym Laplacian  can then be expressed as follows in terms of Prym tau-function $\tau_-$:
\be
\det\Delta_-^{\Ch,|Q|}=const\,(\det \Im\Pi)\, |\tau_-(\Ch,Q)|^2 \,
\la{detPrym1}
\ee
where $\tau_-(\Ch,Q)$ is the Prym tau-function (\ref{taum}) and $\Pi$ is the Prym matrix of $\Ch$ defined by (\ref{shs}). To derive (\ref{detPrym1}) from (\ref{detQ}) and (\ref{detQ1}) one needs to
use the relation (\ref{Area1}) between the areas of $\CC$ and $\Ch$ and the  relation 
\be
\f{\det\Im \widehat{\Omega}}{\det\Im \Omega}=\det \Im\Pi
\ee
which  follows from the definition of the Prym matrix \cite{Fay73}.

\subsection{Tau-functions in the theory of Teichm\"uller flow}

Tau functions on moduli spaces of abelian and holomorphic differentials naturally appear in the theory of 
the Teichm\"uller flow. 
We are going to briefly describe the context  following  \cite{EKZ,Forni}. 

\subsubsection{Lyapunov exponents of Hogde bundle}

Consider  any stratum $\Hcal_g(k_1,\dots,k_n)$ 
of the moduli space of holomorphic differentials.
The Teichm\"uller geodesic flow on $\Hcal_g(k_1,\dots,k_n)$ is defined by the action 
of diagonal matrices of the form $\left(\ba{cc} e^s& 0\\ 0 & e^{-s}\ea\right)$
in the plane of the flat coordinate $z$ which  is
defined as the abelian integral of the holomorphic differential $v$ ($(\CC,v)$ is a point of this stratum).

An arbitrary $SL(2,\R)$ linear transformation
of $z$-coordinate is also acting on the stratum $\Hcal_g(k_1,\dots,k_n)$;
this defines foliation of the stratum into the orbits of $SL(2,\R)$.
Then projectivisation $P\Hcal_g(k_1,\dots,k_n)$ is  foliated into the union 
of Teichm\"uller disks which are isomorphic to  $SL(2,\R)/SO(2)$ i.e. to the upper half-plane.

 The $SL(2,\R)/SO(2)$ matrix can be parametrized by a complex parameter $t$ 
 as follows:
\be
M_t=\f{1}{t+\bar{t}}
\left(\ba{cc} 2              &      i(t-\bar{t})    \\
             i(t-\bar{t})    &      2 t \bar{t}  \ea \right)\;.
             \la{Mdef}
\ee
On the Teichm\"uller disk  parametrized by $t\in \C$ one defines the  hyperbolic "Teichm\"uller" Laplacian \cite{EKZ} which is the hyperbolic Laplacian in the metric of curvature $-4$):
$$
\Delta^{T}= 16(\Re t)^2 \p^2_{t\bar{t}}\;.
$$

For the definition of  the Lyapunov exponents of the Teichm\"uller flow we give the  quote from the section 1.4 of \cite{EKZ}:
\vskip0.3cm
{\it "Informally, the Lyapunov exponents of a vector bundle
endowed with a connection can be viewed as logarithms of mean eigenvalues of
monodromy of the vector bundle along a flow on the base.

In the case of the Hodge bundle, we take a fiber of $H^1_{\R}$
and pull it along a
Teichm\"uller geodesic on the moduli space. We wait till the geodesic winds a lot
and comes close to the initial point and then compute the resulting monodromy
matrix $A(t)$. Finally, we compute logarithms of eigenvalues of $A^TA$, and normalize
them by twice the length $t$ of the geodesic."}
\vskip0.3cm
In this way one gets a set of $2g$ real numbers which is symmetric with respect to the origin; the 
largest of them equals $\l_1=1$ and the other $g-1$  
are ordered in such a way that $\l_1\geq \l_2\leq\dots\geq \l_g$.  The sum $\l_1+\dots+\l_g$ turns out to have many
nice properties (in particular, it is always rational). It  also has a deep geometrical meaning as the  integral
over a connected component of the stratum of the moduli space (this formula was proposed in \cite{Kontsevich} and proved by \cite{Forni2} in more general setting when the connected component of the stratum is replaced by  any subspace of the moduli space invariant under the action of the Teichm\"uller flow, see \cite{EKZ} for the complete history and references).
This relation looks as follows
\be
\l_1+\dots+\l_g=-\f{1}{4}\int_{\Mcal_1} \Delta^{T}\,\log\{\det\, \Im\Omega\}\, d\nu_1
\la{sumLya}
\ee
where $\Mcal_1$ is any connected component (there could be up to 3 of them, see \cite{KonZor}) of $\Hcal_g(k_1,\dots,k_n)$ and $\Omega$ is the period matrix of $\CC$; $\nu_1$ is the natural volume form (defined in the homological coordinates) on the component $\Mcal_1$ normalized such that the total volume of $\Mcal_1$ is 1. 

The formula  (\ref{sumLya}) can be rewritten as the sum of two geometrically important parts by 
representing the  expression $\det \Im\Omega$ as the ratio of two "more complicated" expressions using the
holomorphic factorization formula (\ref{DetA}):
\be
\det \Im\Omega=\frac{\det\Delta^{\CC,|v|^2}}{|\tau(\CC,v)|^2 \, {\rm Area}(\CC,|v|^2)}\;.
\ee

Then, since $SL(2)$ transformation preserves the area of $\CC$, we get 
\be
\Delta^{(h)} \log\det \Im\Omega=\Delta^{(h)}\{\log\det\Delta^{|v|^2}\}- \Delta^{(h)}\{\log(|\tau(\CC,v)|^2)\}
\la{split}
\ee
The last term of (\ref{split}) can be explicitly evaluated using the homogeneity property (\ref{Hom1}) of  the tau-function
$\tau(\CC,v)$. 

Namely, the $SL(2)$ transformation (\ref{Mdef}) acts on each homological coordinate $z_j$ in the same way as it acts  on the flat coordinate $z$:
\be
\left(\ba{c} \Re z^t_j\\ \Im z^t_j\ea\right)=M_t\left(\ba{c} \Re z_j\\ \Im z_j\ea\right)\;,
\ee
or
\be
z_i^t= \f{1}{t+\tbar} \left\{(1+t\tbar) z_i +(1-t+\tbar - t\tbar) \zbar_i\right\}\;.
\la{zit}
 \ee
\be 
\zb_i^t=\f{1}{t+\tbar} \left\{(1+t\tbar) \zb_i +(1+t-\tbar - t\tbar)
z_i \right\} 
\la{zbt}
\ee 

For any function $f(\{z_j,\bar{z}_j\})$ the Teichm\"uller Laplacian is defined by
$$\Delta^{T} f={4}(t+\tbar)^2 \,\p^2_{t\tbar}\{ f(\{z_j(t,\tbar),\bar{z}_j(t,\tbar))\}\big|_{t=1}\;.$$

We have 
\be
f_{t\tbar}=\sum_{i,j} \left(f_{z_i z_j} {z_i}_t {z_j}_{\tbar} +f_{\bar{z}_i \bar{z}_j} {\bar{z}_i}_t {\bar{z}_j}_{\tbar}+
f_{z_i \bar{z}_j} {z_i}_t {\bar{z}_j}_{\tbar}\right)
+\sum_i\left(f_{z_i} {z_i}_{t\tbar}+f_{\bar{z}_i}{\bar{z}_i}_{t\tbar}\right)
\la{fttb}
\ee

From (\ref{zit}) it follows that
$$
\f{\p z_i}{\p\tbar}\big|_{t=1}=\f{\p\zbar_i}{\p t} \big|_{t=1}=0
$$
Moreover, since $\tau$ is a holomorphic function of $\{z_i\}$, we have $(\log|\tau|^2)_{z_i\zbar_j}=0$, and, therefore, all terms in the double sum 
in (\ref{fttb}) vanish if $f=\log|\tau|^2$.

The remaining terms we compute  using (\ref{zit}):
\be
\f{\p^2 z_i}{\p t\,\p\tbar} \big|_{t=1}=\f{z_i}{8}\;,\;\hskip0.7cm
\f{\p \zbar_i}{\p t\,\p\tbar} \big|_{t=1}=\f{\zbar_i}{8}
\ee
and, therefore (we need to multiply by 16 to take care of the the factor $4(t+\tbar)^2$)
\be
\Delta^{T}\log |\tau|^2= {2}\sum_i \left(z_i\p_{z_i}\right)\log\tau+{2}\left(\sum_i \zbar_i\p_{\zbar_i}\right)\log\bar{\tau}\;.
\ee
Thus, taking into account the homogeneity (\ref{Hom1}) of the tau-function, we get the following simple formula:
\be
\Delta^{T}\log |\tau|^2= \f{1}{3}\sum_{i=1}^N \f{m_i(m_i+2)}{m_i+1}\;.
\la{Deltahtau}
\ee

Then the Kontsevich-Zorich-Forni formula (\ref{sumLya}), being combined with (\ref{split}), gives the expression which was first derived  in \cite{EKZ}:
\be
\l_1+\dots+\l_g=\f{1}{12}\sum_{i=1}^N \f{m_i(m_i+2)}{m_i+1} -\f{1}{4}\int_{\Mcal_1} \Delta^{T}\log\det\Delta^{|v|^2}\;.
\la{sumla}
\ee

The last integral turns out also  to have an important geometrical meaning: according to \cite{EKZ} there is the relation
\be
\int_{\Mcal_1} \Delta^{T}\log\det\Delta^{|v|^2}=-\f{4\pi^2}{3} c_{area}(\Mcal_1)
\la{detSW}
\ee
where $c_{area}(\Mcal_1)$ is the  Siegel-Veech constant of $\Mcal_1$.

\subsubsection{Lyapunov exponents of Hodge and Prym bundles on spaces of quadratic differentials}

The Teichm\"uller flow preserves also each stratum $\Qcal_g(k_1,\dots,k_M)$ of the space of quadratic differentials. 
There are two natural vector bundles over such stratum or any of its subspaces which are invariant under the flow:
the Hodge bundle $\Lambda_H$  and the Prym bundle $\Lambda_P$. The fiber of the latter is the space $H^-$ of Prym differentials on the canonical cover
$\Ch$ (we denote $\dim H^-=2g_-$ as before). The fiber of their direct sum
$$
\widehat{\Lambda}_H=\Lambda_H\otimes\Lambda_P
$$
is  the space of holomorphic abelian differentials on $\Ch$.

Introduce now the Hodge Lyapunov exponents $\l^+_1,\dots,\l^+_g$, Prym Lyapunov exponents $\l^-_1,\dots,\l^-_{g_-}$
and Lyapunov exponents $\lh_1,\dots,\lh_{\gh}$ of the bundle $\widehat{\Lambda}_H$; clearly
\be
\lh_1+\dots+\lh_{\gh}=(\l^+_1+\dots+\l^+_g)+(\l^-_1+\dots+\l^-_{g_-})\;.
\la{sumlapm}
\ee

Denote again by $\Mcal_1$ a subspace of $\Qcal_g(k_1,\dots,k_M)$ invariant under the action of the Teichm\"uller flow.
Then the Kontsevich-Zorich-Forni theorem implies
\be
\l^+_1+\dots+\l^+_g=-\f{1}{4}\int_{\Mcal_1}\Delta^{T}\,\log\{\det\, \Im\Omega\}\, d\nu_1
\la{sumlplus}
\ee
or, equivalently, using (\ref{detQ}),
\be
\l^+_1+\dots+\l^+_{g_+}=\f{1}{12}\sum_{i=1}^N \f{k_i(k_i+4)}{k_i+2} -\f{1}{4}\int_{\Mcal_1} \Delta^{T}\log\det\Delta^{\CC,|Q|}\;.
\ee

Applying the Kontsevich-Zorich-Forni theorem to the Hodge bundle over the moduli space of covers $\Ch$ (which is an invariant subspace 
of $\Hcal_g(\{\widehat{k}_i\})$ where $(v)=\sum \widehat{k}_j x_j$ on $\Ch$)
 we get
\be
\lh_1+\dots+\lh_{\gh}=-\f{1}{4}\int_{\Mcal_1}\Delta^{T}\log\{\det\, \Im\widehat{\Omega}\}\, d\nu_1
\la{sumlh}
\ee
where $\widehat{\Omega}$ is the period matrix of $\Ch$. Due to (\ref{detPrym1}) we get from (\ref{sumlapm}) the following formula for the sum of Prym Lyapunov exponents:
\be
\l^-_1+\dots+\l^-_{g_-}=-\f{1}{4}\int_{\Mcal_1}\Delta^{T}\,\log\{\det\, \Im\Pi\}\, d\nu_1\;.
\la{sumlam}\ee
Using (\ref{detPrym1}), we can express this sum via the determinant of Prym Laplacian and the Prym tau-function:
\be
\l^-_1+\dots+\l^-_{g_-}=-\f{1}{4}\int_{\Mcal_1}\Delta^{T}\log\det\Delta_-^{\Ch,|Q|}\,\, d\nu_1 + \f{1}{4}\int_{\Mcal_1}\Delta^{T}\log|\tau_-(\Ch,Q)|^2\,\, d\nu_1\;.
\la{sumlam1}
\ee
Similarly to (\ref{Deltahtau}), using the expression (\ref{kappammin}) for the homogeneity coefficient of $\tau_-$, we get
\be
\l^-_1+\dots+\l^-_{g_-}=\f{1}{24}\sum_{j=1}^M \f{k_j(k_j+4)}{k_j+2} + \f{1}{4}\sum_{k_j\;\; {\rm even}} \f{1}{k_j+2} -\f{1}{4}\int_{\Mcal_1}\Delta^{T}\log\det\Delta_-^{\Ch,|Q|}\,\, d\nu_1\;.
\la{sumlam2}
\ee
It was proved in \cite{EKZ} (Lemma 1.1) that 
the Siegel-Veech constant of $\Ch$ and $\CC$ are related by the factor of 2 which implies
the somewhat unexpected identity
\be
\int_{\Mcal_1}\Delta^{T}\log\det\Delta_-^{\Ch,|Q|}\,\, d\nu_1=\int_{\Mcal_1}\Delta^{T}\log\det\Delta^{\CC,|Q|}\,\, d\nu_1\;.
\la{intint}
\ee
In turn, this gives  the  link between the sums of  Hodge and Prym Lyapunov's exponents which was  originally proved in \cite{EKZ}:
\be
\l^-_1+\dots+\l^-_{g_-}=\l^+_1+\dots+\l^+_{g}+\f{1}{4}\sum_{k_j\;\; {\rm even}} \f{1}{k_j+2}\;.
\la{lplm}
\ee

Concluding, we see that the explicit terms in the sum of the Lyapunov exponents of both Hodge and Prym vector 
bundles have the meaning of homogeneity coefficients of the tau-functions $\tau_+$ and $\tau_-$, respectively.
This observation provides a non-trivial link between the world of integrable dynamical systems and the world of ergodic dynamical systems 
where the theory of Teichm\"uller flows belongs.

\appendix

\section{ Canonical bidifferential and Szeg\"o kernel on Riemann surfaces. Variational formulas}
\la{appsec}
Here we summarize a few facts from the theory of Riemann surfaces and their variations which are used in the main text.

 On  a compact Riemann surface $\CC$ of genus
$g$ introduce a canonical basis of cycles $(a_\a,b_\a)$ in  $H_1(\RS,\Z)$. Denote by $B(x,y)$ for $x,y\in \CC$ the canonical bidifferential,
which is the symmetric bimeromorphic differential on $\RS$ having quadratic pole with biresidue 1 on the diagonal and vanishing $a$-periods. 
The bidifferential $B$ is expressed via the the prime-form $E(x,y)$ as follows: $B(x,y)=d_x d_y \log E(x,y)$. Consider the basis of holomorphic differentials
$v_\a$ on $\RS$ normalized as follows:
\be
\oint_{a_\a}v_{\beta}=\delta_{\a\b}\;.
\ee
The period matrix $\O$ of $\RS$ is given by
\be
\O_{\a\b}=\oint_{b_\a} v_\b\;.
\ee
The Abel map with the base-point $x_0\in\CC$ is defined by 
\be
\Acal_\a(x)=\int_{x_0}^x v_\a\;.
\la{Abel}\ee

Introduce the theta-function with characteristics $\Thpq({\bf z}|\Omega)$, where $\pb,\qb\in\C^g$ are vectors of 
characteristics; ${\bf z} \in\C^g$ is the argument. 
The theta-function satisfies the heat equation:
\be
\f{\p^2\Thpq({\bf z})}{\p z_\a\p z_\b}=4\pi i\f{\p\Thpq({\bf z})}{\p\O_{\a\b}}\;.
\la{heat}\ee
Let us consider some non-singular odd half-integer characteristic $[\pb^*,\qb^*]$. 
The prime-form $E(x,y)$ is defined as follows:
\be
E(x,y)=\f{\Th\left[^{\pb^*}_{\qb^*}\right](\Acal(x)-\Acal(y))}{h(x) h(y)}\;,
\la{prime}\ee
where the square of a section $h(x)$ of a spinor bundle over $\CC$ is given by the following expression:
\be
h^2(x)=\sum_{\a=1}^g \p_{z_\a}\left\{\Th\left[^{\pb^*}_{\qb^*}\right](0)\right\} v_j(x)\;.
\la{hp}\ee
Then  $h(x)$ itself is  a  section of the spinor bundle corresponding to the characteristic
$[^{\pb^*}_{\qb^*}]$.  The automorphy factors of the prime-form along all cycles $a_\a$ are trivial;
the automorphy factor along  cycle $b_\a$ equals to  $\exp\{-\pi i \O_{\a\a}- 2\pi i (\Acal_\a(x)-\Acal_\a(y))\}$. 
The prime-form is independent of the choice of characteristic $[^{\pb^*}_{\qb^*}]$. It has also   the following local behaviour as $x\to y$:
\be
E(x,y)=\frac{\xi(x)-\xi(y)}{\sqrt{d\xi(x)}\sqrt{ d\xi(y)}}(1+ o(1))\;,
\la{asprime}\ee
where $\xi(x)$ is a local parameter.

Choosing some local coordinate $\xi$ near the diagonal $\{x=y\}\subset \RS\times \RS$,  we have the following expansion of $B(x,y)$ as $y\to x$:
\be
B(x,y)=\left(\f{1}{(\xi(x)-\xi(y))^2}+\f{S_B(\xi(x))}{6}
+O((\xi(x)-\xi(y))^2)\right)d\xi(x) d\xi(y)
\la{asW}
\ee
where $S_B$ is a projective connection on $\RS$ called the {\it Bergman projective connection}.

If two canonical bases of cycles on $\RS$, $\{a_\a',b_\a'\}_{\a=1}^g$ and 
$\{a_\a,b_\a\}_{\a=1}^g$
 are related by
a matrix 
\be
\sigma=
\left(\begin{array}{cc} d & c\\
b & a \end{array}\right)\in Sp(2g,\Z)\;,
\la{symtrans}
\ee
then the corresponding canonical bidifferentials are related as follows (see item 4 on page 21 of \cite{Fay73}):
\be
B^{\sigma}(x,y)=B(x,y)-2\pi i\sum_{\a,\b=1}^g  [(c\O+d)^{-1} c]_{\a\b}\; v_\a(x) v_\b(y)\;.
\la{BsB}
\ee
  
The period  matrix  $\Omega^\sigma$, corresponding to the new canonical basis of cycles, is related to $\Omega$ as follows:
\be
\Omega^\sigma=(a\O+b) (c\O+d)^{-1}\;.
\la{sympBper}
\ee

For any two vectors $\pb,\qb\in\C^g$  such that $\Th\left[^\pb_\qb\right](0)\neq 0$ the Szeg\"o kernel is
defined by the formula
\be
S_{pq}(x,y) = \f{1}{\Th\left[^\pb_\qb\right](0)}\f{\Th\left[^\pb_\qb\right](\Acal(x)-\Acal(y))}{E(x,y)}\;.
\la{szego}\ee

Near the diagonal, when $y\to x$, $S_{p,q}$ behaves as follows 
\be
S_{pq}(x,y)=\left(\f{1}{\xi(x)-\xi(y)}+ a_0(x) + O(\xi(x)-\xi(y))\right)\sqrt{d \xi(x)}\sqrt{d \xi(y)}
\la{Sdia}
\ee
where $\xi(x)$ is a local coordinate and coefficient $a_0$ is given by (\cite{Fay92}, p.29)
\be
a_0(x)=\f{1}{d \xi(x)}\sum_{\a=1}^g \p_{\a}\{\log\Thpq(0)\}{v_\a(x)}\;.
\la{a0P}
\ee

\begin{proposition}\la{thvarpq} ({\bf Prop. 1 of \cite{CMP}})
The following variational formulas for the Szeg\"o kernel with respect to components of 
characteristic vectors $p$ and $q$ hold:
\be
\f{d}{d p_\a} S_{pq}(x,y)=  - \oint_{b_\a}  S_{pq}(x,t)S_{pq}(t,y)\;,
\la{varSp}
\ee
\be
\f{d}{d q_\a} S_{pq}(x,y)= - \oint_{a_\a}  S_{pq}(x,t)S_{pq}(t,y)\;.
\la{varSq}
\ee
\end{proposition}

The Szeg\"o kernel is related to  $B(x,y)$ by the following relation \cite{Fay73}:
\be
S_{pq}(x,y)S_{pq}(y,x)=- B(x,y)- \sum_{\a,\b=1}^g \p_\a\p_\b\log\th_{pq}(0)\,
 v_\a(x) v_\b(y)\;.
\la{SSB}\ee

For any two sets   $x_1,\dots,x_n$ and $y_1,\dots,y_n$ of points on the Riemann surface Fay discovered 
the following  identity (see \cite{Fay73}, p.33):
\be\la{ident}
\det\{S(x_j,y_k)\}
= \frac{\Th\left[^\pb_\qb\right]\left(\sum_{j=1}^n (\Acal(x_j)-\Acal(y_j))\right)}
{\Th\left[^\pb_\qb\right](0)}\frac{\prod_{j<k} E(x_j,x_k) E(y_k,y_j)}{\prod_{j,k} E(x_j,y_k)}\;.
\ee
In particular, for $n=2$ (\ref{ident}) is  called {\it Fay trisecant identity}.

Let us also define
\be
\Ccal(x)=\frac{1}{{\mathcal W}(x)}\sum_{\a_1, \dots,
\a_g=1}^g
\partial^g_{\a_1,\dots\a_g}\Theta(K^x)
v_{\a_1}\dots v_{\a_g}(x)
\la{Ccal}\ee
where
\be
{\mathcal W}(x):= {\rm \det}_{1\leq \a, \b\leq g}||v_\b^{(\a-1)}(x)||
\la{Wronks}
\ee
is the Wronskian determinant of the basic holomorphic differentials, and $K^x$ is the vector of Riemann constants with initial point $x$.
The expression (\ref{Ccal}) is a multi-valued $n(1-n)/2$-differential on $\CC$ which  does not have any zeros or poles \cite{Fay92}.


In the case of  genus $1$ the $x$-dependence in (\ref{Ccal}) drops out and $\Ccal(x)$ turns into $\th'((\O+1)/2)$.

Let us now consider some Abelian differential $v$ on $\CC$; it could be a differential of either first, second or third kind on;
it can be also a meromorphic differential of mixed type (i.e having both: poles of higher order and residues).

Introduce the meromorphic differential $Q_v$, which is
 given by the zero order term of the asymptotics of the  Bergman bidifferential on the diagonal: 
$$
Q_v=\frac{  B_{reg}(x,x)}{v(x)}\;,
$$
where the regularization is done using the differential $v$, i.e., 
\be
Q_v(x)=\frac{1}{v(x)}\left( B(x,y)-\f{v(x)v(y)}{(\int_{x}^y v)^2}\right)\Big|_{y=x}\;.
\la{defQv}
\ee

Denote by $\{\cdot,\cdot\}$ the Schwarzian derivative and
define the following meromorphic projective connection associated with the differential $v$:
\be
S_v:= \left\{ \int^x v, \xi(x)\right\}\equiv
\frac{v''}{v}-\frac{3}{2}\left(\frac{v'}{v}\right)^2
\la{defSv}
\ee
where prime denotes the derivative with respect to the local coordinate $\xi$. 

Since   $S_v$ is the meromorphic projective connection on $\CC$, 
the difference $S_B-S_v$ is the meromorphic quadratic differential. 
 Dividing $S_B-S_v$ by $6v$ we also get the differential $Q_v$:
\be
Q_v:= \f{1}{6}\f{S_B-S_v}{v}
\la{defQv1}
\ee
The meromorphic Abelian differential $Q_v$,  constructed from the  (holomorphic or meromorphic) Abelian differential $v$ 
 plays the key role in the construction of the Bergman tau-function. Notice that $Q_v$ is determined by $v$ and depends on  Torelli marking of  $\RS$.

\begin{remark}\rm
Using the definition of the Schwarzian derivative (\ref{defSv})  it is easy to verify that at the pole  $x_i$ ($i=1,\dots,n$) of order $-k_i$  the differential $Q_v$ has 
zero of order $k_i-2$ for poles of order $3$ and higher (i.e. when $k_i\leq -3$). At a pole of $v$ of second order, i.e. when $k_i=-2$, the differential $Q_v$ has zero of order at least 1.
Finally, at a simple pole of $v$ with residue $r$, the differential $Q_v$ also has the simple pole, and its residue equals to $-1/(12r)$. 
At a zero $x_{n+i}$ of order $k_{n+i}$, the differential  $Q_v$ has the pole of order $k_{n+i}+2$. 
\end{remark}

As a corollary of (\ref{BsB}), $Q_v$ transforms as follows under a symplectic transformation $\sigma$:
\be
Q_v^{\sigma}(x)=Q_v(x) - 12\pi i\sum_{\a,\b=1}^g  [(c\O+d)^{-1} c]_{\a\b} \f{v_\a(x) v_\b(x)}{v(x)}\;.
\la{transQv}
\ee

\subsection{Spaces of meromorphic Abelian differentials}

Denote by $\Hgk$ the space of pairs $(\CC,v)$ where $\CC$ is a Riemann surface of genus $g$ and $v$ is a meromorphic differentials on $\CC$ such that
the divisor of $v$ is given by $(v)=\sum_{i=1}^{m+n} k_i x_i$ with $k_1,\dots,k_n<0$ and $k_{n+1},\dots,k_{n+m}>0$.

Considering the homology group of $\RS$ punctured at poles of $v$, relative to the set of zeros of $v$, which is denoted by 
\be
H_1(\RS\setminus\{x_i\}_{i=1}^n;\{x_i\}_{i=n+1}^{n+m})\;,
\la{relhol1}
\ee
we can choose its set of generators $\{s_i\}_{i=1}^{2g+m+n-2}$ as follows:
$$
s_\a= a_\a\;,\;\;  s_{\a+g}= b_\a \hskip0.5cm \a=1,\dots,g\;; \hskip0.5cm
s_{2g+k}=c_{k+1}\;, k=1,\dots,n-1\;,
$$
\be
s_{2g+n-1+k}= l_{n+1+k}\;,\hskip0.5cm k=1,\dots,m-1
\la{defsmerom}
\ee
where 
$c_2,\dots,c_{n}$ are small contours around the poles $x_2,\dots,x_{n}$;
$l_{n+2},\dots,l_{n+m}$ are contours connecting the ``first'' zero $x_{n+1}$ with the other zeros $x_{n+2},\dots,x_{n+m}$.
 
 The homological coordinates on $\Hgk$ are defined as integrals of $v$ over $\{s_i\}$:
\be
z_i=\int_{s_i} v\;,\hskip0.5cm i=1,\dots, 2g+n+m-2
\la{homcoormain}
\ee

The homology group dual to (\ref{relhol1}) is the homology group of $\RS$, punctured at {\it zeros} of $v$, relative to the set of {\it poles} of $v$; it is denoted by 
\be
H_1(\RS\setminus\{x_i\}_{i=n+1}^{n+m};\{x_i\}_{i=1}^{n})\;.
\la{defsstarmer}
\ee
The  set  of generators $\{s_i^*\}_{i=1}^{2g+n+m-2}$ of the group (\ref{defsstarmer}) dual to (\ref{defsmerom}) is given by 
 $$
s^*_\a= -b_\a\;,\;\;  s^*_{\a+g}= a_\a \hskip0.5cm \a=1,\dots,g\;; \hskip0.5cm
s^*_{2g+k}= -\tilde{l}_{k+1}\;,\hskip0.5cm k=1,\dots,n-1\;
$$
\be
s^*_{2g+n-1+k}= \tilde{c}_{n+1+k}\;,\hskip0.5cm k=1,\dots,m-1
\la{dualcont2}
\ee
where $\tilde{l}_2,\dots,\tilde{l}_{n}$ are contours connecting the ``first'' pole $x_1$ with other poles $x_2,\dots,x_n$, respectively; $\tilde{c}_{n+2},\dots,\tilde{c}_{m+n}$ are small
circles around the zeros $x_{n+2},\dots, x_{n+m}$.

The variational formulas for the period matrix, the basic holomorphic differentials,the  Bergman bidifferential and the differential 
$Q_v$
formally look  analogous to the case of the space of
holomorphic differentials (see Theorem 3 of \cite{JDG}). 

To write down these formulas we introduce  on $\RS$ a system of cuts homologous to $a$- and $b$-cycles to get the fundamental polygon ${\RS}_0$; inside
of  ${\RS}_0$ we also introduce branch cuts connecting poles of $v$ with non-vanishing residues; these branch cuts are assumed to
start at $x_1$, i.e. they connect $x_1$ with $x_2,\dots,x_n$; denote them by the same letters $\tilde{l}_2,\dots,\tilde{l}_n$ 
as their homology classes in (\ref{dualcont2}). In this way we get the domain $\tilde{\RS}_0$ where 
the Abelian integral $z(x)=\int_{x_{n+1}}^x v$ is single-valued.

Now we are in a position to formulate the following theorem, which gives variational formulas on $\Hgk$ with respect to homological coordinates 
$\Pcal_{s_i}=\int_{s_i}v$.

\begin{theorem}
\la{thvarfo}
\cite{CMP}
The variational formulas for period matrix  $\O$, the normalized holomorphic differentials $u_\a$, the Bergman bidifferential $B(x,y)$ and the differential $Q_v$
 on the moduli  space $\Hgk$ look as follows (in the cases of $v_\a$, $B(x,y)$ and $Q_v(x)$ the 
coordinates $z(x)$ and $z(y)$ remain fixed under differentiation):
\be
\f{\p \O_{\a\b}}{\p \Pcal_{s_i}} =\int_{s_i^*}\f{v_\a v_\b}{v}\;,
\la{varO2}
\ee 
\be
\f{\p v_{\a}(x)}{\p \Pcal_{s_i}}\Big|_{z(x)=const} =\f{1}{2\pi i}\int_{t\in s_i^*}\f{v_\a(t)B(x,t)}{v(t)}\;,
\la{varva2}
\ee 
\be
\f{\p }{\p \Pcal_{s_i}}\log (E(x,y)\sqrt{v(x)}\sqrt{v(y)})\Big|_{z(x),z(y)=const} 
=-\f{1}{4\pi i}\int_{t\in s_i^*}\f{1}{v(t)}\left[d_t\log\f{E(x,t)}{E(y,t)}\right]^2\;,
\la{varExy2}
\ee 
\be
\f{\p B(x,y)}{\p \Pcal_{s_i}}\Big|_{z(x),z(y)=const} 
=\f{1}{2\pi i}\int_{t\in s_i^*}\f{B(x,t)B(y,t)}{v(t)}\;,
\la{varBxy2}
\ee 
\be
\f{\p Q_v(x)}{\p \Pcal_{s_i}}\Big|_{z(x)=const} 
=\f{1}{2\pi i}\int_{t\in s_i^*}\f{B^2(x,t)}{v(t)}\;,
\la{varQv2}
\ee 
\be
\f{\p}{\p \Pcal_{s_i}}{S_{pq}(x,y)}\Big|_{z(x),z(y)}= \f{1}{4}\int_{t\in s_i^*} \f{W_t[S_{pq}(x,t),\;S_{pq}(t,y)]}{v(t)}
\la{varSzego}\ee
where
$W_t$ denotes the Wronskian with respect to the variable $t$.
\end{theorem}

The variational formulas for $v_\a$, $B$ and $Q_v$ with respect to the relative periods were first given in \cite{Annalen}. The proof of  variational formulas with respect to coordinates
$\int_{a_\a}v$,  $\int_{b_\a}v$ and $\int_{x_{n+1}}^{x_{n+j}}v$ is given by  Theorem 3 of \cite{JDG}. The proof of variational formulas with respect to
residues $r_j$ of $v$  is given in   Theorem 1 of \cite{CMP}.

We notice that the Wronskian $W_t[S_{pq}(x,t),\, S_{pq}(t,y)]$ is a quadratic differential with respect to $t$; dividing it by $v(t)$ we get a meromorphic differential on $\RS$ (with respect to $t$).

\subsection{ Hurwitz spaces}

Let $\mu_1,\dots,\mu_M$ be $M$ partitions of $n$.
Denote by  $Hur_{g,n}(\mu_1,\dots,\mu_M)$ the Hurwitz space of pairs $(\CC,f)$ where $\CC$ is a Riemann surface of genus $g$  and $f$ 
is a meromorphic function on $\CC$ with simple poles and $M$ critical values such that the multiplicities of critical points corresponding to $k$th critical value 
are determined by the  partition $\mu_k$ (such that the simple critical point corresponds to entry $2$ of the partition; entry $1$ of $\mu_k$ corresponds to a regular point of $df$). The genus $g$ is expressed via the numbers of parts $p_1,\dots,p_M$ of the partitions via the Riemann-Hurwitz formula: $g=\sum_{i=1}^M (n-p_i)/2 -n+1$.

The Hurwitz space $Hur_{g,n}(\mu_1,\dots,\mu_M)$ is the subspace of the 
space $\Hgk$ with  $k_1=\dots=k_n=-2$: on the Hurwitz space the differential $v$ is exact: $v=df$. It means in particular that all periods of $v$ over cycles $a_\alpha$, $b_\alpha$ and $c_{k+1}$ in (\ref{defsmerom}) vanish. Moreover, periods over contours $l_{n+1+k}$ are dependent if there is more than one critical point of $f$ corresponding to the same critical value. In fact, the  periods of $v$ over the  cycles (\ref{defsmerom}) reduces to the set of critical values (the {\it branch points}) of the function $f$,
which we denote by $z_1,\dots,z_M$. 

Then theorem (\ref{thvarfo}) implies the following corollary 
\begin{corollary}\cite{Annalen,MPAG}
The variational formulas on  $Hur_{g,n}(\mu_1,\dots,\mu_M)$ with respect to the critical values $z_1,\dots,z_M$ of the function $f$ look as follows
\be
\f{\p \O_{\a\b}}{\p z_i} =2\pi i \sum_{p\in f^{-1}(z_i)}{\rm res}|_{t=p} \f{v_\a v_\b}{d f}\;,
\la{varO3}
\ee 
\be
\f{\p v_{\a}(x)}{\p z_i}\Big|_{z(x)=const} =2\pi i \sum_{p\in f^{-1}(z_i)}{\rm res}|_{t=p} \f{v_\a(t)B(x,t)}{df(t)}\;,
\la{varva3}
\ee 
\be
\f{\p}{\p z_i}\log (E(x,y)\sqrt{df(x)}\sqrt{df(y)})\Big|_{z(x),z(y)=const} 
=-\f{1}{2}\sum_{p\in f^{-1}(z_i)}{\rm res}|_{t=p} \f{1}{df(t)}\left[d_t\log\f{E(x,t)}{E(y,t)}\right]^2\;,
\la{varExy3}
\ee 
\be
\f{\p B(x,y)}{\p z_i}\Big|_{z(x),z(y)=const} 
= \sum_{p\in f^{-1}(z_i)}{\rm res}|_{t=p}\f{B(x,t)B(y,t)}{df(t)}\;,
\la{varBxy3}
\ee 
\be
\f{\p Q_v(x)}{\p z_i}\Big|_{z(x)=const} 
=\sum_{p\in f^{-1}(z_i)}{\rm res}|_{t=p}\f{B^2(x,t)}{df(t)}\;,
\la{varQv3}
\ee 
\be
\f{\p}{\p z_i}{S_{pq}(x,y)}\Big|_{z(x),z(y)}
= \f{1}{2} \sum_{p\in f^{-1}(z_i)}{\rm res}|_{t=p}\f{W_t[S_{pq}(x,t),\;S_{pq}(t,y)]}{df(t)}\;.
\la{varSzego3}\ee
\end{corollary}

\subsection{Spaces of holomorphic quadratic and $n$-differentials}

The moduli spaces of quadratic differentials (both holomorphic and meromorphic) can also be considered as  subspaces of the moduli spaces of 
abelian differentials, and corresponding variational formulas can be obtained by reduction of the variational formulas from Th.\ref{thvarfo}.
Here we consider the simplest and the most important case of the space $\Qcal^0_g$ which is the space of pairs $(\CC,\qd)$ where $\CC$ is a Riemann surface of genus $g$ and $\qd$ is a holomorphic differential on $\CC$ with $4g-4$ simple zeros (we refer \cite{contemp,BKN,BK2} for the general case); $\dim \Q_g^0 =6g-6$.

The periods $\{A_i,B_i\}_{i=1}^{3g-3}$ of the holomorphic Abelian differential $v$ over a basis   $(a_\alpha^-,b_\a^-)$ in odd homologies $H_-$ of   the canonical cover $\Ch$ defined by $v^2=\qd$
can be used as local coordinates on $\Qcal^0_g$. Since $v$ has on $\Ch$ zeros of order 2 at all branch points $\{x_i\}_{i=1}^{4g-4}$, the space $\Qcal^0_g$ 
can be considered as a subspace of the moduli space of Abelian differentials with all double zeros on Riemann surfaces of genus $\gh$. 
Therefore variational formulas on $\Qcal^0_g$ can be obtained from Th.\ref{thvarfo} (see Prop. 3.1 of \cite{BKN}) via an elementary chain rule.
Let us  consider the fundamental polygon $\Ch_0$ which is invariant under the involution $ \mu$ and such that $x_1$ is at the vertex of $\Ch_0$. Define the
coordinate on $\Ch_0$ by $z(x)=\int_{x_1}^x v$. Then we have

\begin{corollary}\la{varfoQ}
The variational formulas 
 on the moduli  space $\Qcal_g^0$ look as follows (in the cases of $v_\a$, $B(x,y)$ and $Q_v(x)$ the 
coordinates $z(x)$ and $z(y)$ remain fixed under differentiation):
\be
\f{\p \O_{\a\b}}{\p \Pcal_{s_i}} =\f{1}{2}\int_{s_i^*}\f{v_\a v_\b}{v}\;,
\la{varO4}
\ee 
\be
\f{\p v_{\a}(x)}{\p \Pcal_{s_i}}\Big|_{z(x)=const} =\f{1}{4\pi i}\int_{t\in s_i^*}\f{v_\a(t)B(x,t)}{v(t)}\;,
\la{varva4}
\ee 
\be
\f{\p }{\p \Pcal_{s_i}}\log (E(x,y)\sqrt{v(x}\sqrt{v(y)})\Big|_{z(x),z(y)=const} 
=-\f{1}{8\pi i}\int_{t\in s_i^*}\f{1}{v(t)}\left[d_t\log\f{E(x,t)}{E(y,t)}\right]^2\;,
\la{varExy4}
\ee 
\be
\f{\p B(x,y)}{\p \Pcal_{s_i}}\Big|_{z(x),z(y)=const} 
=\f{1}{4\pi i}\int_{t\in s_i^*}\f{B(x,t)B(y,t)}{v(t)}\;,
\la{varBxy4}
\ee 
\be
\f{\p Q_v(x)}{\p \Pcal_{s_i}}\Big|_{z(x)=const} 
=\f{1}{4\pi i}\int_{t\in s_i^*}\f{B^2(x,t)}{v(t)}\;.
\la{varQv4}
\ee 
\end{corollary}

The variational formulas for $B(x,y)$ used in \cite{contemp} differ from (\ref{varBxy4}) by the factor of 2 due to the different normalization of homological coordinates.

{\bf Spaces of $N$-differentials.}

Denote by $\Mgn$ the moduli space of pairs $(\CC,W)$ where $\CC$  is a Riemann surface of genus $g$ and
$W$ is a holomorphic $N$-differential on $\CC$;  ${\rm dim} \Mgn =(2N+2)(g-1)$.
 Consider the $N$-sheeted cover $\Ch$  defined by $v^N=W$; $\Ch$ possesses the natural $\Z_N$ symmetry; denote by $\mu$ the 
 degree $N$ isomorphizm acting on $v$ as $\mu^*v= \e v$ (here $\e=e^{2\pi i/N}$).
 
 Denote by $\Mgnp$ the subspace of $\Mgn$ defined by the condition  that all the zeros of $W$ are simple. Then the genus of $\Ch$ equals 
 $\gh=N^2(g-1)+1$.  Then the homology group of $\Ch$ can be decomposed as $H_1(\Ch,\C)=\oplus_{k=0}^{N-1}\Hcal_k$ where $\Hcal_k$ is the $k$th invariant subspace of $\mu_*$:
 for any $s\in \Hcal_k$ we have $\mu_* s=\e^k s$. Then $\dim \Hcal_0=g$ and $\dim \Hcal_k=(2N+2)(g-1)$ for $k=1,\dots,N-1$.
 
 Denote by $\{s_j\}_{j=1}^{(2N+2)(g-1)}$ a basis in $\Hcal_1$; then $\Pcal_j=\int_{s_j}v$ can be used as local coordinates on $\Mgnp$;
the  integrals of $v$ over other cycles from other $\Hcal_k$'s vanish. Let $\{s_j^*\}_{j=1}^{(2N+2)(g-1)}$ be the dual basis in  $\Hcal_{N-k}$ 
 satisfying $s_k^*\circ s_j=\delta_{jk}$.
 
 Denote the fundamental polygon of $\Ch$ by $\Ch_0$ and  define the
coordinate on $\Ch_0$ by $z(x)=\int_{x_1}^x v$. Then we have

\begin{corollary}\la{varfoQN}
The variational formulas 
 on the moduli  space $\Mgnp$ look as follows (in the cases of $v_\a$, $B(x,y)$ and $Q_v(x)$ the 
coordinates $z(x)$ and $z(y)$ remain fixed under differentiation):
\be
\f{\p \O_{\a\b}}{\p \Pcal_{s_i}} =\f{1}{N}\int_{s_i^*}\f{v_\a v_\b}{v}
\la{varO5}
\ee 
\be
\f{\p v_{\a}(x)}{\p \Pcal_{s_i}}\Big|_{z(x)=const} =\f{1}{2N\pi i}\int_{t\in s_i^*}\f{v_\a(t)B(x,t)}{v(t)}
\la{varva5}
\ee 
\be
\f{\p \log E(x,y)}{\p \Pcal_{s_i}}\Big|_{z(x),z(y)=const} 
=-\f{1}{4N\pi i}\int_{t\in s_i^*}\f{1}{v(t)}\left[d_t\log\f{E(x,t)}{E(y,t)}\right]^2
\la{varExy5}
\ee 
\be
\f{\p B(x,y)}{\p \Pcal_{s_i}}\Big|_{z(x),z(y)=const} 
=\f{1}{2N\pi i}\int_{t\in s_i^*}\f{B(x,t)B(y,t)}{v(t)}
\la{varBxy5}
\ee 
\be
\f{\p Q_v(x)}{\p \Pcal_{s_i}}\Big|_{z(x)=const} 
=\f{1}{2N\pi i}\int_{t\in s_i^*}\f{B^2(x,t)}{v(t)}
\la{varQv5}
\ee 
\end{corollary}
{\it Proof.} We give the short derivation of (\ref{varBxy5}) from the
variational formula (\ref{varBxy2}) on the stratum 
$\mathcal{H}_{\hat g}(N,\ldots,N)$ of the moduli space of 
holomorphic 1-differentials of genus $\gh=N^2(g-1)+1$,
 similarly to Lemma 5 of \cite{contemp} and Proposition 3.2 of \cite{BKN}. 

Namely, consider a point $(\Ch,v)$ of the space  $\mathcal{H}_{\hat g}(N,\ldots,N)$. Denote the zeros of $v$ on $\CC$ by $x_1,\dots,x_{N(2g-2)}$; the  same notation will be used for the zeros of $v$ on $\Ch$. The canonical bidifferential on $\Ch$ will be denoted by $\Bh(x,y)$. The homological coordinates on $\mathcal{H}_{\hat g}(N,\dots,N)$
are given by $\int_{s_i}v$ where $\{s_i\}=\{\{a_j,b_j\}_{j=1}^{\gh},\,\{l_i\}_{i=1}^{N(2g-2)-1}\}$ where the path $l_i$ 
connects $x_{N(2g-2)}$ with $x_i$. Consider the flat coordinate on $\Ch$ given by $z(x)=\int_{x_{N(2g-2)}}^x v$.

Under any choice of Torelli marking on $\Ch$ used to define $\Bh$ the variational formulas on the space $\mathcal{H}_{\hat g}(N,\dots,N)$ look as follows  (\ref{varBxy2}):
\be
\f{\p \Bh (x,y)}{\p  \int_{s_i}v}\Big|_{z(x),z(y)}= \f{1}{2\pi \sqrt{-1} }\int_{t\in s_i^*} \f{\Bh(x,t) \Bh(t,y)}{v(t)} 
\la{varforHn}
\ee
The   cycles $s_i^*$ form the  basis in $H_1(\Ch\setminus \{x_i\}_{i=1}^{N(2g-2)-1})$ which is dual to the basis $\{s_i\}$
($s_i^*\circ s_j=\delta_{ij}$); we have $\{s_i\}=\{-b_j, a_j\}_{j=1}^{\gh},\,\{r_i\}_{i=1}^{N(2g-2)-1}$ where $r_i$ is the small 
positively oriented cycle around $x_i$.

Let now  the curve $\Ch$ be defined by the equation $v^N=W$.  
We shall require the following correspondence between Torelli markings of $\Ch$ and $\CC$.

Choosing a Torelli marking of $\CC$ consider the set of 
$g$ contours on $\CC$ representing 
$a$-cycles 
$(a_1,\dots,a_g)$.  Consider their lifts to $\Ch$ (each $f_* [a_j]$ is a system of non-intersecting closed loops on $\Ch$) and require the Torelli marking of $\Ch$ to be chosen such that  each cycle from $f_* [a_j]$ belongs to the Lagrangian subspace in $H_1(\Ch,\Z)$ generated by the cycles $a_1,\dots,a_{\gh}$.
Under such agreement between the Torelli markings we have the identity
\be
f^*_x f^*_y B(x,y)= \Bh(x,y)+\mu^*_y\Bh(x,y)+\dots +(\mu^*_y)^{n-1}\Bh(x,y)
\la{ave1}
\ee
(here $f:\Ch\to \CC$ is the natural projection and $\mu$ is the $\Z_N$ automorphism of $\Ch$)
which can be verified by comparing the singularity structure and normalization of both sides.
Further averaging of (\ref{ave1}) over  the action of $\mu^*$ on $x$ we get
\be
f^*_x f^*_y B(x,y)= \f{1}{N}\sum_{l,k=1}^{N-1} (\mu^*_x)^k (\mu^*_y)^l\Bh(x,y)\;.
\la{ave2}
\ee

Now choosing $s\in \Hcal_{1}$ and the restricting variational formulas (\ref{varforHn}) to  $\Mgn$ we get

$$
\f{\p \Bh (x,y)}{\p  (\int_{s}v)}= \f{1}{2\pi \sqrt{-1} }
\sum_{k=1}^{\gh} \left[-\f{\p (\int_{\ah_k}v)}{\p  (\int_{s}v)}\int_{t\in \bh_k} \f{\Bh(x,t) \Bh(t,y)}{v(t)} \right.
$$
$$
\left.+
\f{\p (\int_{\bh_k}v)}{\p  (\int_{s}v)}\int_{t\in \ah_k} \f{\Bh(x,t) \Bh(t,y)}{v(t)} \right]
$$
\be
+\f{1}{2\pi \sqrt{-1} }\sum_{j=1}^{(N-1)(2g-2)-1}\f{\p z_j}{\p  (\int_{s}v)}\int_{t\in r_j}\f{\Bh(x,t) \Bh(t,y)}{v(t)}\;.
\la{varBhvs}
\ee
Furthermore, averaging both sides of (\ref{varBhvs}) as in (\ref{ave2}) we get
$$
\f{\p B (x,y)}{\p  (\int_{s}v)}= \f{1}{2\pi \sqrt{-1} N}
\sum_{k=1}^{\gh} \left[-\f{\p (\int_{\ah_k}v)}{\p  (\int_{s}v)}\int_{t\in \bh_k} \f{f^*_t[B(x,t) B(t,y)]}{v(t)} \right.
$$
$$
\left.+
\f{\p (\int_{\bh_k}v)}{\p  (\int_{s}v)}\int_{t\in \ah_k}  \f{f^*_t[B(x,t) B(t,y)]}{v(t)} \right]
$$
\be
+\f{1}{2\pi \sqrt{-1} }\sum_{j=1}^{(N-1)(2g-2)-1}\f{\p z_j}{\p  (\int_{s}v)}\int_{t\in r_j} \f{f^*_t[B(x,t) B(t,y)]}{v(t)}\;.
\la{varBhvs1}
\ee
Each integral in the  sum over $j$ in (\ref{varBhvs1}) vanishes since the contour $r_j$ is invariant under the action of $\mu_*$ while the integrand
gets multiplied with $\rho^{-1}$ under the action of $\mu^*$.

The first sum in (\ref{varBhvs1}) can be computed by introducing a symplectic basis $\{\hat{s}_j\}_{j=1}^{2\gh}$ in $H_1(\Ch,\R)$ such that
each $\hat{s}_j$ belongs to some of eigenspaces of $\mu_*$. Since the transformation from the basis $(\ah_j,\bh_j)$ to the basis
$\{\hat{s}_j\}$ is $Sp(2\gh,\R)$  this sum equals to the right hand side of (\ref{varBxy5}).

$\Box$

\end{document}